\documentclass{aa}

\usepackage[version=4]{mhchem}
\usepackage[varg]{txfonts}
\usepackage{graphicx}

\usepackage[colorlinks,bookmarks]{hyperref}
\usepackage{color}
\definecolor{linkblue}{rgb}{0,0,0.8}
\definecolor{linkgreen}{rgb}{0,0.5,0}
\hypersetup{pdfpagemode=UseNone, pdfstartview=FitH, linkcolor=linkblue,citecolor=linkblue, urlcolor=linkblue}

\newcommand{\be}{\begin{equation}}
 \newcommand{\ee}{\end{equation}}
\def \dd{\mathrm{d}}

\usepackage{booktabs}
\newcommand{\angstrom}{\mbox{\normalfont\AA}}

\def\hzero    {{\rm{H^0}}}
\def\hezero    {{\rm{He^0}}}
\def\hi    {{\rm{H \, \textsc{i}}}}
\def\hii   {\rm{H \, \textsc{ii}}}
\def\hisub {\rm{H \scriptscriptstyle I}}
\def\hiisub {\rm{H \scriptscriptstyle II}}
\def \cii {\rm{C \, \textsc{ii}}}
\def \czero  {{\rm{C^0}}}
\def \oi {\rm{O \,  \textsc{i}}}
\def\nh {n_{\rm{H}}}
\def\nhi {n_{\hisub}}
\def\Nhi {N_{\hisub}}
\def\nhii {n_{\hiisub}}
\def \halpha {\rm{H} \alpha}
\def \lyc {\rm{LyC}}
\def \lya {\rm{Ly\alpha}}
\def \lyb {\rm{Ly\beta}}
\def \ciiline {\rm{C \, \textsc{ii} \, \lambda 1334}}
\def \ciistar {\rm{C \,  \textsc{ii}^{\star} \, \lambda 1335}}
\def \oilinesiii {\rm{O \, \textsc{i} \, \lambda 1302}} 
\def \mgiiline {\rm{Mg \, \textsc{ii} \, \rm \lambda \lambda 2796,2803}}
\def \aliiline {\rm{Al \, \textsc{ii} \, \lambda 1670}}
\def \siii {\rm{Si \,  \textsc{ii}}}
\def \siiistar {\rm{Si \,  \textsc{ii}^{\star} \, \lambda 1265}}
\def \siiiline {\rm{Si \,  \textsc{ii} \, \lambda} 1260}
\def \siiilines {\rm{Si \,  \textsc{ii}} \, (\lambda 1190, \, \lambda 1193, \, \lambda 1260, \, \lambda 1304)}
\def \feiilines {\rm{Fe \,  \textsc{ii}} \, (\lambda 2344, \, \lambda 2374, \, \lambda 2382, \, \lambda 2586, \, \lambda 2600)}
\def\Msun {\rm{M}_{\rm{\odot}}}
\def\Zsun {\rm{Z}_{\rm{\odot}}}
\def\percs {\rm{cm}^{-2}}                                       % cm-2
\def\percc {\rm{cm}^{-3}}                                       % cm-3
\def\ev {\rm{eV}}                                                   % eV
\def\kms {\rm{km} \, \rm{s}^{-1}}                           % km/s
\def\Mstar {M_{*}}
\def \Mgas {M_{\rm{gas}}}
\def \Rvir {R_{\rm{vir}}}
\def \ewabs{\rm{EW_{abs}}}
\def \ewfluo{\rm{EW_{fluo}}}
\def \vmax{v_{\rm{max}}}
\def \vcen{v_{\rm{cen}}}
\def \vnine{v_{90}}
\def \fchi{f_C^{\hi}}
\def \tenth{$10^{\rm{th}} \,$}
\def \nineteenth{$90^{\rm{th}} \,$}
\def\fesc {f_{\rm{esc}}}
\def\fesclimit {f_{\rm{esc}}^{900}}
\def \ndust {n_{\mathrm{dust}}}

\usepackage{natbib}
\bibpunct{(}{)}{;}{a}{}{,} % to follow the A&A style

\begin{document}

% \thanks{} \fnmsep 
\title{UV absorption lines and their potential for tracing the Lyman continuum escape fraction} 

%\subtitle{II. An example text with infinitesimal scientific value\\  whose title and subtitle may also be split} 

\author{V. Mauerhofer
  \inst{1,2}\fnmsep\thanks{\email{valentin.mauerhofer@unige.ch}}
  \and A. Verhamme \inst{1,2}
  \and J. Blaizot\inst{2}
  \and T. Garel\inst{1,2}
  \and T. Kimm\inst{3}
  \and L. Michel-Dansac\inst{2}
  \and J. Rosdahl\inst{2}}

 %\offprints{V. Mauerhofer, \email{valentin.mauerhofer@unige.ch}

 \institute{Observatoire de Genève, Université de Genève, Chemin Pegasi 51, 1290 Versoix, Switzerland
  \and Univ Lyon, Univ Lyon1, ENS de Lyon, CNRS, Centre de Recherche Astrophysique de Lyon UMR5574, 69230 Saint-Genis-Laval, France
  \and Department of Astronomy, Yonsei University, 50 Yonsei-ro, Seodaemun-gu, Seoul 03722, Republic of Korea}

\date{Received 16 September 2020  / Accepted 3 December 2020}

\abstract
{The neutral intergalactic medium above redshift $\sim 6$ is opaque to ionizing radiation, and therefore indirect measurements of the escape fraction of ionizing photons are required from galaxies of this epoch. Low-ionization-state absorption lines are a common feature in the rest-frame ultraviolet (UV) spectrum of galaxies, showing a broad diversity of strengths and shapes.
As these spectral features indicate the presence of neutral gas in front of UV-luminous stars, they have been proposed to carry information on the escape of ionizing radiation from galaxies.}
{We aim to decipher the processes that are responsible for the shape of the absorption lines in order to better understand their origin. We also aim to explore whether the absorption lines can be used to predict the escape fraction of ionizing photons.}
 {Using a radiation-hydrodynamical cosmological zoom-in
 simulation and the radiative transfer postprocessing code $\rm{\textsc{RASCAS}}$
 we generated mock $\ciiline$ and $\lyb$ lines of a virtual galaxy at $z = 3$ with $M_{1500} = -18.5$
 as seen from many directions of observation. We also computed the escape fraction of ionizing photons in those directions and looked for correlations between the escape fraction and properties of the absorption lines, in particular their residual flux.}
 {We find that the resulting mock absorption lines are comparable to observations and that the lines and the escape fractions vary strongly depending on the direction of observation. 
 The effect of infilling due to the scattering of the photons and the use of different apertures of observation both result in either strong or very mild changes of the absorption profile.
 Gas velocity and dust always affect the absorption profile significantly.
 We find no strong correlations between observable $\lyb$ or $\ciiline$ properties and the escape fraction. 
 After correcting the continuum for attenuation by dust to recover the intrinsic continuum, the residual flux of the $\ciiline$ line correlates well with the escape fraction for directions with a dust-corrected residual flux larger than $30\%$. For other directions, the relations have a strong dispersion, and the residual flux overestimates the escape fraction for most cases. Concerning $\lyb$, the residual flux after dust correction does not correlate with the escape fraction but can be used as a lower limit.
 }
   {}

\maketitle 

%%%%%%%%%%%%%%%%%%%%%%%%%%%%%%%%%%%%%%%%%%%%%%%%%%%%%%%%%%%%%%%%%%%%%%%%%%%%%%%%%%%%
\section{Introduction}

Decades of observations show that the intergalactic medium (IGM) was completely ionized by redshift $z \sim 6$, about one billion years after the Big Bang \citep[e.g.,][]{Fan06, Schroeder13, 2015MNRAS.453.1843S, 2018Natur.553..473B, 2018PASJ...70...55I, 2018PASJ...70S..13O,Bosman18,Kulkarni19}. This implies that, before this redshift, sources of ionizing radiation must have emitted enough photons that managed to escape their galactic environment and reach the IGM to photoionize it. The main candidates for the source of reionization are active galactic nuclei (AGNs) and massive stars. Although recent work suggests AGNs could be an important driver of reionization \citep{Madau15,Giallongo19}, a consensus seems to be emerging on the idea that their contribution is negligible compared to that of stars \citep{Grissom14, Parsa18, Obelisk}.

Models based on observations show that stars alone could drive reionization under reasonable assumptions on the Lyman continuum ($\lyc$) production efficiency and on the escape fraction of ionizing photons ($f_{\rm{esc}}$), for example in \cite{Atek15}, \cite{Gnedin16} and \cite{Livermore17}. The escape fraction is the least constrained quantity, and is often fixed to a value of between 10 and 20$\%$ \citep[e.g.,][]{Robertson15}. \cite{Finkelstein19} argue that reionization can also be explained when assuming lower escape fractions. Additionally, simulations have successfully reionized the Universe with stellar radiation, either at very large scales \citep[e.g.,][]{Iliev14,Ocvirk16}, or at smaller scales with high enough resolution to resolve the escape of ionizing radiation from galaxies \citep[e.g.,][]{Kimm17,Sphinx}.

In order to better understand the process of reionization and to be able to compare observations and simulations, it is essential that we obtain accurate constraints on the $\lyc$ escape fractions of observed galaxies. Recently, several low-redshift galaxies were found to be leaking ionizing photons \citep[e.g.,][]{Bergvall06,Leitet13,Leitherer16,Puschnig17}. When this ionizing radiation flux is detected, it is possible to use a direct method to infer the escape fraction \citep[e.g.,][]{Borthakur14,Izotov16a,Izotov16b}. The total $\lyc$ production is computed as the addition of the observed $\lyc$ photons and the ones absorbed in the galaxy, deduced from the flux of a (dust corrected) Balmer line, assuming that this line is produced by recombinations of photoionized hydrogen ions. The ratio of the observed ionizing flux to the assessed intrinsic $\lyc$ production gives the escape fraction.

However, the ionizing radiation escaping galaxies during the epoch of reionization is not directly observable because of IGM absorption, and therefore this method cannot be used.
To circumvent this limitation, different ways to infer the escape fractions indirectly have been proposed in the literature. The connections between Lyman $\alpha$ ($\lya)$ and Lyman continuum emission have been studied \citep{Verhamme15,Dijkstra16,Kimm19} with promising results at $z \sim 3$ \citep{Steidel18} or in the local Universe \citep{Verhamme17,Izotov18b}. Above $z \sim 6$, the intergalactic neutral hydrogen may alter $\lya$ too much, except for galaxies that have such big  ionized bubbles that $\lya$ is redshifted enough before reaching the neutral IGM \citep{Dijkstra14,Mason18b,Gronke20}. A possible alternative is to use $\mgiiline$ \citep{Henry18}, a resonant emission line doublet which is a candidate proxy of $\lya$ and has the advantage of not being absorbed by the neutral IGM of the epoch of reionization.
A second method is to use O32, which is defined as the line ratio $[\mathrm{O} \, \textsc{iii} \, \lambda 5007]/[\mathrm{O} \, \textsc{ii} \, \lambda \lambda 3726,3729]$, and is thought to correlate with the escape fraction \citep{Jaskot13, Nakajima14, Izotov16b}. However, recent studies shed doubt on this method \citep{Izotov18b, Bassett19, Katz20}. A high O32 ratio seems to be a good indicator of a nonzero escape fraction, but the correlation between the two quantities is weak.

A third way to infer the escape fractions
indirectly is to use down-the-barrel absorption lines from low-ionization states (LISs) of metals, such as $\cii$, $\oi$ or $\siii$ lines.
The shapes of these lines are used as an indicator of the covering fraction of the absorber.
Indeed, column densities of metals in galaxies typically lead to saturated absorption lines (i.e., absorption profiles where the minimum flux reaches zero). When nonsaturated lines are observed, the residual flux is interpreted as being due to partial coverage of the sources by the absorbing material, meaning that the ratio of this residual flux to the continuum is used to deduce an escape fraction. As the LISs of metals are thought to be good tracers of neutral hydrogen, this escape fraction is then linked to the escape fraction of ionizing photons \citep[e.g.,][]{Erb15, Steidel18, Gazagnes18, Chisholm18}

There have not yet been observations of absorption lines at the epoch of reionization because it is challenging to detect the stellar continuum with a sufficiently high signal-to-noise ratio. However, there has been impactful progress in the study of absorption lines for two decades thanks to the observation of galaxies at ever higher redshifts and of local analogs of high-redshift galaxies. Detections of down-the-barrel absorption lines of individual galaxies with the highest redshifts, with $z \sim$ 3--4, are carried out using gravitational lenses \citep[e.g.,][]{Smail07,Mirka10,Jones13,Patricio16}. Alternatively, without gravitational lensing, it is possible to study absorption lines at these redshifts by stacking spectra of many galaxies \citep[e.g.,][]{Shapley03,Steidel18,Feltre20}. The most common lines observed at these high redshifts are $\siiilines$, $\oilinesiii$ or $\ciiline$, which are strong lines whose observed wavelengths fall in the optical range. There are also observations of slightly redder absorption lines, such as $\feiilines$ or $\mgiiline$, at intermediate redshifts between 0.7 and 2.3 \citep[e.g.,][]{Finley17,Feltre18}. Finally, there are many studies of absorption lines at $z<0.3$ \citep[e.g.,][]{Rivera15,Chisholm17,Chisholm18,Jaskot19}.

Because most atoms and ions in the interstellar medium (ISM) are in their ground state, absorption lines are typically transitions from the ground state to a higher level. The energy of these transitions translates into lines which are mostly in the rest-frame ultraviolet (UV). Some lines are resonant (e.g., $\lya$, $\mgiiline$, $\aliiline$), but most are not completely resonant because the ground state is split into different spin levels, and de-excitation may produce photons with different wavelengths. If the wavelength of the so-called fluorescent channel(s) is different enough from the resonant wavelength the emitted photon will leave the resonance.

Observed absorption lines show a large diversity of spectral profiles, for example in terms of depth, width, and complexity. A common feature is that absorption is often blueward of the systemic velocity, with a minimum of the absorption occurring between around -400 and 100 $\kms$. There are also detections of fluorescent emission redward of the absorption \citep[e.g.,][]{Shapley03,Rivera15,Finley17,Steidel18,Jaskot19}. 

There are several physical processes that challenge the use of absorption lines to measure the covering fraction of neutral gas and the escape fraction of ionizing photons. One of them is the distribution of the velocity of the gas at galaxy scales. The gas in front of the sources moves at different velocities, leading to an absorption that is spread in wavelength. This can lead to an increase of the flux at the line center compared to a case where all the gas has the same velocity \citep{Rivera15}. Another process is scattering: when photons are absorbed by an ion they are not destroyed but scattered. Thus, the flux in absorption lines can be increased by photons that were initially going away from the observer but that end up in the line of sight because of resonant scattering \citep{Prochaska11,Scarlata15}.
 
This paper investigates the relation between the escape fraction of ionizing photons and LIS absorption lines using mock spectra constructed from a radiation-hydrodynamic (RHD) simulation. There have been studies of various emission lines by post-processing simulations \citep[e.g.,][]{Barrow17,Barrow18,Katz19,Pallottini19,Katz20,Corlies20}, and there are also studies on absorption lines in simulations, but these latter focus on quasar sightline absorption instead of down-the-barrel absorption \citep[e.g.,][]{Hummels17,Peeples19}. Concerning down-the-barrel lines, studies have been done using Monte-Carlo methods in idealized geometries \citep{Prochaska11} or using semi-analytic computations \citep{Scarlata15,Carr18}. Finally, there have been studies of mock down-the-barrel absorption lines in simulations \citep[]{Kimm10}, but this present work is to our knowledge the first time that such lines are produced in a high-resolution RHD simulation and including resonant scattering. We focus in particular on the $\ciiline$ line, which is among the strongest LIS absorption redward of $\lya$, is not contaminated by absorption lines of other abundant elements (unlike $\oi$ lines), and is frequently used in the literature \citep[e.g.,][]{Heckman11,Rivera15,Steidel18}. We also show results using the $\siiiline$ line. In addition, we study $\lyb$ to see if it is in general a better tracer
of the escape fraction, even though it is not observable at the epoch of reionization because its wavelength is $1026 \, \angstrom$.

The paper is structured as follows: In $\rm{Sect.} \,  \ref{sec:method}$ we provides details of our method, from the simulation that we use to the post-processing that is necessary to make mock observations and to compute escape fractions of ionizing photons. In $\rm{Sect.} \,  \ref{sec:spectra}$ we present the resulting spectra and we assess their robustness against changes in modeling parameters. We also study the effect that various physical processes have on the spectra. In $\rm{Sect.} \,  \ref{sec:fesc}$ we show the values of the escape fractions, and discuss the effects of helium and dust. In $\rm{Sect.} \,  \ref{sec:interp}$ we compare the line properties with the escape fractions to try to find correlations. We also explain the different sources of complexity of the relations we find. Finally, we summarize our results in $\rm{Sect.} \, \ref{sec:summary}$.

%%%%%%%%%%%%%%%%%%%%%%%%%%%%%%%%%%%%%%%%%%%%%%%%%%%%%%%%%%%%%%%%%%%%%%%%
\section{Methods} \label{sec:method}

In this section we provide details of the simulation that we use and the steps that are necessary to build the mock observations. In this paper we focus on the $\ciiline$ absorption line but the method can be applied to any absorption line. In $\rm{Sect.} \, \ref{sec:siii}$ we show the main results of the paper for $\siiiline$. Here we use one galaxy from a zoom-in simulation, and in future work we will study a statistical sample of galaxies from various simulations like \textsc{Sphinx}\footnote{\url{https://sphinx.univ-lyon1.fr}} \citep{Sphinx} or Obelisk \citep{Obelisk}.

%-----------------------------------------------------------------------
\subsection{Simulation} \label{subsec:sim}

We use a zoom-in simulation run with the adaptive mesh refinement (AMR) code \textsc{Ramses-RT} \citep{Ramses,Joki2013,RamsesRT}, including a multi-group radiative transfer algorithm using a first-order moment method which employs the M1 closure for the Eddington tensor \citep{Levermore84}. The collisionless dark matter and stellar particles are evolved using a particle-mesh solver with cloud-in-cell interpolation \citep{Guillet11}. We compute the gas evolution by solving the Euler equations with a second-order Godunov scheme using the HLLC Riemann solver \citep{Toro94}.
The initial conditions are generated by MUSIC \citep{MUSIC} and chosen such that the resulting galaxy at $z=3$ has a stellar mass of around $10^9 \, \Msun$.
The physics of cooling, supernova feedback, and star-formation are the same as in the \textsc{Sphinx} simulations \citep{Sphinx}.
Radiative transfer allows for a self-consistent and on-the-fly propagation of ionizing photons in the simulation, which provides an accurate nonequilibrium ionization state of hydrogen and helium, as well as radiative feedback. 
We use three energy bins for the radiation: The first bin contains photons with energies between $13.6  \, \ev$ and $24.59  \, \ev$, which ionize $\hzero$, the second bin between $24.59  \, \ev$ and $54.42  \, \ev$, which ionize $\hzero$ and $\hezero$, and the last above $54.42  \, \ev$, which ionize $\hzero$, $\hezero$ and $\rm{He}^+$. The ionizing radiation is emitted from the stellar particles following version 2.0 of the BPASS\footnote{\url{https://bpass.auckland.ac.nz/index.html}} stellar library spectral energy distributions (SEDs) \citep{BPASS1, BPASS2}. In this version of BPASS the maximum mass of a star is $100 \, \rm{\Msun}$, and all stars are considered to be in binaries. 
To account for the ionizing radiation produced by external galaxies that are not present in the zoom-in simulation, we add an ionizing UV background (UVB) to all cells of the simulation with $\nh < 10^{-2} \, \percc$, following \cite{UVB_simu}.
The outputs of the simulation have a time resolution of 10 Myr, going down to redshift $z = 3$. The maximum cell resolution around $z=3$ is 14 pc. The halo mass, stellar mass, stellar metallicity, and gas metallicity of the galaxy at $z = 3$ are $M_h=5.6 \times 10^{10} \, \Msun$, $\Mstar=2.3 \times 10^9 \, \Msun$, $Z_{\star}=0.42 \, \Zsun,$ and $Z_{\rm{gas}}=0.43 \, \Zsun$, respectively.

\begin{figure}
  \resizebox{\hsize}{!}{\includegraphics{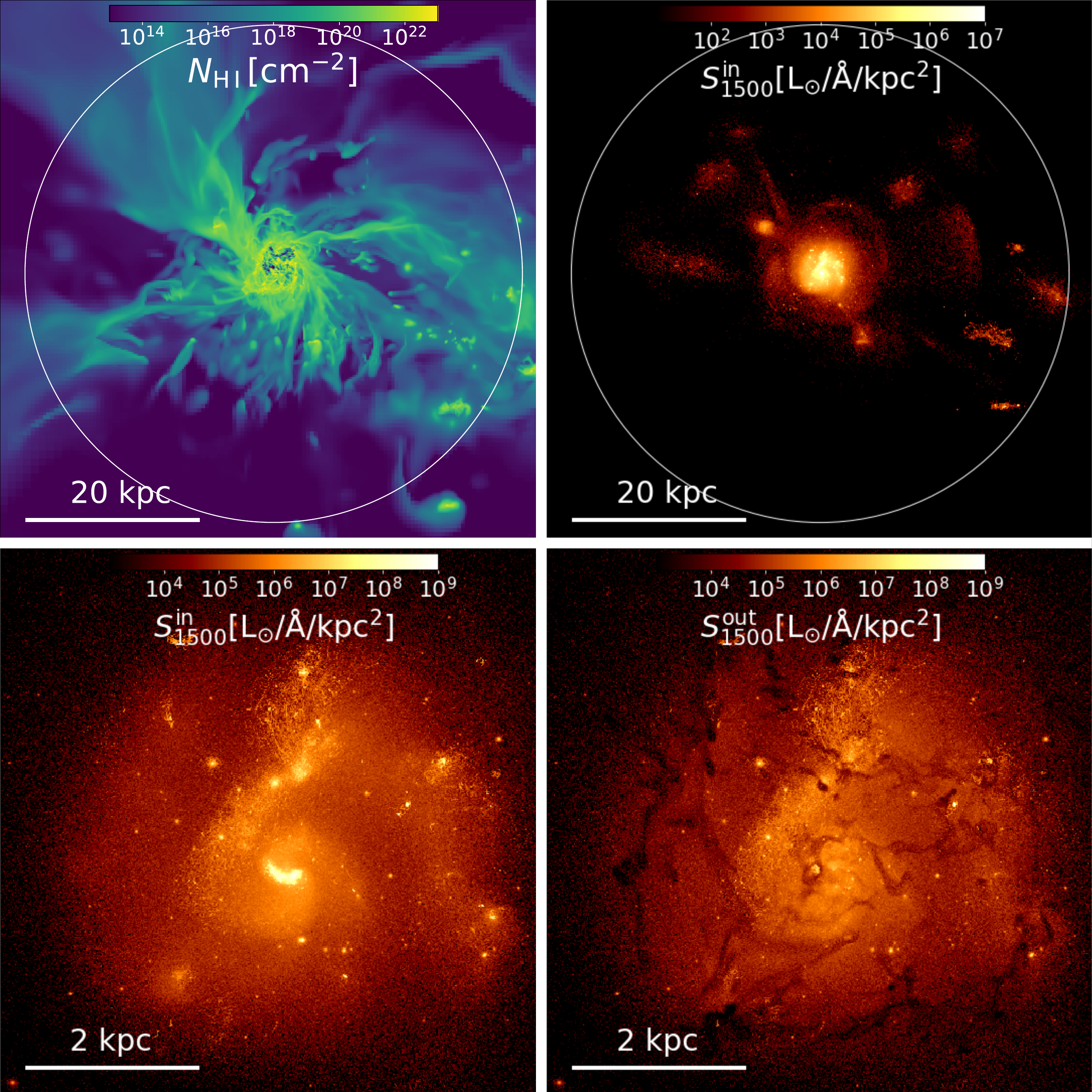}}
  \caption{Maps of neutral hydrogen column density and surface brightness at $1500 \, \angstrom$ for a snapshot of the simulated galaxy at $z=3$. Upper left: Column density of neutral hydrogen atoms, with the virial radius highlighted by the white circle. Upper right: Intrinsic surface brightness at $1500 \, \angstrom$. Lower left: Zoom-in to a tenth of the virial radius. Lower right: Surface brightness at $1500 \, \angstrom$ after extinction by dust.}
  \label{fig:164_visu}
\end{figure}

In this paper we focus on three outputs of the simulation that span a range of ionizing photon escape fractions; we call these outputs A, B, and C. There are roughly 80 Myr of separation between each output. Output C is the last of the simulation, at $z=3$. Properties of the outputs, including the star formation rate (SFR), are shown in $\rm{Table} \, \ref{table:outputs}$. In $\rm{Fig.} \, \ref{fig:164_visu}$ we plot the neutral hydrogen column density and the stellar surface brightness at 1500 $\angstrom$ at the virial radius scale and the ISM scale (a tenth of the virial radius), for output C. The column density map in the upper left panel shows gas that is being accreted from the IGM, and regions where the supernova feedback disrupts the gas flows. The upper right panel shows that there are many small satellites orbiting the galaxy. The lower panels highlight the high resolution of the simulation, where small star-forming clusters and dust lanes are well resolved.

\begingroup
\setlength{\tabcolsep}{6pt} % Default value: 6pt
\renewcommand{\arraystretch}{1.3} % Default value: 1
\begin{table*}
  \caption{Properties of the simulated galaxy at three different times. All the masses are computed inside the virial radius and the metallicities in a sphere of a tenth of the virial radius. The SFR is averaged over the last 10 Myr. $ \lyc \; f_{\rm{esc}}$ is the escape fraction of hydrogen ionizing photons computed at the virial radius ($\rm{Sect.} \, \ref{sec:compute_fesc}$). $\rm{M}_{1500}$ is the absolute magnitude of the galaxy at 1500 $\angstrom$, including dust attenuation. $A_{1500}$ represents the attenuation by dust, following the equation: $F_{1500}^{\rm{observed}} = 10^{-0.4 \, A_{1500}} \times F_{1500}^{\rm{intrinsic}}$. The escape fractions, magnitudes, and attenuation by dust, which are the only three quantities depending on the direction of observation, were computed for 1728 isotropically distributed directions. We show the mean of the results and the \tenth and \nineteenth percentiles.}
  \label{table:outputs}
   \centering
  \begin{tabular}{l|ccccccccccc}
    \toprule
    Output  & $z$ & $\Rvir$ & $M_h$ & $\Mgas$ & $\Mstar$ & $Z_{\rm{gas}}$ & $Z_{*}$ & $\rm{SFR}$ &  $\lyc \; f_{\rm{esc}}$ & $\rm{M}_{1500}$  & $A_{1500}$ \\ 
            &      & [kpc]   & [$\Msun$]   & [$\Msun$]   & [$\Msun$] & [$\Zsun$] & [$\Zsun$] & [$\Msun / \rm{yr}$] & [$\%$] & [mag] & [mag]   \\
    \midrule
    A   & 3.2 & $27$    & $5.09 \times 10^{10}$  & $5.15 \times 10^9$ & $1.88 \times 10^9$ & 0.40 & 0.36 & $5.0$ &  $11.0^{+23.9}_{-10.5}$ & $-19.0^{+1.0}_{-0.5}$ & $1.3^{+ 1.1}_{-0.5}$ \\
    B   & 3.1 & $28$    & $5.30 \times 10^{10}$  & $5.13 \times 10^9$ & $2.08 \times 10^9$ & 0.43 & 0.40 & $2.2$ & $3.8^{+8.2}_{-3.7}$  & $-18.2^{+0.6}_{-0.4}$ & $1.6^{+0.6}_{-0.4}$ \\
    C   & 3.0 & $29$    & $5.56 \times 10^{10}$  & $4.83 \times 10^9$ & $2.27 \times 10^9$ & 0.43 & 0.42 & $4.2$  & $1.0^{+2.2}_{-1.0}$ & $-18.5^{+0.5}_{-0.4}$ & $1.5^{+0.5}_{-0.4}$ \\
    \bottomrule
  \end{tabular}
\end{table*}
\endgroup

%-------------------------------------------------------------------------------
\subsection{Computing densities of metallic ions} \label{sec:comp_densities}

To be able to make mock observations of metallic absorption lines in the simulation, we have to compute the density of the ions of interest in post-processing. The cosmological zoom-in simulation does not trace densities of elements other than hydrogen and helium, but it follows the value of the metal mass fraction $Z$ in every cell. Assuming solar abundance ratios, we obtain the carbon density via the equation $n_{C} = n_{H} A_{C}^{\odot} Z/Z_{\odot}$, where $n_H$ is the hydrogen density, $A_{C}^{\odot} = 2.69 \times 10^{-4}$ is the solar ratio of carbon atoms over hydrogen atoms, and $Z_{\odot} = 0.0134$ is the solar metallicity \citep{Grevesse10}. We do not consider the depletion of carbon into dust grains, which could alter the densities in the gas phase. However, \cite{Jenkins09} finds that, in the Milky Way (MW), not more than $20$--$50 \%$ of the carbon is depleted into dust. Moreover, \cite{DeCia18} shows that the depletion of metals into dust is less important in high-redshift galaxies than in the MW, which is why we can assume that the depletion does not change the gas-phase carbon density significantly. In order to assess the impact of the uncertainties of abundance ratios and dust depletion on our conclusions, we test the effect of dividing and multiplying the density of $\rm{C}^+$ by two in $\rm{Sect.} \,  \ref{sec:robust}$. The effect is visible but mild.

What remains is to compute the ionization fractions of carbon to get the densities of $\rm{C}^+$ . To do so we use \textsc{Krome}\footnote{\url{http://kromepackage.org}} \citep{2014MNRAS.439.2386G}, which is a code that computes the evolution of any chemical network implemented by the user. As this computation is done in post-processing, we need to assume equilibrium values for the ionization fractions of carbon; we therefore use a routine of \textsc{Krome} that evolves the chemical network until it reaches equilibrium. For hydrogen and helium we use the (nonequilibrium) ionization fractions from the simulation.

The ionization fractions are determined by three main processes: recombination, collisional ionization, and photoionization. There can also be charge-transfer reactions between carbon ions and hydrogen or helium ions, but we omit them in this work, as we explain below. One of the major advantages of \textsc{Krome} is that the chemical network is completely customizable. The user chooses all the reactions and their rates as a function of temperature. We use the following rates:
\begin{itemize}
\item Recombination rates: \cite{BadnellRR}.
\item Collisional ionization rates: \cite{Voronov97}.\item Photoionization cross-sections: \cite{Verner96}.
\end{itemize}
We choose recombination and collisional ionization rates to reproduce the carbon ionization fractions as a function of temperature for a collisional ionization equilibrium setup in Cloudy\footnote{\url{https://nublado.org}} \citep{Cloudy}, which is the state-of-the-art tool to compute ionization fractions (among other things), but is too slow to use on a full simulation. There are works that use Cloudy on Ramses simulations, but these adopt strategies to avoid running it on every cell. \cite{Katz19} use Cloudy on a subset of cells and then extrapolate on the other cells using machine learning algorithms. \cite{Pallottini19} create Cloudy grids that are then interpolated for each cell of the simulation. We opt to use \textsc{Krome} because it is fast enough to directly compute the ionization fractions in every cell of the simulation, although it does not compute line emissivities of the gas.
The collisional ionization rates that we use are slightly different from the ones in Cloudy \citep{Dere07}. We show in $\rm{Appendix} \, \ref{app:comp_fractions}$ that we recover the same ionization fractions of carbon as a function of temperature as Cloudy, despite the fact that we lack charge-transfer reactions and have different collisional ionization rates.

Concerning photoionization, the cross-sections that we use are the same as in Cloudy. To obtain a photoionization rate, those cross-sections have to be multiplied by a flux of photons. Taking advantage of the radiation-hydrodynamics in the simulation, we directly use the inhomogeneous radiation field self-consistently computed by the simulation at energies higher than 13.6 eV. Below 13.6 eV, the simulation does not track the radiation field or its interaction with gas. We nevertheless need to account for this lower energy range as it may ionize metals (e.g., the ionizing potential of C is 11.26 eV). For this, we add a simple model for radiation in the Habing band, from 6 eV to 13.6 eV \citep{Habing68}, in post-processing, using the UVB as inferred by \cite{HM12}.
In $\rm{Sect.} \,  \ref{sec:uvb}$ we look at the effect of dividing the UVB by two or multiplying it by 100, and we see that this does not affect the results of the mock observations. We also use the \cite{HM12} UVB at all wavelengths instead of using the radiation field of the simulation, for comparison. In this case there can be drastic changes, which demonstrates that it is important to use a simulation that treats radiative transfer on-the-fly. In $\rm{Appendix} \, \ref{app:photorate}$ we show in more detail how we compute the photoionization rates using the \cite{Verner96} photoionization cross-sections and the number density of photons in the simulation.

Another advantage of \textsc{Krome} is that it is possible to fix the state of hydrogen and helium. If we compute the ionization fractions with Cloudy, the convergence to equilibrium of hydrogen and helium ionization fractions modifies the electron density compared to the value of the simulation. This different electron density would lead to different carbon ionization fractions than with the simulation electron density. To summarize, we compute the ionization fractions of carbon at equilibrium, but with the hydrogen and helium ionization fractions fixed at the simulation values. Therefore we set all rates of reactions containing hydrogen and helium to zero in \textsc{Krome}. This is why we do not include charge transfer reactions in \textsc{Krome}, as they would also change the simulation values of hydrogen and helium ionization fractions.

%-------------------------------------------------------------------------------
\subsection{Fine structure of the ground state} \label{sec:hyperfine}

\begin{figure}
  \resizebox{\hsize}{!}{\includegraphics{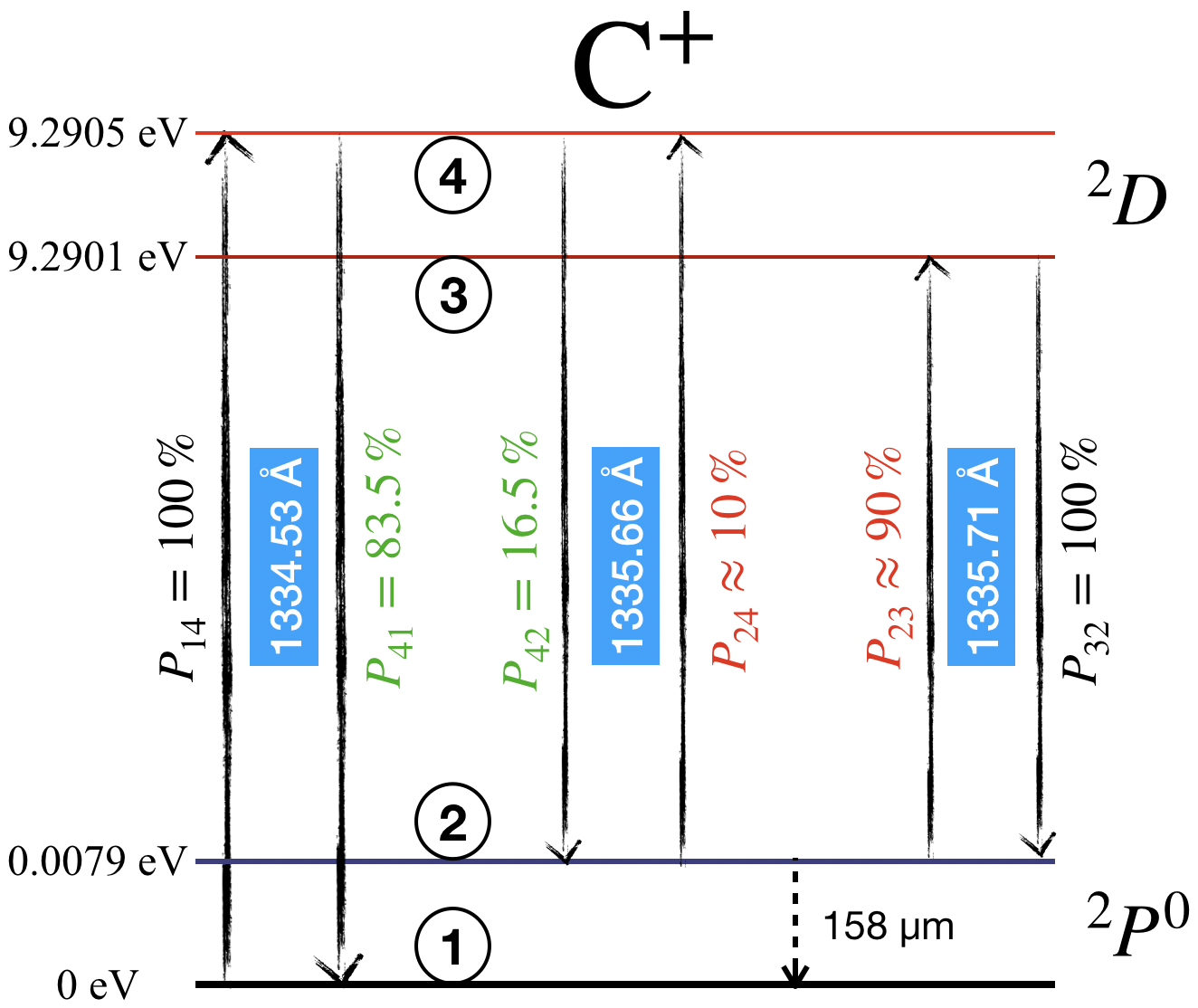}}
  \caption{Energy levels of the $\rm{C}^+$ ion. $P_{14}=100\%$ indicates that a $\rm{C}^+$ ion in level 1 can be photo-excited only to level 4. The green numbers are the probabilities that a $\rm{C}^+$ ion in level 4 radiatively de-excites to level 1 or 2. The red numbers are the probabilities that an incoming photon with a wavelength between $1335.66 \, \angstrom$ and $1335.71 \, \angstrom$ hitting a $\rm{C}^+$ ion in level 2 excites it to level 3 or 4. $P_{32}=100\%$ indicates that a $\rm{C}^+$ ion in level 3 can only radiatively de-excite to level 2.}
  \label{fig:CII_levels}
\end{figure}

Almost all ions have a ground state that is split into several spin states. The $\rm{C}^+$ ion has one such spin level, which we call level 2, just above the ground state (level 1), as illustrated in $\rm{Fig.} \, \ref{fig:CII_levels}$. A fraction of $\rm{C}^+$ ions are populating level 2 due to collisions with electrons, and they are absorbing photons at $1335.66 \, \angstrom$ or $1335.71 \, \angstrom$ instead of $1334.53 \, \angstrom$ from level 1. We take this effect into account by computing the percentage of ions that are in level 2 using PyNeb\footnote{\url{https://pypi.org/project/PyNeb/}} \citep{Pyneb}. For every cell of the simulation we give PyNeb the temperature (assumed to be both the gas temperature and the electron temperature) and the electron density taken from the results of \textsc{Krome}. The fraction of $\rm{C}^+$ ions populating level 3 and 4 is extremely small, due to the short lifetime of these levels, and so we consider that the ions only populate levels 1 and 2.

The three lines at 1334.53, 1335.66, and $1335.71 \, \angstrom$ are considered in our radiative transfer post-processing, as explained in $\rm{Sect.} \,  \ref{sec:RASCAS}$. The effects of these different channels on the $\ciiline$ line are studied in $\rm{Sect.} \,  \ref{sec:process}$.

%-------------------------------------------------------------------------------
\subsection{Radiative transfer of the lines} \label{sec:RASCAS}

The radiative transfer post-processing is computed with the code RAdiation SCattering in Astrophysical Simulations (\textsc{RASCAS}\footnote{\url{http://RASCAS.univ-lyon1.fr}}) \citep{Rascas}. The stellar luminosity of the simulated galaxy is sampled with one million so-called photon packets, each having a given wavelength and carrying one millionth of the total luminosity in the wavelength range considered (a few angstr\"oms on each side of the line). We launch the photon packets from the stellar particles of the simulation with the following initial conditions:

  \begin{itemize}
  \item Direction: The initial direction of every photon packet is randomly drawn from an isotropic distribution.
  \item Position: Each photon packet is randomly assigned to a stellar particle, with a weight proportional to the stellar luminosity in the wavelength range considered. The initial position of the packet is the position of the stellar particle. 
\item Frequency: The SED of the stellar particle, given by BPASS using the age and metallicity of the particle, is fitted by a power-law approximation, removing any stellar absorption lines. The frequency is then drawn randomly from this power law.
  \end{itemize}
Photon packets are then propagated through the AMR grid of the simulation. The total optical depth in each cell is the sum of four terms, corresponding to four channels of interaction: 
\be
\tau_{\rm{cell}} = \tau_{\ciiline} + \tau_{\cii^{\star} \, \lambda 1335.66} + \tau_{\cii^{\star} \, \lambda 1335.71} + \tau_{\rm{dust}}.
\ee
The optical depth of a line is the product of the cross-section and the column density, $\tau_{\rm{line}} = \sigma_{\rm{line}}  N_{\cii}$. The column density is the product of the ion density and the distance to the border of the cell\footnote{We use the density of $\rm{C}^+$ in level 1 for the line $\ciiline$, and the density of $\rm{C}^+$ in level 2 for $\cii \, \lambda 1335.66$ and $\cii \, \lambda 1335.71$ ($\rm{Sect.} \, \ref{sec:hyperfine}$).}, $N_{\cii} = n_{\cii} d$.
The cross-section for a given line of the ion is given by the formula
\be
\sigma_{\rm{line}} = \frac{\sqrt{\pi} e^2 f_{\rm{line}}}{m_e c} \frac{\lambda_{\rm{line}}}{b} \rm{Voigt}(x,a),
\ee
where $f_{\rm{line}}$ is the oscillator strength of the line, $\lambda_{\rm{line}}$ is the line wavelength, and $b$ is the Doppler parameter, which is defined below ($\rm{Equation} \, \ref{eq:vth}$).
The variable $a$ is defined by $a = A_{\rm{line}} \lambda_{\rm{line}} / (4 \pi b),$ where $A_{\rm{line}}$ is the Einstein coefficient of the line. The variable $x$ is defined by $x = (c/b) (\lambda_{\rm{line}} - \lambda_{\rm{cell}})/\lambda_{\rm{cell}},$ where $\lambda_{\rm{cell}}$ is the wavelength of the photon packet in the frame of reference of the cell. Finally, the Voigt function is computed with the approximation of \cite{Smith_Colt}.

When an interaction occurs, one of the four channels is randomly chosen with a probability $\tau_{\rm{channel}}/\tau_{\rm{cell}}$.  If the photon packet interacts with dust, it can either be scattered or absorbed (see $\rm{Sect.} \,  \ref{sec:tau_dust}$). If the photon packet is absorbed by $\rm{C}^+$ in any channel, it will be re-emitted in a direction drawn from an isotropic distribution. If the absorption channel is $1334.53 \, \angstrom$, the photon packet will be re-emitted either via the same channel, which is called ``resonant scattering'', or via the $1335.66 \, \angstrom$ channel, which is the fluorescent channel. The probabilities of the two channels are determined by their Einstein coefficients, and are shown on $\rm{Fig.} \, \ref{fig:CII_levels}$. Similarly, if the photon packet is absorbed by the channel $1335.66 \, \angstrom$, it can be re-emitted at $1334.53 \, \angstrom$ or at $1335.66 \, \angstrom$. Finally, if the photon packet is absorbed by the channel $1335.71 \, \angstrom$, it will always be re-emitted at $1335.71 \, \angstrom$. The transfer ends when all the photon packets either escape the virial radius or are destroyed by dust. More details, for example on the position of interaction or on the frequency redistribution after scattering, are given in \cite{Rascas}.

Mock observations are made from different directions of observation thanks to the peeling-off method \citep[e.g.,][]{Dijkstra_Saas}. This method allows us to make mock images, spectra, or data cubes. We can choose a spatial and spectral resolution and an aperture radius to make a mock observation of the whole galaxy or only of a part of it.

We also make mock $\lyb$ absorption lines. For this we use the density of neutral hydrogen directly from the simulation. No fine-structure level is taken into account because the resulting differences on the $\lyb$ wavelength would be of only $0.5 \, \rm m\angstrom$, which is too small for the photon to leave resonance. $\lyb$ is also a ``semi-resonant'' line: Photons at the $\lyb$ wavelength can scatter on $\hzero$, leading to potential infilling effects, and, like in the case of $\ciiline$, there is a way to leave the resonance, namely via an emission of an $\halpha$ photon. The probability for $\lyb$ to be resonantly scattered is $88.2\%$.

In this paper, the Doppler parameter b, which enters in the computation of the optical depth, depends on the thermal velocity and the turbulent velocity $v_{\rm{turb}}$:
\be \label{eq:vth}
b = \sqrt{\frac{2 k_b T}{m_{\rm{ion}}} + v^2_{\rm{turb}}}.
\ee
The turbulent velocity increases the probability of interaction of photons with wavelengths away from line center due to gas moving along or opposite to the direction of propagation, and therefore broadening the effective cross-section. This is important due to two limitations of the simulations that challenge a completely self-consistent treatment of absorption studies. The first limitation is the fact that the simulated galaxy is discretized on a grid, which means there are discrete jumps in the gas velocity from one cell of the grid to the next. Therefore, for some conditions, a photon packet with a wavelength corresponding to the velocity $v + \frac{\Delta v}{2}$ could go through two cells with projected velocities $v$ and $v + \Delta v$ without being absorbed, while it would be absorbed at intermediate velocities if the grid was more refined. The turbulent velocity is a way to smooth out those discontinuities. It is possible to estimate this quantity in the simulation by comparing the velocity in a cell with the velocities of neighboring cells, as is done for the star formation criteria of \cite{Kimm17} or \cite{Sphinx}. Doing so in the three outputs of $\rm{Table} \, \ref{table:outputs}$ yields an average $n_{\hi}$-weighted turbulent velocity of approximately $12 \, \kms$.

The second limitation of simulations is the fact that there are always gas motions at scales unresolved by the simulation, which would also increase the probability that photons get absorbed. This is called ``subgrid turbulent velocities'', which is not taken into account in our simulation or in the previous computation of the turbulent velocity based on neighboring cells. Some other \textsc{Ramses} simulations quantify those subgrid motions on-the-fly \citep[e.g.,][]{Agertz15,Kretschmer20}. 
In this work we assume a uniform turbulence in every cell of the simulation, accounting for those two kinds of turbulence. We choose a fiducial value of $v_{\rm{turb}} = 20 \, \kms$.  In $\rm{Sect.} \,  \ref{sec:robust}$ we investigate the effects of changing this parameter.

%......................................................................................
\subsection{Dust model} \label{sec:tau_dust}

As described in \cite{Rascas}, we follow \cite{Laursen09b} to model the effect of dust on radiation. This is a post-processing method, because the dust is not followed directly in the simulation. In this model, the optical depth of dust is the product of a pseudo-density and a cross-section. The pseudo-density is
\be \label{eq:ndust}
\ndust = \frac{Z}{Z_0} \times (\nhi + f_{\rm{ion}} \nhii), 
\ee
where $Z_0$ is a reference metallicity and $f_{\rm{ion}}$ sets the quantity of dust that is put in ionized regions.
Throughout this paper we use $Z_0 = 0.005 $ and $f_{\rm{ion}} = 0.01$, corresponding to the calibration on the Small Magellanic Cloud (SMC) of \cite{Laursen09b}.

For the cross-section, we use the fits of \cite{Gnedin08} in the SMC case. This cross-section accounts for both absorption and scattering events. \textsc{RASCAS} has an albedo parameter, representing the probability that a photon-packet that interacts with dust is scattered instead of destroyed. For this albedo we use the value of \cite{Li01}, which is 0.276 for $\lyb$ and 0.345 for $\ciiline$. Finally, the asymmetry parameter $g$ of the Henyey-Greenstein function \citep{Henyey41}, which influences the probability distribution of the direction in which a scattered photon goes, is also taken from \cite{Li01}, and is $g=0.721$ for $\lyb$ and $0.717$ for $\ciiline$.

%......................................................................................
\subsection{Computation of escape fractions} \label{sec:compute_fesc}

To compute the escape fractions of ionizing photons, we first create mock spectra of the galaxy between $10 \, \angstrom$ and $912 \, \angstrom$ using the method described in $\rm{Sect.} \,  \ref{sec:RASCAS}$. The only changes are that, first, we use BPASS but without a power-law approximation of the SEDs, and second, the optical depth is computed differently. The optical depth for ionizing photons is a sum of contributions from H, He, $\rm{He}^+$,  and dust. The first three are given by \cite{Verner96} and dust is treated exactly as in $\rm{Sect.} \,  \ref{sec:tau_dust}$, including the scattering, with an albedo of 0.24. The mock spectra are then used to compute the total number of ionizing photons going out of the virial radius, which is divided by the intrinsic number to get the escape fraction. We also compute the escape fraction at $900 \, \angstrom$ by averaging the spectra between $890 \, \angstrom$ and $910 \, \angstrom$. Results are shown in $\rm{Table} \, \ref{table:outputs}$ and in $\rm{Sect.} \, \ref{sec:fesc}$.

%%%%%%%%%%%%%%%%%%%%%%%%%%%%%%%%%%%%%%%%%%%%%%%%%%%%%%%%%%%%%%%%%%%%%%%%%
\section{Mock spectra} \label{sec:spectra}

\begin{figure}
  \resizebox{\hsize}{!}{\includegraphics{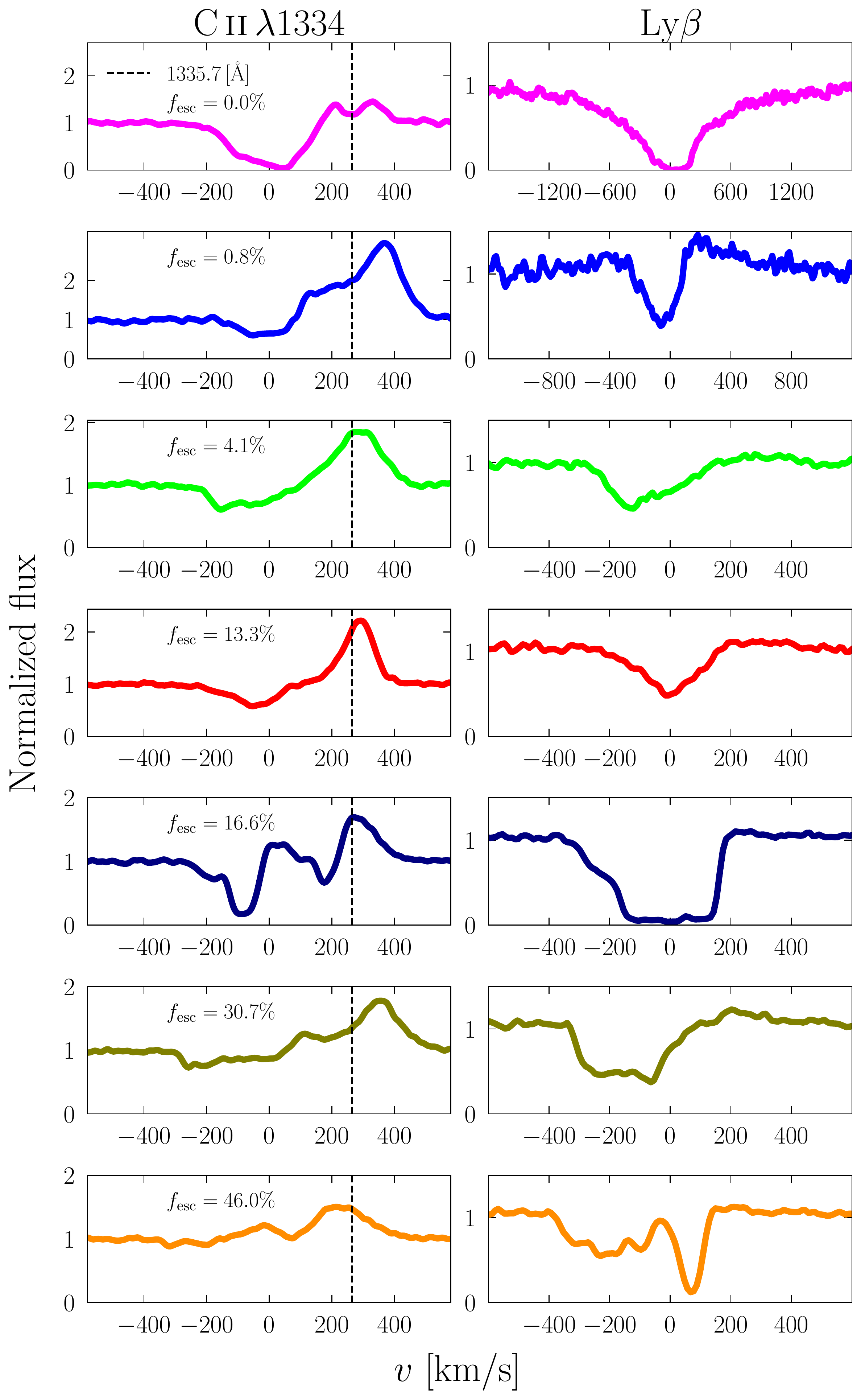}}
  \caption{$\ciiline$ and $\lyb$ absorption lines for seven directions of observation. The vertical dashed lines highlight the wavelength of the fluorescent channels. The spectra are arranged in increasing order of escape fraction of ionizing photons in the direction of observation, $\fesc$ (as computed in $\rm{Sect.} \, \ref{sec:compute_fesc}$). The noise is Monte-Carlo noise, due to a sampling of the luminosity with a finite number of photon packets. The spectra are smoothed with a Gaussian kernel with $\rm{FWHM}=20 \, \kms$. The aperture diameter for the mock observation is a fifth of the virial radius, which corresponds to about 5.6 kpc, or 0.75" at $z=3$. No observational effects such as noise, spectral binning, Earth atmosphere, or MW alterations are considered. We note that the x-axis in the first two rows of the second column differ from the others.
  }
  \label{fig:spectra_fid}
\end{figure}

In this section we show a selection of $\ciiline$ and $\lyb$ spectra. From the 5184 mock observations that we made (i.e., 1728 directions of observation for the three outputs A, B and C), we select seven directions that have absorption profiles that are as diverse as possible, shown in $\rm{Fig.} \, \ref{fig:spectra_fid}$. We highlight that those spectra do not represent the average of the galaxy over time and direction of observation, but they rather depict the large range of possible shapes and properties for one galaxy. Most lines are similar to the one in the top row of $\rm{Fig.} \, \ref{fig:spectra_fid}$, that is to say saturated and wide. 

There is a very wide range of shapes of absorption lines for a single galaxy seen from different directions of observation and at time differences of 80 Myr. This shows that the distribution of gas in the galaxy is highly anisotropic. There is however an axis along which the galaxy is more flat, resulting in an irregular disk-like structure. There are slight trends between the properties of the lines and the inclination of this disk: Edge-on angles have somewhat stronger absorption than face-on ones. We will explore those relations in a following paper.

The main properties of the lines on which we focus in this paper are the residual flux and the equivalent width (EW). The residual flux is 
\be \label{eq:resflux}
R = \frac{F^0_{\rm{line}}}{F_{\rm{cont}}^{\rm{observed}}},
\ee 
where $F^0_{\rm{line}}$ is the flux at the deepest point of the absorption and $F_{\rm{cont}}^{\rm{observed}}$ is the flux of the continuum next to the absorption line. The EW corresponds to the area of the absorption line after the continuum has been normalized to one. The EW of the fluorescence corresponds to the area of the emission part above the continuum. Both EWs are defined as positive. We highlight that in several spectra the EW of fluorescence is larger than that of the absorption. This is because fluorescence is an emission process through which the gas emits in all directions. There are directions of observation with almost no absorption but with a strong fluorescence, which is emitted by gas that is excited by photons traveling in other directions. Lastly, we note that to compensate the movement of the galaxy in the simulation box we add $\vec{v_{\rm{CM}}} \cdot \vec{k_{\rm{obs}}}$ to the velocity axis of $\rm{Fig.} \, \ref{fig:spectra_fid}$, where $\vec{v_{\rm{CM}}}$ is the velocity of the center of mass of the stars and $\vec{k_{\rm{obs}}}$ is the unit vector pointing toward the observer.

\subsection{Comparison with observations} \label{sec:comp_obs}

In comparison with observed $z \sim 3$ galaxies, our $\lyb$ and $\ciiline$ lines are slightly less wide. In our mock spectra we do not find absorption at very negative velocities. The median velocity at which the absorption reaches 90\% of the continuum is $-200 \, \kms$, and the minimum is $-450 \, \kms$. This is to be contrasted with studies such as those of \cite{Steidel10}, \cite{Heckman15} and \cite{Steidel18} where this velocity is often  $-800 \, \kms$. We note however that these observations are averages over many galaxies and typically obtained with lower spectral resolution, meaning that a direct comparison is not really indicative. Furthermore, these absorptions at very negative velocities are produced in outflowing gas, and so an alternative explanation is that our simulation probes galaxies with weaker outflows. This is expected because our galaxy has a much lower mass and SFR than the Lyman Break Galaxies of those studies. However, it is possible that the simulation does not produce enough outflows to be realistic, as is argued in \cite{Mitchell20}, where they use a similar simulation. We will study the gas flows of our galaxy in a forthcoming paper.

In contrast, we find $\ciiline$ profiles that are very similar to high-resolution observations of galaxies in the local Universe, as in \cite{Rivera15} and \cite{Jaskot19}. These latter authors find similar absorption velocities as in our mock spectra, and a large variety of shapes, residual fluxes, and strengths of fluorescence. They even detect fluorescent lines with a P-Cygni profile (i.e., a blue part in absorption and a red part in emission), as in the fifth row of $\rm{Fig.} \, \ref{fig:spectra_fid}$, due to the presence of $\rm{C}^+$ in the excited fine-structure level.

Our $\lyb$ lines are rather different from current observations of either high or low redshift galaxies. \cite{Steidel18} and \cite{Gazagnes20} show $\lyb$ absorption lines that have a large EW and a significant residual flux. Our lines are either wide with a small residual flux, or have a small EW and a large residual flux. This could be due to low signal-to-noise ratios, spectral resolution, or stacking effects. Alternatively, these effects could come from stellar absorption features or $\lyb$ emission from the gas, neither of which is included in this work.

We note that some $\lyb$ spectra display some emission redward of the absorption, especially in the second row of $\rm{Fig.} \, \ref{fig:spectra_fid}$, and very little in the last five rows. This is due to scattering effects, and has not been observed yet because it is rare (the first row, which is the most common, has no emission) and requires an excellent signal-to-noise ratio.

% -----------------------------------------------------------------------------
\subsection{Robustness of the method} \label{sec:robust}

\begin{figure*}
  \resizebox{\hsize}{!}{\includegraphics{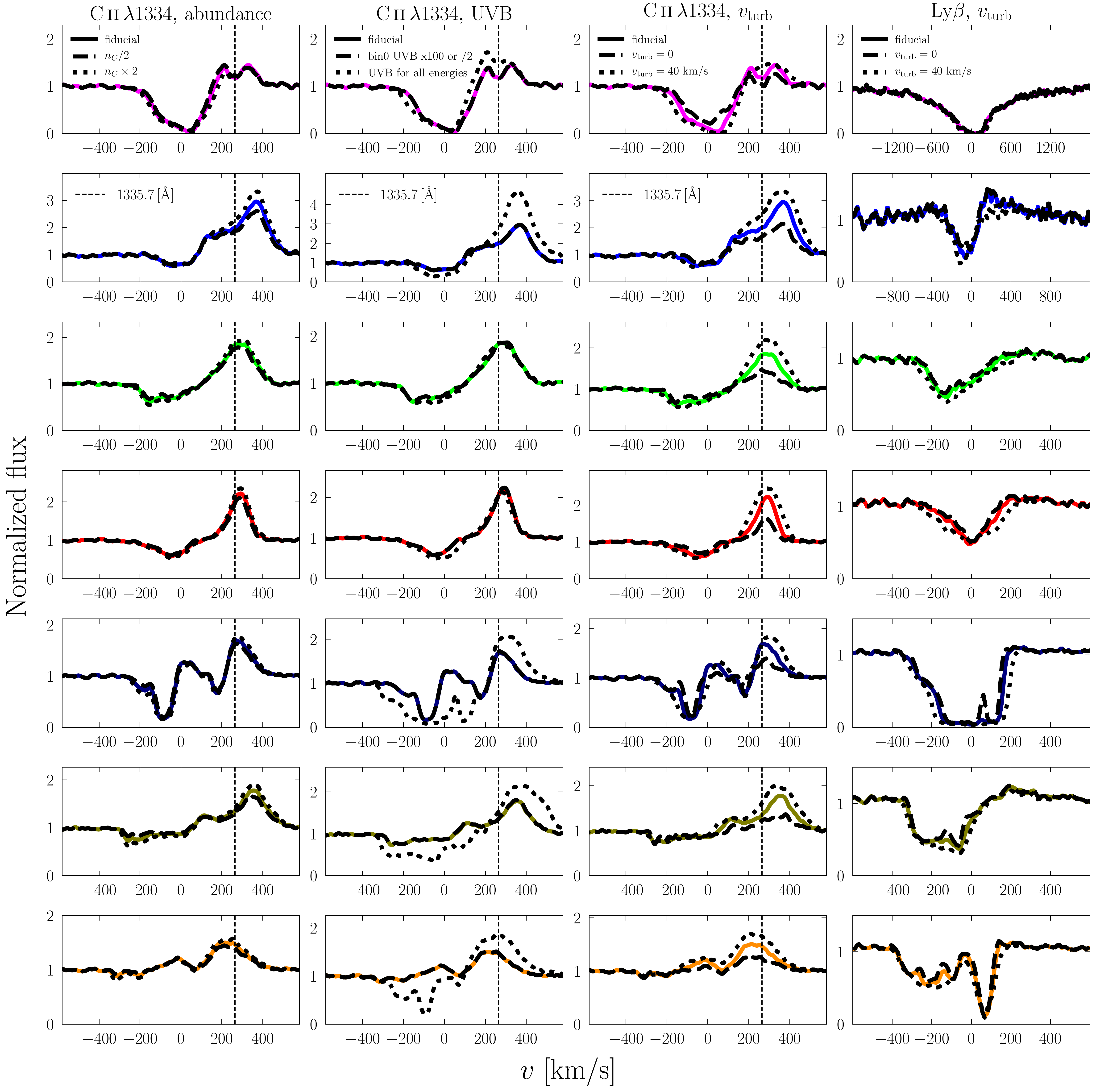}}
  \caption{Test of the robustness of the spectra against changing free parameters. The seven rows show the same seven directions as in $\rm{Fig.} \, \ref{fig:spectra_fid}$. The vertical dashed lines highlight the wavelength of the fluorescent channels. First column: Test of the effect of multiplying and dividing the fiducial carbon density by two. Second column: Test of multiplying (dividing) the Habing band UVB by 100 (2) and comparison with the result with only UVB and no radiation field from the simulation. Third and fourth columns: Test of changing the turbulent velocity for $\ciiline$ and $\lyb$ respectively. The fiducial turbulent velocity is 20 $\kms$.}
  \label{fig:spectra_robust}
\end{figure*}

\begin{figure*}
  \resizebox{\hsize}{!}{\includegraphics{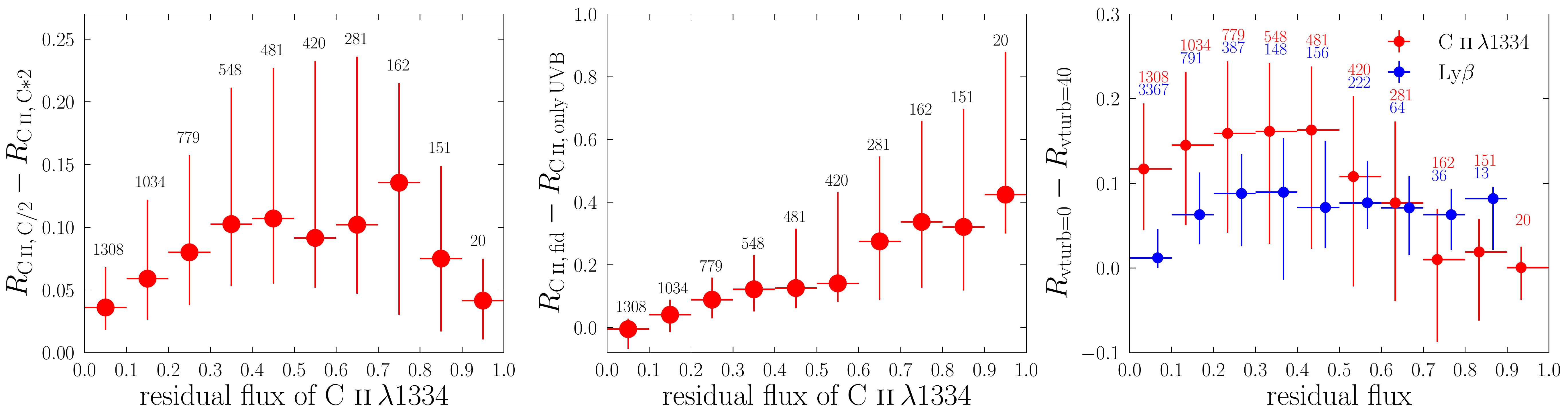}}
  \caption{Effect of different parameters of the modeling on the residual fluxes. The y-axes display differences of residual fluxes between two setups. The dots are the median of the differences in a given bin of residual flux, and the bars show the \tenth and \nineteenth percentiles. The numbers indicate how many spectra are in the corresponding bin of residual flux. The left panel shows the effects of carbon density, which is divided or multiplied by two. The middle panel shows the effect of UVB radiation for ionizing photons instead of the radiation field from the simulation. The right panel shows the effects of turbulent velocity. Turbulent velocities are varied between 0 and 40 $\kms$.
  }
  \label{fig:resflux_robust}
\end{figure*}

Our procedure to make mock observations involves several free parameters. Here we test the extent to which the spectra are affected if we change their values. We focus in particular on the effect on residual flux, which is the quantity that we use the most in this paper. $\rm{Figure} \, \ref{fig:spectra_robust}$ shows the same spectra as in $\rm{Fig.} \, \ref{fig:spectra_fid}$ but with different alterations, which we explain individually below. $\rm{Figure} \, \ref{fig:resflux_robust}$ shows the statistics of the change of residual flux for the 5184 directions of observation, for every setup that we try.

%............................................................................
\subsubsection{Carbon abundance} \label{sec:abundance}

As discussed in $\rm{Sect.} \, \ref{sec:comp_densities}$, the simulation does not trace carbon directly and we have to assume an abundance ratio for carbon to infer it in post-processing. This setup incurs uncertainties due to the poorly constrained carbon-to-hydrogen ratio and dust depletion factor in $z = 3$ galaxies, and therefore we vary the density of carbon to see if the result is sensitive to that variation. The first column of $\rm{Fig.} \, \ref{fig:spectra_robust}$ shows the effects of dividing and multiplying the density of carbon by a factor of two. Overall the effect is small. The fluorescence of the second and sixth spectra is affected the most, while the absorption part is never drastically changed.

For a more global view, the left panel of $\rm{Fig.} \, \ref{fig:resflux_robust}$ shows the effects on the 5184 directions. We see that directions with a very small or a very large residual flux are not significantly affected by the change in carbon density. This is because when the residual flux is either very small or very large the line photons have gone through mostly optically thick or thin media, respectively, which are not strongly affected by a change of a factor of two. Alternatively, when the residual flux is between $\sim 0.2$ and $0.8$, the gas through which the photons went is neither completely thick nor thin, and therefore a change in density of a factor of two has more of an  effect, with a difference of residual flux of up to 0.25. As the carbon-to-hydrogen ratio is poorly constrained, we use solar abundance ratios and test the relations between the absorption lines and the escape fraction of ionizing photons under this assumption.

%.........................................................................
\subsubsection{Implementation of the radiation below 13.6 eV} \label{sec:uvb}

As the simulation was done with only hydrogen- and helium-ionizing photons, we have to estimate in post-processing the flux of photons between 6 eV and 13.6 eV, which ionize neutral carbon, among other metals.
Our fiducial model is to use the \cite{HM12} UVB in every cell of the simulation, except in cells where $\nhi > 10^2 \, \percc$, because the optical depth of dust in those cells is enough to extinguish the UVB. We note that removing the UVB in those cells does not impact the absorption lines at all because those cells are so dusty that they hide all the stellar continuum. Additionally, as explained in $\rm{Sect.} \,  \ref{sec:comp_densities}$, we take the ionization fractions of hydrogen and helium from the simulation, and therefore we remove all reactions including hydrogen or helium from our \textsc{Krome} chemical network.

In the second column of $\rm{Fig.} \, \ref{fig:spectra_robust}$ we compare the fiducial spectra with three other setups:
\begin{enumerate}
\item The same as fiducial, but dividing the flux between 6 eV and 13.6 eV by two in every cell. This is done to evaluate the effect of reduction of the UVB due to absorption in the galaxy.
\item The same as fiducial, but multiplying the flux between 6 eV and 13.6 eV by 100 in every cell. This is done to simulate the contribution of a young metal-poor stellar population of $1000 \, \Msun$ at a distance of 10 pc (which is roughly the resolution in the ISM) at this energy range.
\item Using the \cite{HM12} UVB at all energies, ignoring the radiation field from the simulation. We let \textsc{Krome} compute the ionization fractions of every species, including hydrogen and helium, unlike in the fiducial setup. We compute the photoionization rate that the UVB imposes on $\hzero$, $\hezero$, $\rm{He}^+$, $\czero$ and $\rm{C}^+$ at the redshift of the output, and we apply this photoionization rate in every cell with a hydrogen density smaller than a certain threshold.
This threshold is $10^{-2} \, \percc$ for $\hzero$ photoionization, $4.3 \times 10^{-2} \, \percc$ for $\hezero$, $3.3 \times 10^{-1} \, \percc$ for $\rm{He}^+$, $10^{2} \, \percc$ for $\czero$ and $4.3 \times 10^{-2} \, \percc$ for $\rm{C}^+$.
For each ion this threshold is chosen so that a cell with a size of 10 pc having this density would have an optical depth of around one for photons having the smallest ionizing energy of the ion. 
We test this setup to assess the importance of using an RHD simulation instead of a hydrodynamics simulation where only a UVB can be used for the photoionization.
\end{enumerate}

The two first setups are represented by the dashed lines in the second column of $\rm{Fig.} \, \ref{fig:spectra_robust}$. They are identical to the fiducial spectra. This is because the \cite{HM12} UVB below 13.6 eV is high enough to completely photoionize $\czero$ to $\rm{C}^+$ (in all the regions with $\nhi <10^2 \, \percc$), and therefore multiplying it by 100 does not change the result, and dividing it by two is not enough to stop photoionizing $\czero$. To confirm this we multiplied the UVB by a factor 10 000, because locally the radiation field below 13.6 eV can be even stronger than 100 times the UVB, and the spectra are still not affected. On average, over the 5184 directions the residual flux is changed by only $0.002$, both when we divide the UVB by two or multiply it by 100.

In the third setup, the dotted lines in the second column of $\rm{Fig.} \, \ref{fig:spectra_robust}$, the effects are more noticeable. The first, third, and fourth rows are not affected significantly, while the second and the last three are drastically changed. The middle panel of $\rm{Fig.} \, \ref{fig:resflux_robust}$ shows the differences for the 5184 directions. We see that the directions with small residual fluxes are not affected significantly, such as the purple spectrum in the first row of $\rm{Fig.} \, \ref{fig:spectra_robust}$, but that the directions with high fiducial residual fluxes tend to have a much deeper absorption line when we use the UVB instead of the radiation field of the simulation, which is the case for the last three rows of the second column of $\rm{Fig.} \, \ref{fig:spectra_robust}$. This shows that using a UVB at all energies instead of a self-consistent ionizing radiation field simulated on-the-fly does not photoionize $\rm{C}^+$ to $\rm{C}^{++}$ efficiently enough, and therefore globally overestimates the density of $\rm{C}^+$, which leads to deeper absorption lines. Another factor that explains the change of $\rm{C}^+$ density is the modification of the $\rm{C}^+$ recombination and collisional ionization rates due to the change of electron density when we let \textsc{Krome} compute the equilibrium ionization fractions of hydrogen and helium instead of using the values from the simulation. Those results show that it is important to use an RHD simulation to accurately compute the ionization fractions of metals in post-processing.

%..................................................................................
\subsubsection{Turbulent velocity} \label{sec:turb}

Turbulent velocity (with a fiducial value of $20 \, \kms$) is a parameter we add to smooth out the discontinuities of the velocity field in the simulation and to model the subgrid motion of the gas, as we introduced in $\rm{Sect.} \,  \ref{sec:RASCAS}$. The third and fourth columns of $\rm{Fig.} \, \ref{fig:spectra_robust}$ show the effect on $\ciiline$ and $\lyb$ of changing the turbulent velocity to $0$ and $40 \, \kms$. We see that $\lyb$ is almost unchanged, and for $\ciiline$, the absorption parts of our seven spectra are not very sensitive to the turbulent velocity either, except the one in the first row. However, fluorescence grows quickly with turbulent velocity, indicating that there is more absorption overall in the galaxy, leading to a stronger fluorescent emission in every direction of observation.

The right panel of $\rm{Fig.} \, \ref{fig:resflux_robust}$ shows the effects of changing the turbulent velocity on our 5184 mock spectra. The $\ciiline$ line shows large differences of residual fluxes, of up to 0.25. An average difference of 0.15 in residual flux is seen for directions with a low residual flux, and a smaller difference is seen for directions with a high residual flux. There are even directions where the residual flux is larger for a turbulent velocity of $40 \, \kms$, which may seem counter-intuitive. This is because a larger turbulent velocity results in more absorption globally in the galaxy, and so there are also more scattering events and more infilling effects in some directions of observation. Some directions are sensitive to infilling, as we show in $\rm{Sect.} \,  \ref{sec:spectra_infilling}$, and this infilling can increase the residual fluxes. $\rm{Figure} \, \ref{fig:resflux_robust}$ confirms that $\lyb$ is less affected by the turbulent velocity parameter than $\ciiline$. The difference of residual fluxes with a turbulent velocity of 0 and 40 $\kms$ is on average smaller than 0.1. The turbulence affects $\lyb$ less than $\ciiline$ because the thermal velocity of hydrogen is about $ 3.5$ times larger than that of carbon.

% -----------------------------------------------------------------------
\subsection{Effects of various physical processes} \label{sec:process}

\begin{figure*}
  \resizebox{\hsize}{!}{\includegraphics{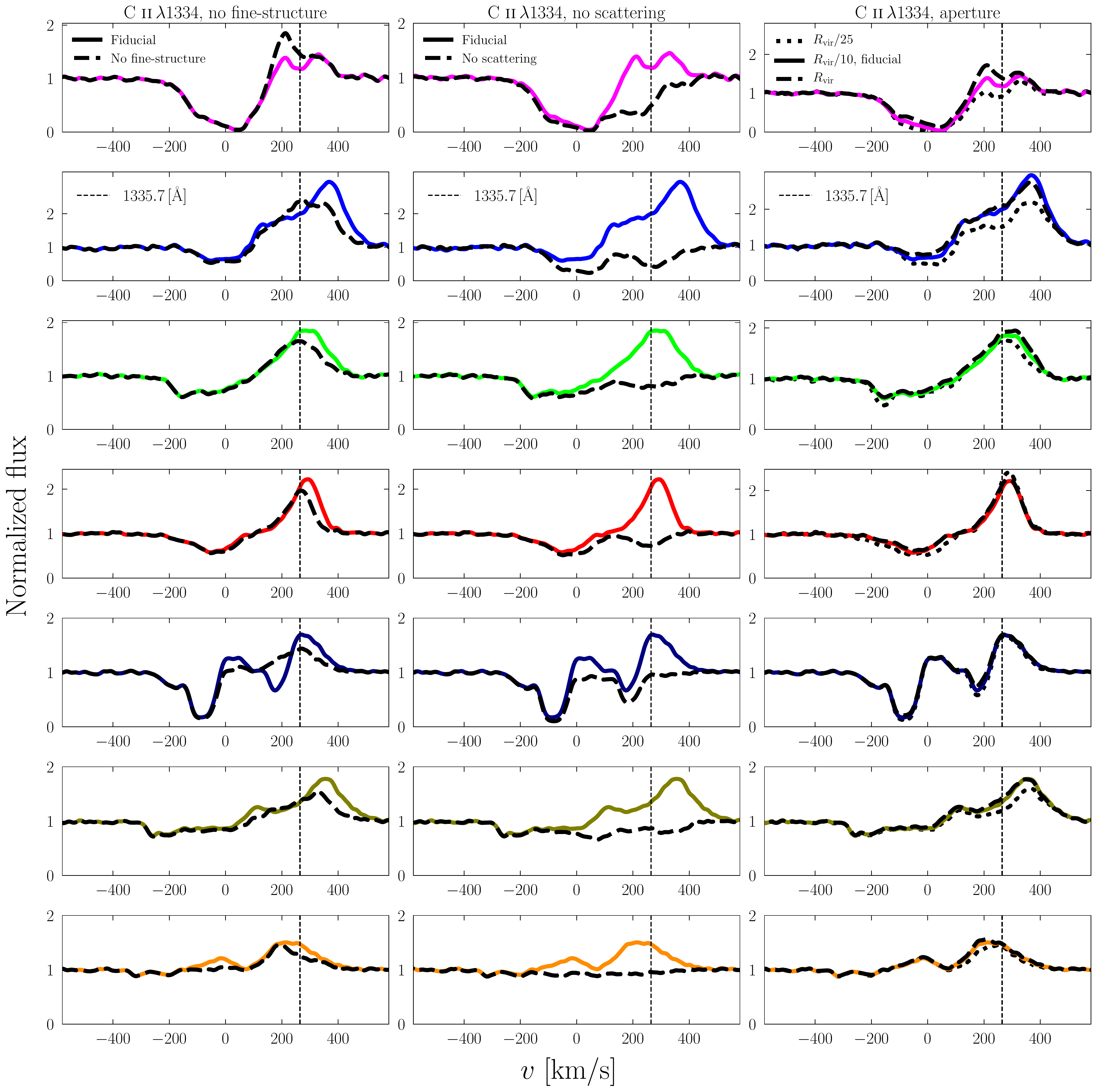}}
  \caption{Illustration of the effects of different physical processes on the same spectra as $\rm{Fig.} \, \ref{fig:spectra_fid}$. The vertical dashed lines highlight the wavelength of the fluorescent channels. 
  First column: Comparison with and without a fraction of $\rm{C}^+$ ions populating their ground-state fine-structure level. Second column: Effect of removing the scattering of photons. Third column: Effect of changing the aperture of the observation.
  }
  \label{fig:spectra_CII_process_1}
\end{figure*}

\begin{figure*}
  \resizebox{\hsize}{!}{\includegraphics{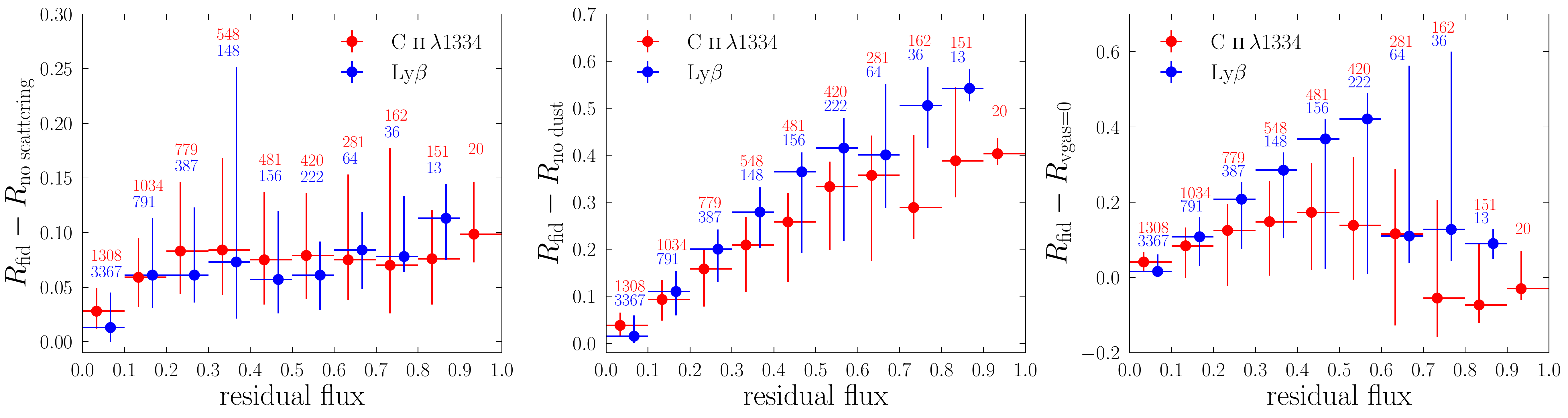}}
  \caption{Same as $\rm{Fig.} \, \ref{fig:resflux_robust}$, to show the effects of different physical processes.
  The left, middle, and right panels show the effects of scattering, dust, and gas velocity, respectively.}
  \label{fig:resflux_processes}
\end{figure*}

\begin{figure}
  \resizebox{\hsize}{!}{\includegraphics{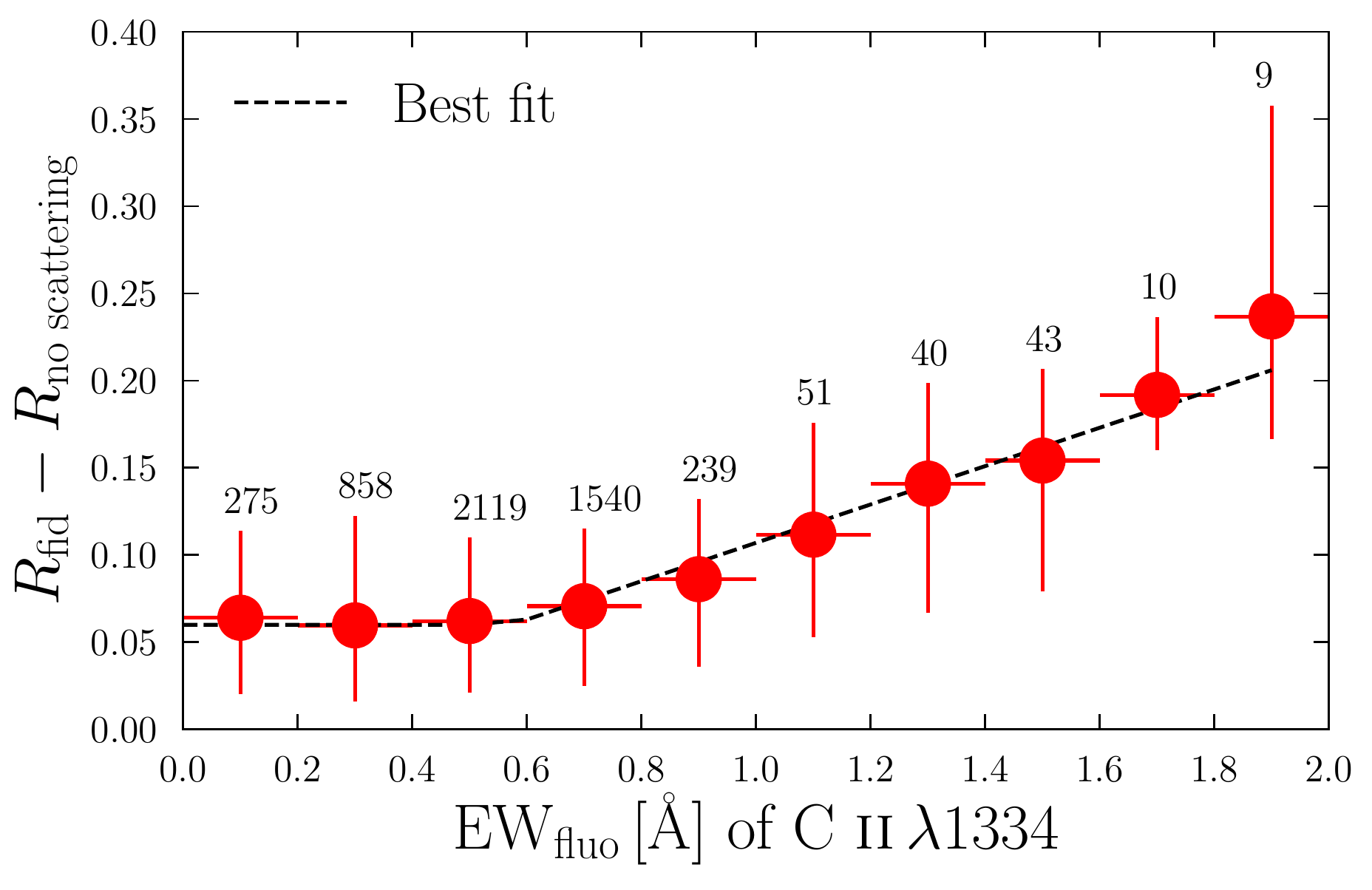}}
  \caption{Same as left panel of $\rm{Fig.} \, \ref{fig:resflux_processes}$, in bins of EW of the fluorescent line instead of residual flux. The best fit is 0.06 for $\rm{EW_{fluo}} < 0.6 \, \angstrom$ and $0.11 \rm{EW_{fluo}} - 0.003$ otherwise.}
  \label{fig:noscatter_EW_fluo}
\end{figure}

\begin{figure*}
  \resizebox{\hsize}{!}{\includegraphics{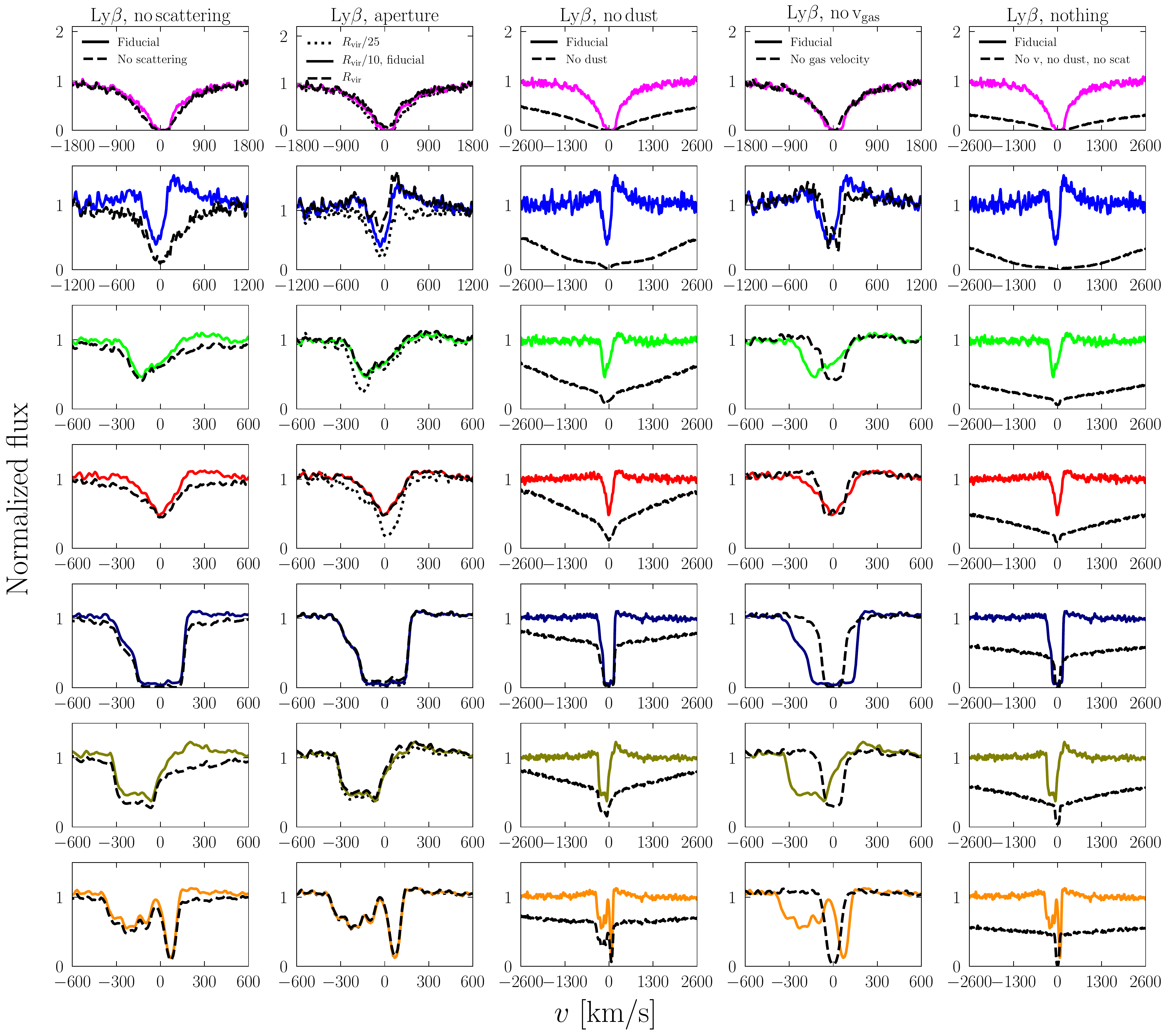}}
  \caption{Same as $\rm{Fig.} \, \ref{fig:spectra_CII_process_1}$ and \ref{fig:spectra_CII_process_2} but for $\lyb$. No fine-structure level is considered here.}
  \label{fig:spectra_lyb_process}
\end{figure*}

\begin{figure*}
  \resizebox{\hsize}{!}{\includegraphics{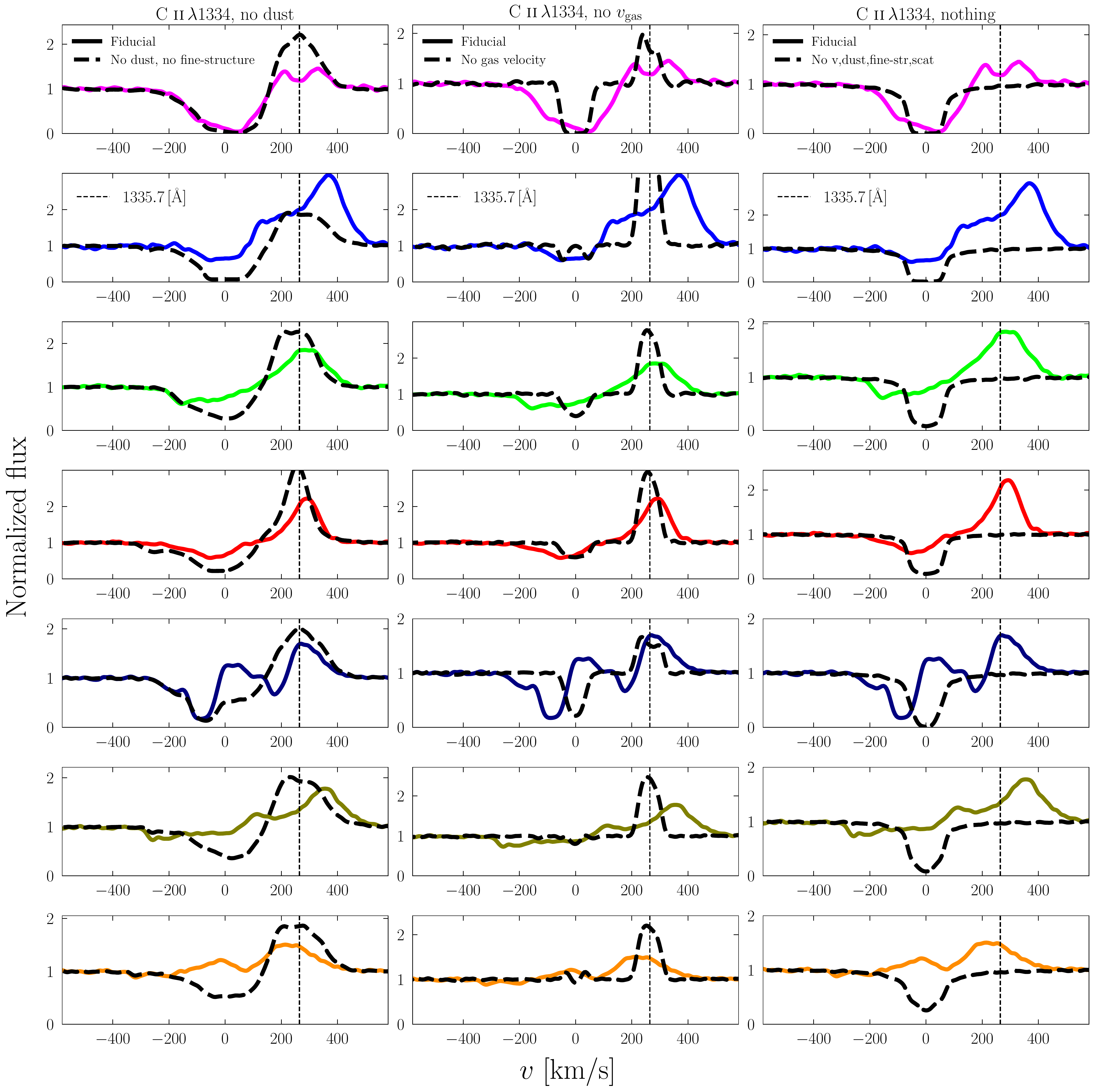}}
  \caption{Same as $\rm{Fig.} \, \ref{fig:spectra_CII_process_1}$, with other processes. First column: Effect of removing dust. We also remove the fine-structure level to simplify the interpretation. Second column: Spectra with and without gas motion in the galaxy. Infilling effects give complex shapes to the second and seventh rows. Third column: Comparison of fiducial spectra with spectra with neither gas motion, dust, scattering, nor fine-structure level.}
  \label{fig:spectra_CII_process_2}
\end{figure*}

In this section we analyze the effects of various physical processes on the spectra to show that the formation of the line is complex. In $\rm{Figs.} \, \ref{fig:spectra_CII_process_1}, \, \ref{fig:spectra_lyb_process},$ and  $\ref{fig:spectra_CII_process_2}$ we show the same spectra as in $\rm{Fig.} \, \ref{fig:spectra_fid}$ but altered by removing a number of processes that are taken into account in the fiducial method. $\rm{Figures} \, \ref{fig:resflux_processes} \, \rm{and} \, \ref{fig:noscatter_EW_fluo}$ show the statistics of the change of residual flux for the 5184 directions of observation for some of the processes that we analyze. Below we explain these processes one by one.

%.................................................................................
\subsubsection{Fine-structure level} \label{sec:spectra_fine}

Around $5\%$ of $\rm{C}^+$ ions in the ISM of our galaxy are excited in the fine-structure level of the ground state, leading to absorption at a different wavelength from the true ground state, as explained in $\rm{Sect.} \,  \ref{sec:hyperfine}$ and $\ref{sec:RASCAS}$. One of the resulting differences is that $\rm{C}^+$ ions in the excited state can resonantly scatter fluorescent photons at $1335.66 \, \angstrom$ and $1335.71 \, \angstrom$.
The first column of $\rm{Fig.} \, \ref{fig:spectra_CII_process_1}$ shows how spectra change if we assume instead that all $\rm{C}^+$ ions are in the true ground state. 

The absorption parts of the spectra are almost unaffected, with an average difference of residual flux of less than 0.01. However, there is significant modification of the fluorescent emission. The distribution of EW of fluorescence is redistributed among the different directions of observations when ions in the fine-structure level are considered because of scattering of resonant photons on those ions. This leads to some  directions having a larger EW of fluorescence and other directions a smaller one.

Furthermore, the peak of the fluorescence is redshifted when the fine structure is present, once more due to the scattering of photons in the $1335.66 \, \angstrom$ and $1335.71 \, \angstrom$ channels. Those photons are trapped by $\rm{C}^+$ ions in the fine-structure level; they can only escape if their wavelength is redshifted. The ones that are blueshifted are absorbed by $\rm{C}^+$ ions in their true ground state. This behavior is reflected in the statistics of the number of scattering events of the photons. Without fine-structure level splitting there is up to a maximum of 65 scattering events per photon, while this number becomes $\sim 27000$ when the fine-structure level is added. Finally, the presence of the fine-structure level can lead to fluorescence with a P-Cygni profile, as illustrated in the first column and fifth row of $\rm{Fig.} \, \ref{fig:spectra_CII_process_1}$.

%................................................................................
\subsubsection{Infilling} \label{sec:spectra_infilling}

The scattering of $\ciiline$ or $\lyb$ photons can lead to an effect called infilling. Without scattering, all the photons that are absorbed are destroyed. However, when scattering is taken into account, the photons are re-emitted, either as an $\rm H \alpha$ photon for $\lyb$ or a fluorescent photon for $\ciiline$, or at the wavelength of the line. In this case, photons that are absorbed can still contribute to the spectra if they are re-emitted resonantly and manage to escape the galaxy. Furthermore, scattered photons can contribute to the spectrum of a different direction of observation than the initial emission. This is the infilling effect. 
In order to see how much infilling occurs in our mock spectra, we generate spectra in which any absorption leads to the destruction of the photon. This means that the resulting spectra are the sum of Voigt absorption profiles for every cell in front of every stellar particle in the simulation. In this case, no fluorescent photons are emitted, and so the fluorescent $\ciistar$ line disappears.

The second column of $\rm{Fig.} \, \ref{fig:spectra_CII_process_1}$ for $\ciiline$ and the first column of $\rm{Fig.} \, \ref{fig:spectra_lyb_process}$ for $\lyb$ show the fiducial spectra and the spectra without scattering. For $\cii$, we see a second absorption around $250 \, \kms$ coming from the presence of $\rm{C}^+$ ions in their fine-structure level. The infilling in the absorption is strong for the second row, with a residual flux that is doubled when scattering is included. For the other spectra, the residual flux is only increased by a few percent.

The left panel of $\rm{Fig.} \, \ref{fig:resflux_processes}$ shows the differences of residual flux with and without scattering for the 5184 directions. Both $\ciiline$ and $\lyb$ show the same trends. For residual fluxes below 0.1 the median of the difference is small (less than 0.03). For very high residual fluxes, the difference is larger, with a median of 0.12 for $\lyb$ and 0.1 for $\ciiline$. In between, for residual fluxes from 0.1 to 0.8, the median of the difference varies between 0.05 and 0.1. Overall, the effect of infilling is not very significant on average, but there are extreme directions where the infilling increases the residual flux by more than 0.3.

 $\rm{Figure} \, \ref{fig:noscatter_EW_fluo}$ shows the differences of residual flux with and without scattering, similar as in $\rm{Fig.} \, \ref{fig:resflux_processes}$, but in bins of EW of fluorescence. We choose these bins to show that the effect of infilling is more important for directions with a strong fluorescence. This highlights the fact that there are directions where many scattered photons escape the galaxy. Indeed, both fluorescent photons and photons creating the infilling effect must have scattered at least once. $\rm{Figure} \, \ref{fig:noscatter_EW_fluo}$ shows that fluorescent photons and resonantly scattered photons both escape the galaxy in preferential directions. This can be used to get a rough estimate of the infilling effect based on the fluorescent line (see $\rm{Fig.} \, \ref{fig:noscatter_EW_fluo}$).

%........................................................................
\subsubsection{Aperture size} \label{sec:spectra_aperture}

The effect of infilling depends on the aperture used for the observation, because photons can potentially travel far before being scattered and redirected toward the direction of observation \citep{Scarlata15}. The third column of $\rm{Fig.} \, \ref{fig:spectra_CII_process_1}$ for $\ciiline$ and the second column of $\rm{Fig.} \, \ref{fig:spectra_lyb_process}$ for $\lyb$ compare the fiducial spectra with different aperture radii: $\Rvir /25$, $\Rvir /10$ and $\Rvir$, which are approximately 1.1, 2.8, and 28 kpc, or angular diameters of about 0".3, 0".7, and 7" at $z=3$. Our fiducial spectra are made with the $\Rvir /10$ aperture, because it includes as much as $\approx 95 \%$ of the stellar continuum emission. The $\Rvir$ aperture is used to assess the effect of adding the light scattered in the circum-galactic medium (CGM) and the $\Rvir /25$ aperture, which contains $\approx 40 \%$ of the stellar continuum, is useful to show what happens if we do not enclose the whole stellar-continuum -emitting region in the mock observation. We highlight that the aperture is a property of the mock observation, and it does not affect the radius of propagation of the photon packets, which is always one virial radius ($\rm{Sect.} \, \ref{sec:RASCAS}$).

The result is that the residual flux is often barely affected by the aperture of observation, except for the small aperture in the second, third, and fourth rows of $\rm{Fig.} \, \ref{fig:spectra_lyb_process}$. It is noticeable in general that smaller apertures lead to smaller residual fluxes. 
This is in part because a smaller aperture misses some of the scattered photons, but  is mainly because a smaller aperture misses flux from stars that are facing optically thin gas. The light from those stars is not absorbed, and therefore it increases the residual flux if the stars are in the aperture of the observation. Similarly, the fluorescence increases for larger apertures, as does the emission part of $\lyb$, because there are photons scattering far away that are missed with smaller apertures. Nevertheless, the differences when taking a larger aperture are small, and the aperture size has little importance as long as it encompass more than $\approx 90 \%$ of the stellar continuum. For example, an aperture corresponding to 1" gives a result very similar to our fiducial aperture. As discussed in \cite{Mitchell20}, where a simulation similar to ours is used, the small effect of increasing the aperture may be due to a lack of neutral gas in the CGM, possibly due to imperfect supernova feedback modeling.

%..................................................................................
\subsubsection{Dust} \label{sec:spectra_dust}

Now we study the effects of dust on the spectra. The third column of $\rm{Fig.} \, \ref{fig:spectra_lyb_process}$ for $\lyb$ and the first column of $\rm{Fig.} \, \ref{fig:spectra_CII_process_2}$ for $\ciiline$ compare the fiducial spectra with the spectra made without dust. To avoid obtaining a complex shape for the fluorescence, which can be misleading in this case, we omit here the fine-structure of $\rm{C}^+$ . Every $\rm{C}^+$ ion is assumed to be in the ground state. All the spectra that we plot are normalized, and so we do not see the difference in the continuum when dust is removed, which would be significantly attenuated. Averaging over the 5184 spectra, the continuum is reduced by $74\%$ at $1330 \, \angstrom$ and by $83 \%$ at $1020 \, \angstrom$. 

One striking difference in the spectra is that the $\lyb$ lines have especially large EWs when dust is removed. This is because there are stars embedded in optically thick regions, with $\Nhi > 10^{22} \, \percs$, causing absorption up to high velocities, yet those stars do not contribute to the spectra when dust is present because they are completely attenuated. Another result is that the absorption lines have smaller residual fluxes when dust is absent. This is because the flux at the deepest point of the line does not change significantly with or without dust, whereas the flux of the continuum does. Thus the residual flux, which is the ratio of those two fluxes, is larger with dust than without. The flux of the line center does not change significantly because dust and $\hzero$ or $\rm{C}^+$ all reside in the neutral regions, and therefore the line center photons have generally a higher probability of being absorbed by the gas than by dust.

The middle panel of $\rm{Fig.} \, \ref{fig:resflux_processes}$ shows the difference in residual flux between the spectra with and without dust for the 5184 directions. For very small fiducial residual fluxes, the difference is almost zero. However, for directions with higher residual fluxes, the spectra without dust have a much smaller residual flux than the ones with dust. 

%.................................................................................
\subsubsection{Gas velocity} \label{sec:spectra_vel}

The gas in front of the stars moves at different velocities, leading to an absorption that is spread in wavelength. This can lead to an increase of the residual flux compared to a case where all the gas has the same velocity. For example, even if each star is facing a column density high enough to saturate the absorption line, the residual flux can be larger than zero if the gas is not going at the same velocity for each star. For the line to be saturated, with zero residual flux, there must be at least one velocity regime for which all the stars are facing a sufficiently high column density of gas going at this (projected) velocity.

To show that the spectra in our simulation are sensitive to the velocity of the gas, the fourth column of $\rm{Fig.} \, \ref{fig:spectra_lyb_process}$ for $\lyb$ and the second column of $\rm{Fig.} \, \ref{fig:spectra_CII_process_2}$ for $\ciiline$ show the spectra with and without gas velocity. The EW is generally smaller without gas velocity, because the absorption is not spread in wavelength. Moreover, there are scattering effects that give complex shapes to the absorption profiles, especially in the second row for both lines and in the seventh row for $\ciiline$.

The right panel of $\rm{Fig.} \, \ref{fig:resflux_processes}$ shows the difference of residual flux between the spectra with and without gas velocity. As for the case without dust, the difference is negligible for very small fiducial residual fluxes and increases for higher residual fluxes, especially for $\lyb$. For very high residual fluxes, above 0.6, the difference becomes smaller again. For $\ciiline$, there are even directions where the residual flux is larger when the gas velocity is zero, because of scattering effects.

%.................................................................................
\subsubsection{All processes removed} \label{sec:spectra_nothing}

Finally, we show in the last column of $\rm{Fig.} \, \ref{fig:spectra_lyb_process}$ for $\lyb$ and of $\rm{Fig.} \, \ref{fig:spectra_CII_process_2}$ for $\ciiline$ the spectra with all four processes removed: fine-structure level, scattering, dust absorption, and gas velocity. Those spectra are more saturated because all four processes that we remove tend to increase the residual flux. The spectra still do not resemble Voigt profiles because of the fact that there are many sources, each facing a different column density of gas.

% % ----------------------------------------------------------------------
\subsection*{}

To summarize this section, we find that our predictions for the shape of the $\ciiline$ and $\lyb$ lines are rather robust against the free parameters of our modeling, such as metal abundances, nonionizing UVB estimations, and subgrid turbulent velocity. However, there are different processes that are crucial for the formation of the absorption lines. The absorption by dust and the velocity field are the most important, while the absorption from the fine-structure level, the infilling, and the aperture effects alter the lines to a lesser extent.

Additionally, dust has a counter-intuitive impact on the shape of absorption lines: young and luminous stars are often not contributing to the spectrum because of selective suppression of the stellar continuum emerging from the dense and dusty star-forming regions (see also $\rm{Sect.} \, \ref{sec:sources}$), which is why dust increases the residual flux.

%%%%%%%%%%%%%%%%%%%%%%%%%%%%%%%%%%%%%%%%%%%%%%%%%%%%%%%%%%%%%%%%%%%%%%%%%%%%%%%%%
\section{Escape fractions of \texorpdfstring{$\lyc$}{Lg}} \label{sec:fesc}

We now focus on the escape fractions of ionizing photons in the simulated galaxy. In order to compare the absorption lines and the escape fractions of ionizing photons, we compute the escape fraction from the galaxy as seen from different directions in post-processing, as explained in $\rm{Sect.} \, \ref{sec:compute_fesc}$. The correlations between the mock absorption lines and the escape fractions are studied in $\rm{Sect.} \, \ref{sec:interp}$.

% -----------------------------------------------------------------------
\subsection{Distribution of escape fractions}

The resulting escape fractions averaged over all directions are $11\%$ for output A ($z=3.2$), $3.8\%$ for output B ($z=3.1$) and $1\%$ for output C ($z=3.0$). Output A has the largest escape fraction of all outputs with $z<7$ in this simulation. Around $10\%$ of the last 35 outputs, at $z<3.5$, are similar to output B, with an average $\fesc \sim 3$--$4\%$. The vast majority, around $90\%$ of outputs, have similar escape fractions to that of output C, with an average $\fesc \leq 1\%$. 
Such large variations of the escape fraction with time are expected because the escape of ionizing photons is regulated by small-scale processes, essentially the disruption of star-forming clouds by radiation and supernova explosions, which have typical timescales of the order of a few million years \citep[e.g.,][]{Kimm17,Trebitsch17}. Also, at output A, the galaxy has been forming stars relatively intensely over the past $\sim$10 Myr, at around $5 \, \Msun \, \rm{yr}^{-1}$, relative to the average of around $3 \, \Msun \, \rm{yr}^{-1}$ from z=3.5 to z=3, and is thus undergoing strong feedback.  As can be seen from $\rm{Sect.} \, \ref{sec:cd-ionizing}$ below, this results in lowering the covering fraction of neutral gas in front of stars, and therefore increasing the escape of ionizing radiation. We choose these outputs, in particular output A, to study the largest variety of escape fractions possible, despite the fact that output A is not representative of the typical state of the galaxy.

\begin{figure}
  \resizebox{\hsize}{!}{\includegraphics{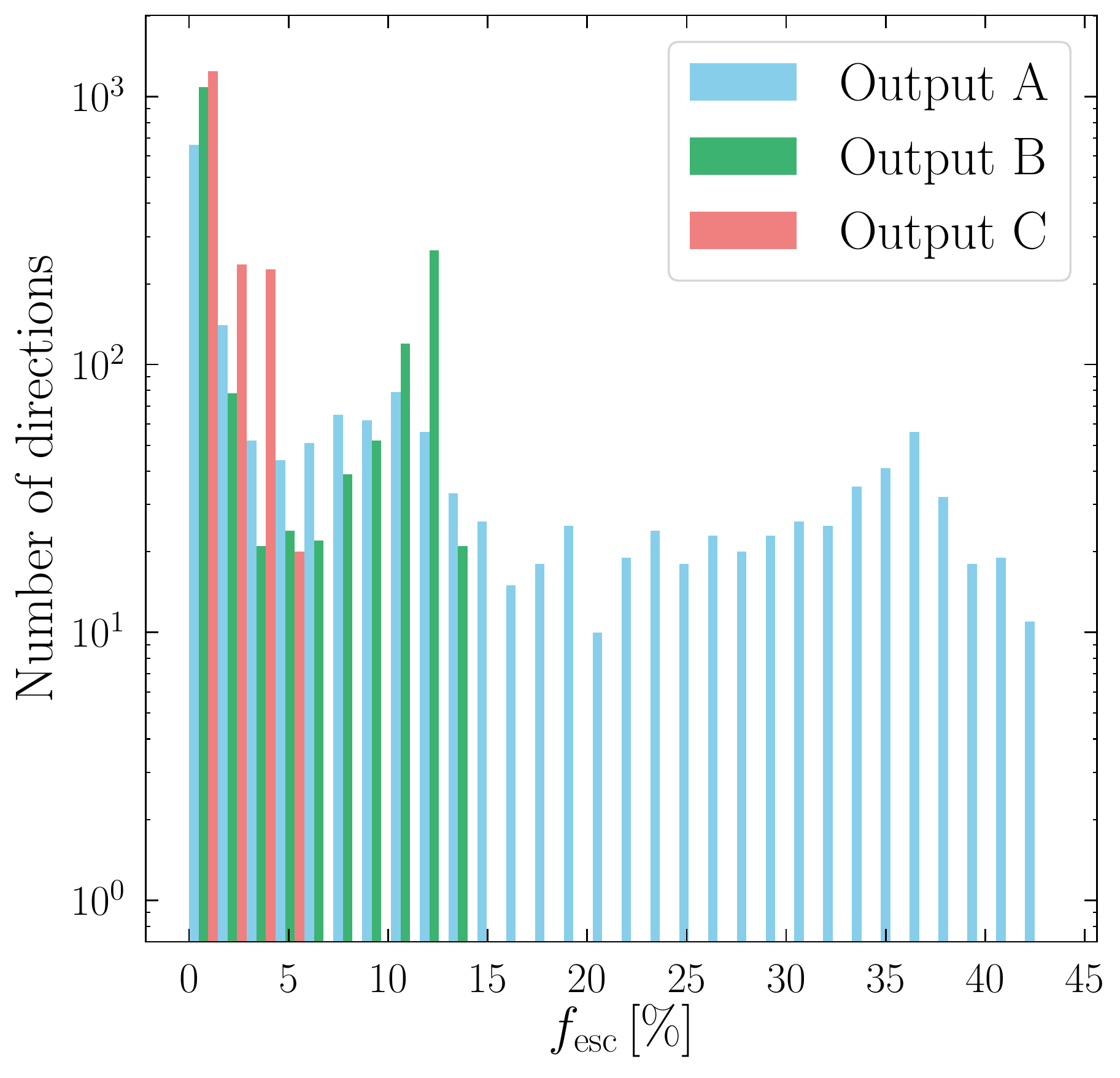}}
  \caption{Histogram of $\lyc$ escape fractions for the three outputs, computed as in $\rm{Sect.} \,  \ref{sec:compute_fesc}$.}
  \label{fig:histo_fesc}
\end{figure}

$\rm{Figure} \, \ref{fig:histo_fesc}$ shows the histogram of escape fractions in the 1728 directions of observation of the galaxy, for the three epochs A, B, and C. The three outputs show very different distributions of escape fractions, even though they are the same galaxy with 80 Myr between the outputs. As for the spectra in $\rm{Sect.} \,  \ref{sec:spectra}$, there is a large variety of escape fractions depending on the direction of observation of the galaxy. This large variety implies that a measurement of escape fraction of a galaxy does not necessarily give the correct contribution of the galaxy to the ionizing photon background \citep[e.g.,][]{Cen15}. One needs a statistical study of a large sample of galaxies to get information on the potential of galaxies to reionize the Universe.

\begin{figure}
  \resizebox{\hsize}{!}{\includegraphics{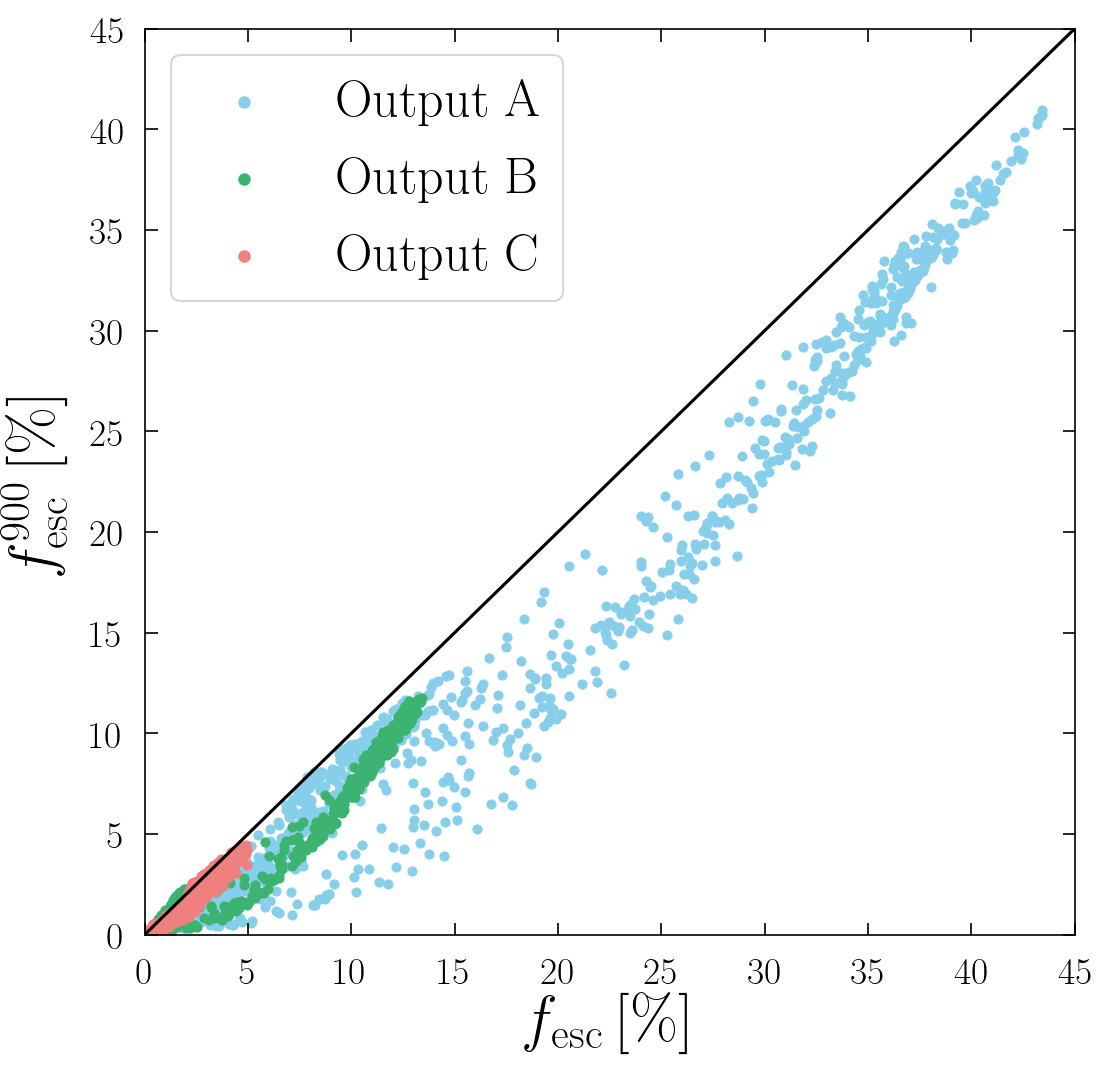}}
  \caption{Comparison of the escape fraction at $900 \, \angstrom$ and the global escape fraction of ionizing photons.
  }
  \label{fig:fesc_900}
\end{figure}

Additionally, we show in $\rm{Fig.} \, \ref{fig:fesc_900}$ the comparison between the global escape fraction of ionizing photons and the escape fraction at $900 \, \angstrom$, $\fesclimit$. We find that $\fesclimit$ is mostly smaller than $\fesc$, from a few percent to around 12$\%$ lower. There are a few points with small escape fractions where $\fesclimit$ is slightly larger than $\fesc$ because of the effect of helium absorption.

% -----------------------------------------------------------------------
\subsection{The effect of helium and dust} \label{sec:hedust_fesc}

As described in $\rm{Sect.} \,  \ref{sec:compute_fesc},$ the computation of escape fractions includes helium and dust in addition to hydrogen. We want to highlight that in our simulation the role of helium and dust is very subdominant, as also found by \cite{Kimm19}. If we remove helium, the escape fractions smaller than $5\%$ increase on average by only about $0.15 \%$ (absolute difference), while the larger escape fractions increase by around $1$--$3\%$. Those differences are small because helium absorbs only ionizing photons with energies larger than $24.59  \, \ev$, which are less numerous than the ones between $13.6  \, \ev$ and $24.59  \, \ev$.

When dust is removed, the change is even smaller, with an average increase of escape fractions of $0.05 \%$, and a maximum increase of $1.1 \%$. This small effect comes from our dust distribution model ($\rm{Equation} \, \ref{eq:ndust}$), where the density of dust is proportional to the density of neutral hydrogen. This implies that for ionizing photons the optical depth of dust is always much smaller than the optical depth of $\hzero$. In other words, dust alone would have a strong impact on the ionizing photons, but neutral hydrogen absorbs them much more efficiently, such that dust becomes ineffective. This is in agreement with other studies that find little or no effect of dust on the escape fraction \citep{Yoo20,Ma20}. However, in \cite{Ma20} the escape fractions in the most massive galaxies are substantially affected by dust, but these galaxies are more massive than ours. Additionally, their dust modeling puts dust in gas with a temperature of up to $10^6 \, \rm{K}$, whereas in our simulation dust is proportional to $n_{\hi} + 0.01 n_{\hii}$, and so the dust is almost completely absent in gas hotter than $3 \times 10^4 \, \rm{K}$. We note that a model of dust creation and destruction in the simulation could yield a different dust distribution, for example with more dust in ionized regions, which could then affect ionizing photons more, but this is beyond the scope of this paper.

%%%%%%%%%%%%%%%%%%%%%%%%%%%%%%%%%%%%%%%%%%%%%%%%%%%%%%%%%%%%%%%%%%%%%%%%%%%%%%%%%
\section{The link between escape fractions and absorption lines} \label{sec:interp}

\begin{figure*}
  \resizebox{\hsize}{!}{\includegraphics{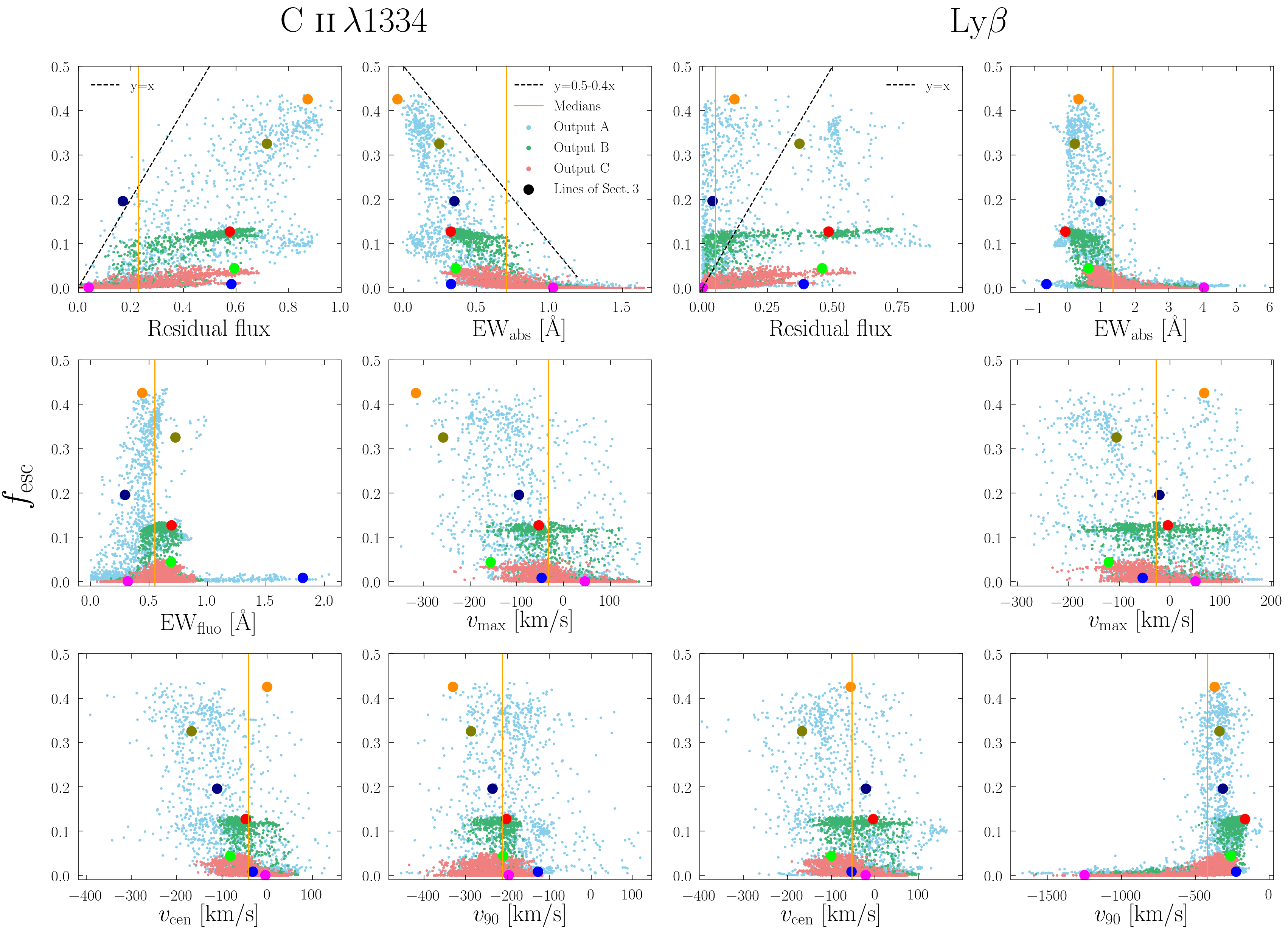}}
  \caption{Scatter plots comparing the escape fractions for the three outputs in the 1728 directions of observation to the properties of the $\cii$ absorption line on the left and the $\lyb$ line on the right. The seven colored dots show the position of the spectra displayed in $\rm{Sect.} \,  \ref{sec:spectra}$. $\ewabs$ is the EW of the absorption line (negative $\ewabs$ means more emission than absorption), $\ewfluo$ is the EW of the fluorescent emission, $\vmax$ is the velocity of the deepest point in the spectrum, $\vcen$ is the velocity at which the absorption line has half its EW, and $\vnine$ is the velocity where the blue side of the absorption line reaches $90\%$ of the continuum value. The orange vertical lines show the median value of the x-axis variable over the 5184 directions. There is no fluorescent channel for $\lyb$, and so one panel is missing.}
  \label{fig:scatter_fesc_all}
\end{figure*}

Now we turn to the search of correlations between properties of absorption lines and the escape fraction of ionizing photons. $\rm{Figure} \, \ref{fig:scatter_fesc_all}$ shows the relations between the escape fractions and some properties of $\ciiline$ on the left and $\lyb$  on the right, for the three outputs and the 1728 directions of observation of each output. We find that none of the observables give a clear indication of escape fraction in all ranges of $\fesc$. We can however see some general trends. 

For $\cii$, the escape fraction is almost always smaller than the residual flux, which may therefore be used as an upper limit for the escape fraction. This means that a zero residual flux implies $\fesc=0$. The EW of absorption may also be used to get an upper limit because the following relation holds approximately (for our simulated low-mass galaxy): $\fesc < 0.5 - 0.4 \times \ewabs [\angstrom]$. In particular, directions of observation with $\ewabs > 1.25 \, \angstrom$ have $\fesc < 2 \%$. The EW of the fluorescence is a poorer indicator of $\fesc$ than the EW of the absorption, but a high fluorescent EW also implies small $\fesc$. More precisely, $\ewfluo > 1 \, \angstrom$ implies $\fesc < 2 \%$. This is compatible with \cite{Jaskot14}, who find a strong fluorescent line in a galaxy whose $\lya$ profile hints at a nonzero escape of $\lyc$. Finally, the three velocity indicators, $\vmax$, $\vcen$, and $\vnine$ show the least correlation with $\fesc$. One potential relation is that directions with $\vcen < -250 \, \kms$ have $\fesc \gtrsim 10\%$, which is a similar behavior to what is observed by \cite{Chisholm17}. However we do not have enough data points in this velocity regime to be confident that $\fesc$ could not be smaller.

For $\lyb,$ the relations are even more scattered. The residual flux does not seem to trace $\fesc$ at all. A high EW  again implies a small escape fraction. Indeed, we find that $\fesc < 2\%$ when $\ewabs > 3 \, \angstrom$. The directions with $\ewabs < -0.6 \, \angstrom$, that is to say with $\lyb$ dominantly in emission, have $\fesc < 2\%$. However, there are not many points, and so this regime might be undersampled. As in the case of $\cii$, a very small $\vcen$ implies a high $\fesc$, but again the number of points is too small to be confident. Finally, a very small $\vnine$ is a sign of low $\fesc$: $\vnine < -700 \, \kms$ implies $\fesc < 2\%$, except for two outliers which are above this relation. However, those rare values of $\vnine$ are arising only when the EW is very large, and therefore they do not bring new information.

In summary, there is one regime where the absorption lines can provide reliable information about $\fesc$ . When the lines are saturated and wide, with one attribute generally accompanying the other, $\fesc$ is smaller than $2\%$. For $\lyb$, being saturated is not a sufficient condition to have a low escape fraction. In addition, the EW has to be larger than $3 \, \angstrom$. A large EW of fluorescence also implies a low $\fesc$, and is sometimes not linked with a large EW of absorption, as is the case for the large blue dot in $\rm{Fig.} \, \ref{fig:scatter_fesc_all}$. For the other regimes, there is always significant scatter and no predictive relations. To gain more confidence in the relations we find here, they must be verified in other simulated galaxies with different masses and metallicities, which we postpone to future work.

It might be surprising that there is little correlation between the escape fractions and the properties of the absorption lines, especially for the residual flux. In contrast, it is common in the literature to use the residual flux of LIS absorption lines to evaluate the covering fraction of neutral gas and deduce an estimate of the escape fraction of ionizing photons \citep[e.g.,][]{Heckman01,Shapley03,Grimes09,Jones13,Borthakur14,Alexandroff15,Reddy16,Vasei16,Chisholm18,Steidel18}. This is because there are idealized geometries where the escape fraction of ionizing photons and the residual flux of the LIS absorption lines are equal. If we imagine a plane of stars covered by a slab of gas composed of optically thick clouds and empty holes, the ionizing photons and the line photons can only escape through the holes, and not through the thick clouds. In this case, the escape fraction of ionizing photons and the residual flux of the lines are the same, and are equal to the fraction of the surface of the sources that is behind holes, which is referred to as the covering fraction. This is usually called the picket-fence model, implemented with some variations \citep[e.g.,][]{Reddy16,Steidel18,Gazagnes18}. Several authors have warned that the residual flux can only give a lower limit on the covering fraction, for example due to the velocity distribution of the gas \citep[e.g.,][]{Heckman11,Jones12,Rivera15}. We argue that there are other effects that complicate the relations between the lines and the escape fractions of ionizing photons, that we detail in this section.

\begin{figure*}
  \resizebox{\hsize}{!}{\includegraphics{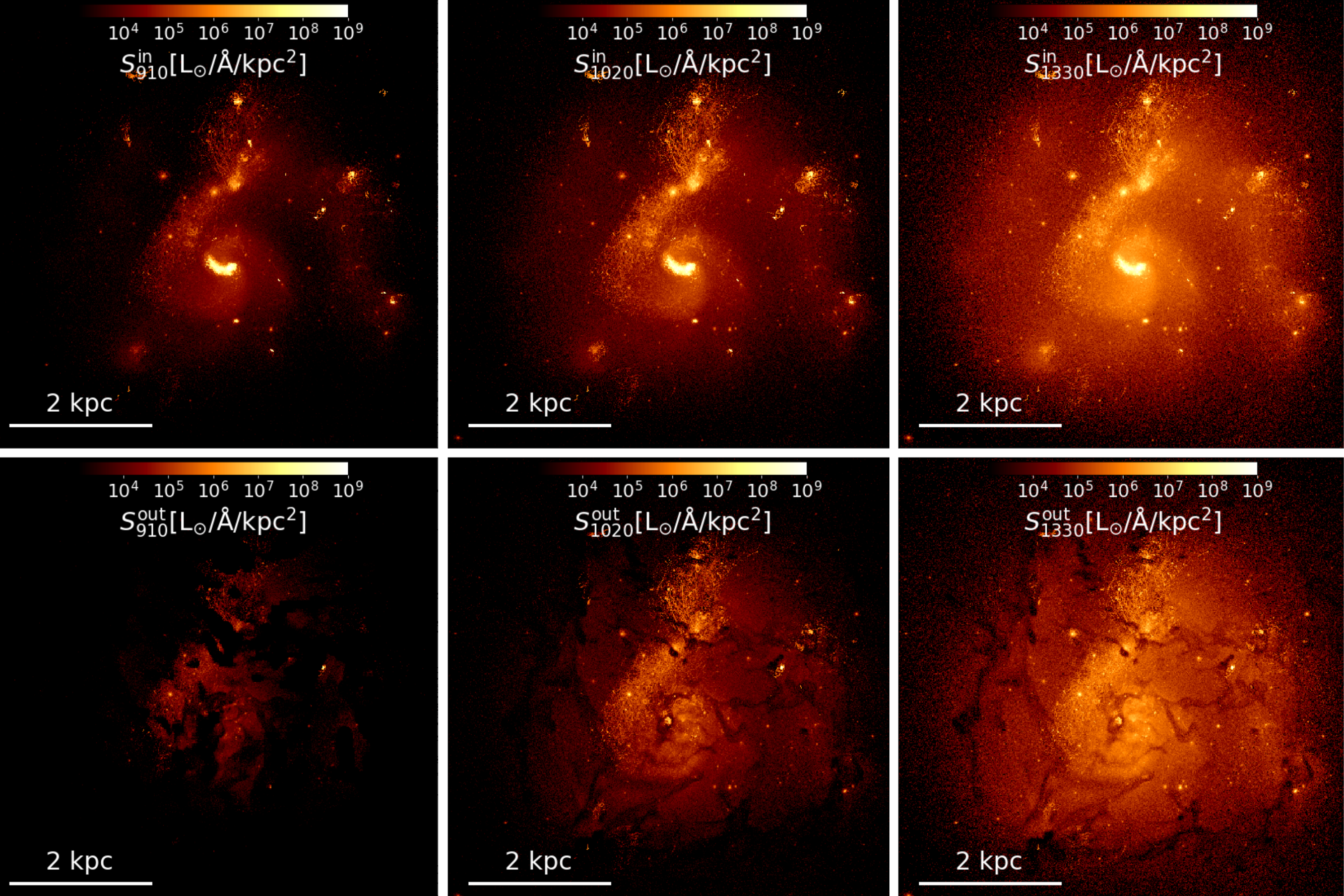}}
  \caption{Surface brightness maps of output C (z = 3.0). Upper left: Intrinsic surface brightness at $910 \, \angstrom$. Upper middle: Intrinsic surface brightness at $1020 \, \angstrom$. Upper right: Intrinsic surface brightness at $1330 \, \angstrom$. Lower left: Surface brightness at $910 \, \angstrom$ after $\hzero$ and dust absorption. Lower middle: Surface brightness at $1020 \, \angstrom$ after dust extinction. Lower right: Surface brightness at $1330 \, \angstrom$ after dust extinction.}
  \label{fig:map_lum_all}
\end{figure*}

There are several assumptions linked with the picket-fence model.
The sources of ionizing photons and nonionizing UV continuum have to be the same so that they trace the same screens of gas. If the continuum of the lines and  the ionizing photons are passing through different media,   it is impossible for the absorption lines to be a perfect tracer of the escape fraction of ionizing photons. 
Concerning dust, there are different treatments in the literature. Some studies do not apply dust corrections, similarly as in $\rm{Fig.} \, \ref{fig:scatter_fesc_all}$. The assumption for the residual flux of absorption lines to be equal to the escape fraction of ionizing photons without dust correction is based on dust being homogeneous, inside and outside of the holes, and the idea that the ionizing photons are affected by dust in the same way as the nonionizing UV continuum. In other studies, the residual flux is corrected under the assumption that the holes have no dust, or to account for the different effects of dust on ionizing and nonionizing photons \citep[e.g.,][]{Steidel18, Gazagnes18, Chisholm18}. Additionally, one assumption is that the distribution of column densities has to have optically thick parts and holes empty enough to be optically thin for both the ionizing photons and the line photons. In this setup, it is straightforward to define a covering fraction, to measure it with the residual flux of absorption lines, and to deduce the escape fraction as one minus the covering fraction. If there are column densities such that the optical depth is around one, the concept of covering fraction is less well defined, and the escape fraction is no longer one minus the covering fraction.
A final assumption is that the infilling effects and the gas velocities do not significantly affect the absorption lines, which would alter the measurement of the covering fraction.

In the following, we compare our simulation with the picket-fence model, addressing these assumptions one by one to decipher the similarities and differences and to explain the large dispersion in the relations between the escape fraction of ionizing photons and the residual flux of the absorption lines. In particular, we apply a dust correction to the residual fluxes using observable quantities, in order to improve the correlations.

% -----------------------------------------------------------------------
\subsection{Distribution of the sources} \label{sec:sources}

In this section we assess whether the ionizing photons and the $\lyb$ or $\ciiline$ photons are emitted from the same place in order to decipher whether or not they probe the same regions of gas. From stellar models, we know that only stars younger than ${\sim} 10$ Myr are bright at $<910 \, \angstrom$ (i.e., ionizing energies), whereas older stars can still be bright at $1020 \, \angstrom$ and $1330 \, \angstrom$, corresponding to the continuum next to the absorption lines. This is the main driver for the difference between sources of ionizing photons and nonionizing photons. The first row of $\rm{Fig.} \, \ref{fig:map_lum_all}$ illustrates those differences. It shows surface brightness maps for one particular direction of observation of the output C, at different wavelengths. Compared to the $910 \, \angstrom$ map, we see that at $1020 \, \angstrom$ there are more stars shining in an extended disk around the central part of the galaxy, and even more so at $1330 \, \angstrom$.

To quantify those differences, we note that the brightest $11\%$ of the stellar particles emit $99\%$ of the $910 \, \angstrom$ photons, while they emit around $98\%$ of the $1020 \, \angstrom$ photons and $85\%$ of the $1330 \, \angstrom$ photons. More detailed results for the three outputs are provided in the first two columns of $\rm{Table} \, \ref{table:starlum}$. While the sources of $1020 \, \angstrom$ photons are almost the same as the $910 \, \angstrom$ photon sources, the distribution of sources of $1330 \, \angstrom$ photons is noticeably different from that of $910 \, \angstrom$ photons. Around $15 \%$ of the luminosity at $1330 \, \angstrom$ is emitted by stars that do not contribute to the ionizing photon budget.

\begingroup
\setlength{\tabcolsep}{6pt} % Default value: 6pt
\renewcommand{\arraystretch}{1.3} % Default value: 1
\begin{table}
  \caption{Fraction of the light that is produced by the stellar particles responsible for $99\%$ of the emission of $910 \, \angstrom$ photons. ``Visible'' means after dust extinction. The results with error bars indicate the mean and the \tenth and \nineteenth percentiles.}
  \label{table:starlum}
   \centering
  \begin{tabular}{l|cccc}
    \toprule
    Output  & $1020 \, \angstrom$ & $1330 \, \angstrom$ & Visible $1020 \, \angstrom$ & Visible $1330 \, \angstrom$ \\ 
    \midrule
    A   & $97.8 \%$ & $85.1 \%$    & $95.1^{+3.4}_{-9.6}\%$  & $76.8^{+10.8}_{-26.9}\%$ \\
    B   & $97.9 \%$ & $87.5 \%$    & $87.9^{+6.4}_{-9.7}\%$  & $68.4^{+7.3}_{-10.1}\%$ \\
    C   & $98.0 \%$ & $85.6 \%$    & $84.1^{+7.8}_{-13.2}\%$  & $58.9^{+8.8}_{-13.9}\%$ \\
    \bottomrule
  \end{tabular}
\end{table}
\endgroup

However, it is the dust-attenuated $1020 \, \angstrom$
or $1330 \, \angstrom$ photons that are directly observable, and not the intrinsic luminosities. Here, we refer to the UV nonionizing photons after dust extinction as the "visible" photons. Once dust is added, the most luminous stars, namely the young stars in dense, neutral regions, are strongly attenuated. As those stars are the main sources of ionizing photons, the distribution of sources of visible photons becomes very different from the distribution of ionizing radiation sources. The middle and right panels of the second row of $\rm{Fig.} \, \ref{fig:map_lum_all}$ illustrate the effect of dust on $1020 \, \angstrom$ and $1330 \, \angstrom$ photons. Various bright regions are strongly attenuated, leading to a substantially different distribution of sources than for $910 \, \angstrom$ photons. The bottom left plot of $\rm{Fig.} \, \ref{fig:map_lum_all}$ shows the $910 \, \angstrom$ surface brightness after $\hzero$ and dust absorption for illustration.

\begin{figure*}
  \resizebox{\hsize}{!}{\includegraphics{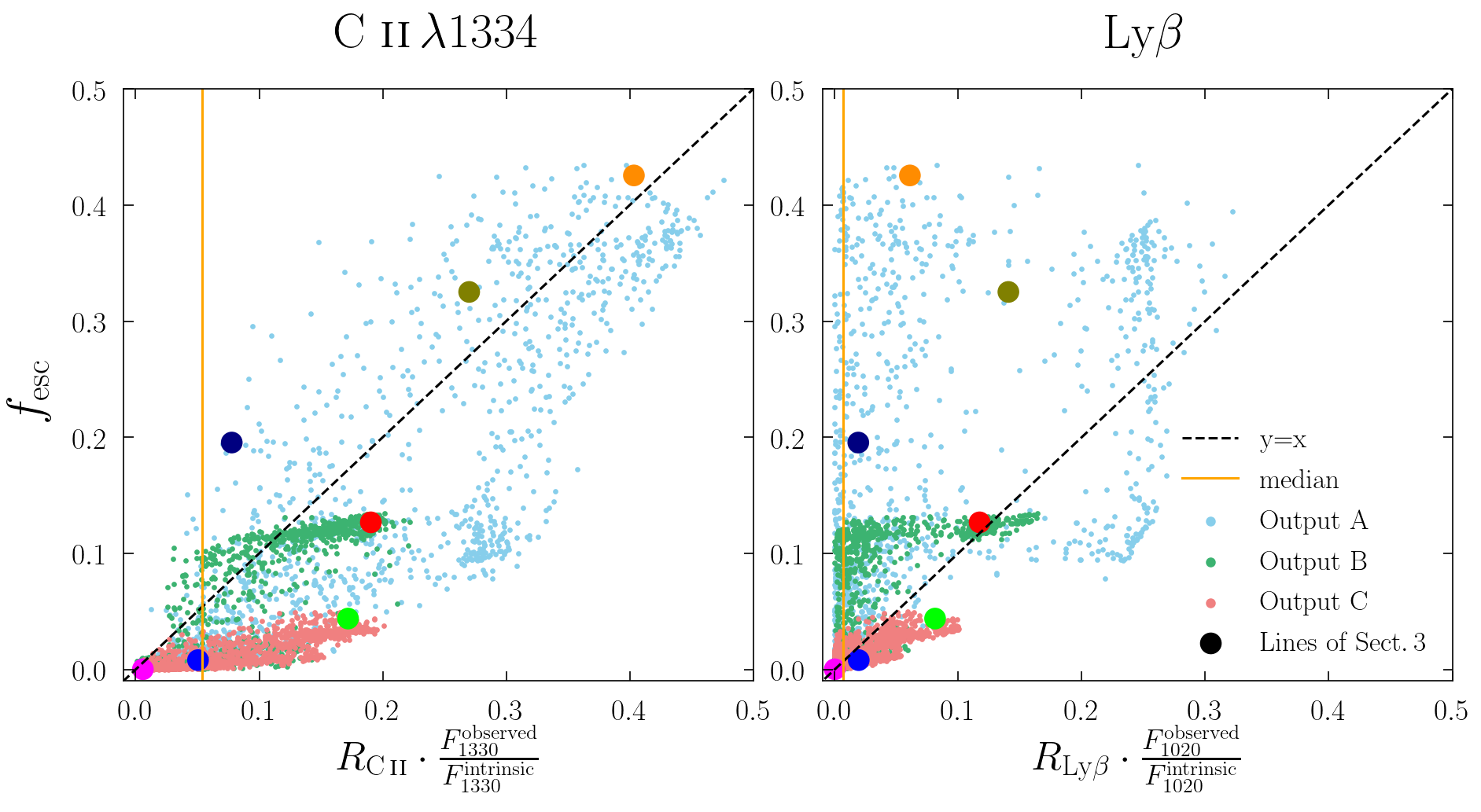}}
  \caption{Scatter plot of the escape fraction of ionizing photons and the dust-corrected residual flux for $\ciiline$ on the left and $\lyb$ on the right. The residual fluxes are multiplied by the ratio  between the flux of the continuum with dust extinction and the flux of the continuum without dust extinction.
  }
  \label{fig:scatter_all_dust}
\end{figure*}

We now compute the fraction of visible $1020 \, \angstrom$ and $1330 \, \angstrom$ photons emitted by the stars responsible for $99\%$ of ionizing photon emission to assess whether or not the visible photons are tracing the gas screens through which ionizing photons travel. The results are listed in the last two columns of $\rm{Table} \, \ref{table:starlum}$. When the attenuation by dust is included, the result depends on the direction of observation, and so we present the mean, and the \tenth  and \nineteenth percentile of the results. In summary, we find that around $15 \%$ ---but sometimes up to more than $30 \%$--- of visible $1020 \, \angstrom$ photons are emitted by stars that do not contribute to the budget of ionizing photons, and therefore those photons are affected by screens of gas that do not influence the escape fraction of ionizing photons. For visible $1330 \, \angstrom$ photons, the effect is even more pronounced, with around $30\%$ and up to $60\%$ of photons passing through gas that does not screen $\lyc$ sources. This is one reason why the residual flux of absorption lines before dust correction of the continuum cannot be a good proxy of the escape of ionizing photons. It is also one motivation to introduce dust correction of the UV nonionizing continuum, as is often done in the literature \citep[e.g.,][]{Steidel18,Gazagnes18,Chisholm18}.

% -------------------------------------------------------------------------------
\subsection{Dust correction} \label{sec:dust_correction}

In addition to what we see in $\rm{Sect.} \, \ref{sec:sources}$, another motivation to apply a dust correction is the different effects that dust has on both ionizing photons and on the nonionizing UV continuum in our galaxy. We highlight in $\rm{Sect.} \, \ref{sec:hedust_fesc}$ that dust absorbs very few ionizing photons because the optical depth of neutral hydrogen is always much larger than the optical depth of dust. However, the flux of the continuum next to the absorption line, which enters in the definition of the residual flux of the line, significantly depends on dust. This suggests that a better proxy of the escape fraction of ionizing photons is the ratio of the flux at the bottom of the line to the flux of the intrinsic continuum, rather than the residual flux we plot in $\rm{Fig.} \, \ref{fig:scatter_fesc_all}$, which uses the continuum after dust extinction. To obtain this new ratio, that we refer to here as the dust corrected residual flux, we multiply the residual flux (see $\rm{Equation} \, \ref{eq:resflux}$) by the dust extinction factor of the continuum:
\be
R^{\rm{dust \, corrected}} = \frac{F^0_{\rm{line}}}{F_{\rm{cont}}^{\rm{intrinsic}}} = R \times \frac{F_{\rm{cont}}^{\rm{observed}}}{F_{\rm{cont}}^{\rm{intrinsic}}},
\ee
where $F_{\rm{cont}}^{\rm{intrinsic}}$ is the value of the continuum next to the absorption line before dust attenuation. Fortunately this dust extinction factor $F_{\rm{cont}}^{\rm{observed}}/F_{\rm{cont}}^{\rm{intrinsic}}$ can be estimated by SED fitting, and therefore the dust correction is feasible in practice. The extinction factor is usually parametrized as $10^{-0.4 E_{B-V} k_{\lambda}}$, where $E_{B-V}$ models the strength of the dust attenuation and $k_{\lambda}$ is a law of dust attenuation as a function of wavelength. However, the extinction factor can be obtained directly from the simulation after the transfer of the stellar continuum with \textsc{RASCAS}, which is why we do not use the observational method to apply the dust correction. This dust correction is very similar to the ones in \cite{Steidel18, Gazagnes18} and \cite{Chisholm18}. The only difference is that we use the extinction factor at the wavelength of the continuum next to the absorption lines, while these latter authors use the extinction factor at $912 \, \angstrom$. This is conceptually different but gives similar results.

$\rm{Figure} \, \ref{fig:scatter_all_dust}$ shows the new relations between the escape fraction of ionizing photons and the dust-corrected residual flux of $\ciiline$, on the left, and $\lyb$, on the right. The results are more promising than before dust correction, in particular for $\ciiline$. The corrected residual flux still shows significant scatter but now follows the one-to-one relation more
closely, especially for directions with high escape fraction. Very low corrected residual fluxes still indicate very low escape fractions. For directions with dust-corrected residual fluxes of between $0.02$ and $0.3$, the prediction of $\fesc$ using the residual flux is unfortunately often too high. Only around $18\%$ of those directions have an escape fraction that is at least equal to $80\%$ of the corrected residual flux. Above a corrected residual flux of $0.3$, the predictions are more successful, with around $80 \%$ of the directions having an escape fraction at least equal to $80\%$ of the corrected residual flux.

After dust correction, $\lyb$ still traces $\fesc$ more poorly than $\ciiline$ does, but we see that the dust-corrected residual flux of $\lyb$ now provides a good lower limit for $\fesc$, whereas before dust extinction, in the right panel of $\rm{Fig.} \, \ref{fig:scatter_fesc_all}$, the points were extremely dispersed.

% -------------------------------------------------------------------------------
\subsection{Covering fractions and escape fractions} \label{sec:cd}

We now turn to the question of the distribution of the column densities and the concept of covering fractions. At the beginning of $\rm{Sect.} \,  \ref{sec:interp}$ we explain that in order to have equality between the residual flux of absorption lines and the escape fraction of ionizing photons, the screen of gas must be split in two parts: some optically very thick regions and some empty holes. The escape fraction is then equal to one minus the covering fraction $f_C$, which is the fraction of sources that are covered by optically thick gas. If there is gas with a column density such that $\tau_{910} \approx 1,$ the covering fraction becomes harder to define and it is no longer true that $\fesc = 1-f_C$.

In the first part of this section, we study the distribution of neutral hydrogen column densities that the ionizing photon sources are facing. We then define the covering fraction in the simulated galaxy and see how well it traces $\fesc$ in the same three outputs and 1728 directions of observation as in the previous sections.

An additional difficulty arises when using absorption lines to probe this covering fraction: $\lyb$ does not have the same optical depth as ionizing photons for a given column density of $\hzero$ and $\rm{C}^+$ is not a perfect tracer of $\hzero$, which is why $\ciiline$ also has a distribution of optical depths that is different from that of ionizing photons. We discuss these issues in $\rm{Sect.} \, \ref{sec:interpretation_lyb}$ and \ref{sec:interpretation_CII}.

\subsubsection{Covering of ionizing radiation sources} \label{sec:cd-ionizing}

To study the distribution of column densities in a given direction of observation we first compute the column density in front of every stellar particle. As we want to study the coverage of photons rather than the coverage of stars, we give a weight to each stellar particle equal to its luminosity at $910 \, \angstrom$. We then compute the fraction of photons behind a given amount of column density. A distribution perfectly matching the picket-fence model would show two distinct populations (for example, $30\%$ of the ionizing photons facing less than $\Nhi=10^{15} \, \percs$ and $70\%$ facing more than $\Nhi=10^{19} \, \percs$).

\begin{figure}
  \resizebox{\hsize}{!}{\includegraphics{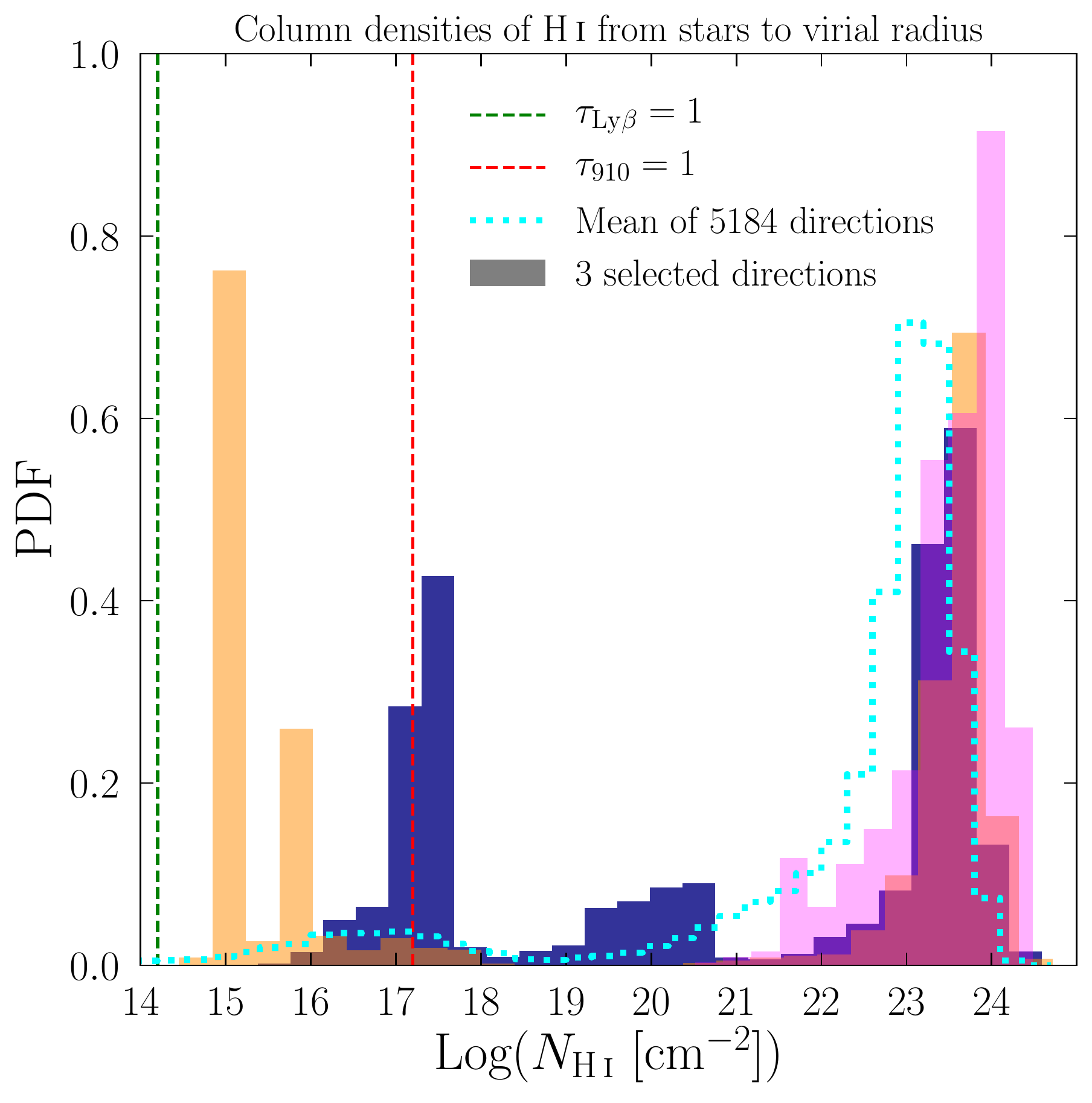}}
  \caption{Probability distribution of column densities faced by $910 \, \angstrom$ photons. The three filled histograms correspond to the three directions that have the same color in $\rm{Sect.} \, \ref{sec:spectra}$. The cyan dashed line is the average distribution over the 1728 directions of the three outputs. It has a covering fraction of $93\%$. Molecular hydrogen chemistry is not considered in the simulation, and so the real atomic hydrogen column densities should saturate around $10^{23} \, \percs$.}
  \label{fig:histo_CD}
\end{figure}

We show in $\rm{Fig.} \, \ref{fig:histo_CD}$ distributions of column densities from three of the seven directions of observation that we studied in $\rm{Sect.} \, \ref{sec:spectra}$. The pink histogram is typical of the majority of our directions, where the escape fraction is zero. All the photons are emitted behind optically thick gas, with $\Nhi>10^{20} \, \percs$. As there are no photons with $\tau_{910} \approx 1$, the covering fraction is well defined: it is equal to one and the escape fraction is zero. The orange histogram shows a clear bimodality, with a (double)-peak of the curve around $\Nhi = 10^{15}-10^{16} \, \percs$, which is the ``holes'' part, and another peak at high column densities, which is the ``cloud'' part. There is only a tiny fraction of $910 \, \angstrom$ photons that are facing gas with $\tau_{910} \approx 1$, and so this is also a case where the picket-fence model is a good description. However, the dark-blue histogram is an example of a situation where there is a significant fraction of photons facing gas such that $\tau_{910} \approx 1$. The photons are partially absorbed in this regime, which makes the notion of the covering fraction unclear. The cyan dashed line corresponds to the average of all 5184 directions, and it shows that the most common configurations have almost exclusively high column densities, and thus very low escape fractions.

\begin{figure}
  \resizebox{\hsize}{!}{\includegraphics{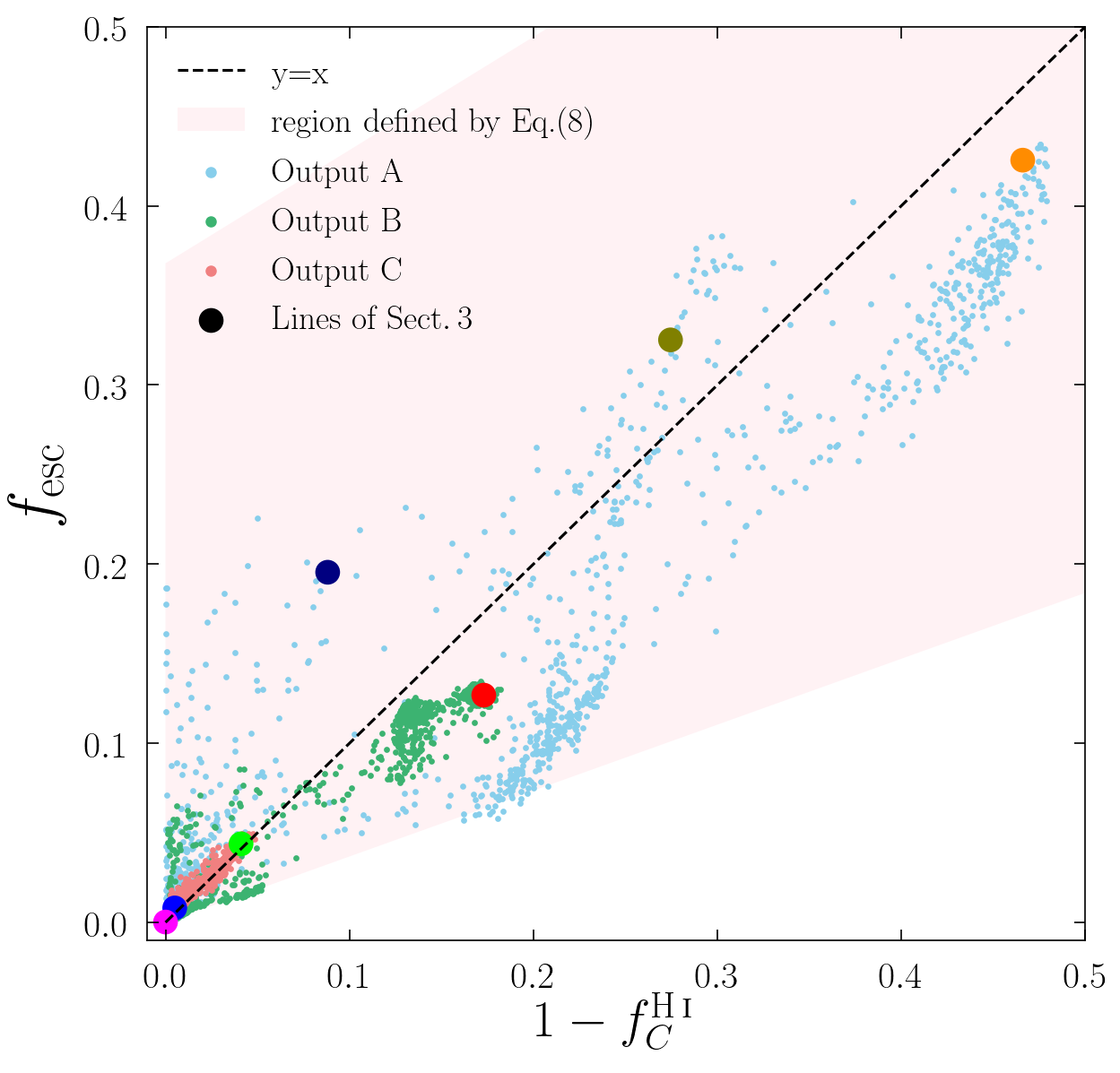}}
  \caption{Relation between the covering fraction of ionizing photons, defined by $\rm{Equation} \, \ref{eq:CF_def_HI}$, and the escape fraction of ionizing photons for the 1728 directions of observation of the three outputs, A, B and C. The seven colored dots are the directions that we study in $\rm{Sect.} \, \ref{sec:spectra}$. The pink area highlights the region of the figure that is allowed by our definition of covering fraction.}
  \label{fig:fesc_vs_CF}
\end{figure}

Let us define the covering fraction as the fraction of $910 \, \angstrom$ photons facing more than $\Nhi=10^{17.2} \, \percs$, which is the column density for which $\tau_{910} = 1$:
\be \label{eq:CF_def_HI}
\fchi = \frac{ \sum \limits_{N_{\hi}^{\rm{star}} \geq 10^{17.2} \, \percs} L_{\rm{star}}^{910} } { \sum \limits_{\rm{all \, stars}} L_{\rm{star}}^{910} },
\ee
where $N_{\hi}^{\rm{star}}$ is the column density of neutral hydrogen in front of a stellar particle and $L_{\rm{star}}^{910}$ is the luminosity of the stellar particle at $910 \, \angstrom$. We note that $\fchi$ depends on the direction of observation of the galaxy.
In an idealized picket-fence geometry with empty holes and optically thick clouds, we have the relation $\fesc = 1 - \fchi$. In the general case, photons that are covered in the sense of $\rm{Equation} \, \ref{eq:CF_def_HI}$ can still have a transmission factor of up to $e^{-1}=37\%$. Indeed, if all the gas has exactly $\Nhi = 10^{17.2}$, the photons are considered as covered, but the optical depth in this case is $\tau_{910}=1$, and so $37\%$ of photons can still escape. Similarly, photons that are not covered can still be absorbed; we can only say that their transmission factor is larger than $37\%$. The equation for the allowed region in the relation $\fesc \leftrightarrow (1-\fchi)$ is:
\be \label{eq:fesc_fchi}
(1-\fchi) \cdot e^{-1} < \fesc < \fchi \cdot e^{-1} + (1-\fchi).
\ee

$\rm{Figure} \, \ref{fig:fesc_vs_CF}$ displays the relation between $\fesc$ and $ 1 - \fchi$ for our 5184 mock observations and shows that indeed not all the directions are perfectly described by the picket-fence model, in the sense that we do not always have $\fesc = 1 - \fchi$. We note that a few points are below the relation given by $\rm{Eq.} \, \ref{eq:fesc_fchi}$ because of absorption by helium and dust that was not considered in this formula. The directions of observations corresponding to the pink and the orange histograms in $\rm{Fig.} \, \ref{fig:histo_CD}$ are close to the one-to-one relation in $\rm{Fig.} \, \ref{fig:fesc_vs_CF}$, as in the picket-fence geometry. This was expected because those directions have almost no stars facing gas such that $\tau_{910} \approx 1$. However, the dark-blue dot sits further from the one-to-one relation, which is explained by the significant fraction of photons facing gas with $\tau_{910} \approx 1$ in the corresponding histogram of $\rm{Fig.} \, \ref{fig:histo_CD}$. We see that overall the directions are close to the one-to-one relation. Therefore, the picket-fence geometry is on average a reasonable model of the covering of ionizing photons in our galaxy.

Here we considered the description of the escape of ionizing photons in the picket-fence framework and compared to the simulation, but the goal is to use absorption lines to infer this escape fraction. This leads to additional limitations due to the different optical depths of $\lyb$ and $\ciiline$ compared to $\lyc$, and the difference in distribution of $\rm{C}^+$ compared to $\hzero$. We now explain those new limitations for the two lines.

%...............................................................................
\subsubsection{The case of \texorpdfstring{$\lyb$}{Lg}} \label{sec:interpretation_lyb}

In order for the residual flux of $\lyb$ to be an indicator of the covering fraction, the ``holes'' in the picket-fence have to be optically thin for both $\lyc$ and $\lyb$. However the column density at which $\tau_{\lyb} =1$, with a turbulent velocity of $20 \, \kms$, is $\Nhi = 10^{14.22} \, \percs$, which is three orders of magnitude below the column density at which $\tau_{\lyc} = 1$. For example, the orange distribution of $\rm{Fig.} \, \ref{fig:histo_CD}$ represents a case where the picket-fence model works well to describe the escape of ionizing photons, in the sense that $\fesc = 1-\fchi$, but $\lyb$ residual flux is not a good indicator of this $\fesc$ because the part that is optically thin for $\lyc$ has $\Nhi$ of around $10^{15} \, - \, 10^{16} \, \percs$, which is already optically thick for $\lyb$. This is confirmed in the right panel of $\rm{Fig.} \, \ref{fig:scatter_all_dust}$, where the orange dot has a residual flux of only $7\%$, while the escape fraction is as high as $47\%$. The difficulty of the optical depth of $\lyb$ being larger than that of $\lyc$ cannot be circumvented. A saturated $\lyb$ line can very well correspond to a direction with a high escape fraction of ionizing photons. This is why $\lyb$ is a good lower limit of $\fesc$, but not a direct tracer.

%..............................................................................
\subsubsection{The case of \texorpdfstring{$\ciiline$}{Lg}} \label{sec:interpretation_CII}

We now analyze the optical depth of $\ciiline$ in comparison with the one of $910 \, \angstrom$ photons. The column density of $\rm{C}^+$ at which the optical depth of the center of $\ciiline$ is one for a turbulent velocity of $20 \, \kms$ is $N_{\cii}=10^{13.89} \, \percs$. For $910 \, \angstrom$ photons, it is $N_{\hi}=10^{17.19} \, \percs$, which means that if $N_{\cii}/N_{\hi} = 5 \times 10^{-4}$, the two optical depths are the same. This ratio is close to the ratio of carbon to hydrogen in the Sun, but metallicity and ionization effects can shift $N_{\cii}/N_{\hi}$ further from this value. 

\begin{figure}
  \resizebox{\hsize}{!}{\includegraphics{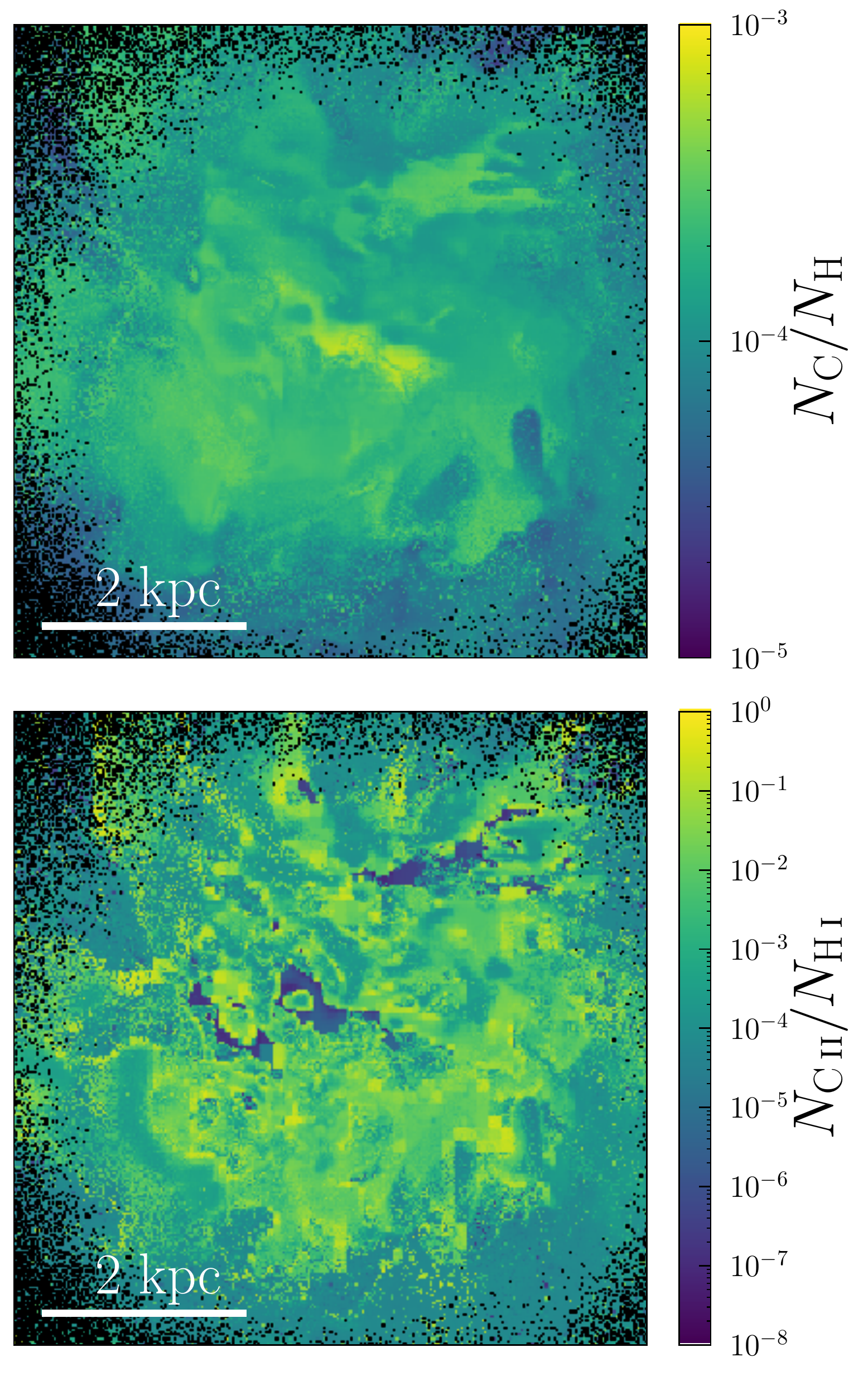}}
  \caption{Ratio of column densities in the same output and direction as in Figs.\ref{fig:164_visu} and \ref{fig:map_lum_all}. Upper panel: Ratio of carbon to hydrogen column densities. Lower panel: Ratio of $\rm{C}^+$ to$\hzero$ column densities. The column densities are computed from the stellar particles to the virial radius. In every pixel, the ratio is averaged over all stellar particles in this pixel and weighted by the luminosity of the stellar particle at $1334 \, \angstrom$. The black pixels have no stellar particles in them.}
  \label{fig:ratio_CII_HI}
\end{figure}

In $\rm{Fig.} \, \ref{fig:ratio_CII_HI},$ we show the ratio of carbon to hydrogen column densities in the upper panel and of $\rm{C}^+$ over $\hzero$ column densities in the lower panel for one direction of our simulated galaxy. We see that the $\rm{C}^+$ over $\hzero$ ratio varies by about eight orders of magnitude from region to region. However, the carbon-to-hydrogen ratio is more uniform, with an average of $1.5 \times 10^{-4}$, and varying by less than two orders of magnitude. Thus, the variation of the $\rm{C}^+$-to-$\hzero$ ratio comes primarily from ionization effects.
There are regions abundant in $\hzero$ but lacking $\rm{C}^+$, because most carbon is in the neutral state. In other regions, $\rm{C}^+$ is overabundant compared to the $N_C/N_H$ ratio, which occurs when hydrogen is mostly ionized and carbon is mainly in the $\rm{C}^+$ state, which is possible because the ionization energy of $\rm{C}^+$ is higher than that of $\hzero$.

For the continuum of stars behind screens with $N_{\cii}/N_{\hi} < 5 \times 10^{-4}$, the $\ciiline$ absorption is weaker than the absorption of ionizing photons, which leads to the residual flux of $\ciiline$ underestimating $\fesc$. Conversely, when stars are facing gas such that $N_{\cii}/N_{\hi} > 5 \times 10^{-4}$, the residual flux of $\ciiline$ overestimates $\fesc$. Therefore, the variation of the value of $N_{\cii}/N_{\hi}$ in the galaxy causes inevitable dispersion in the relation between the residual flux of $\ciiline$ and the escape fraction of ionizing photons. This dispersion changes with time and orientation, depending on the alignment between the brightest stars and the screens of gas with different $N_{\cii}/N_{\hi}$ ratios.

% -------------------------------------------------------------------------------
\subsection{Processes affecting the residual flux} \label{sec:interpretation_processes}

\begin{figure}
  \resizebox{\hsize}{!}{\includegraphics{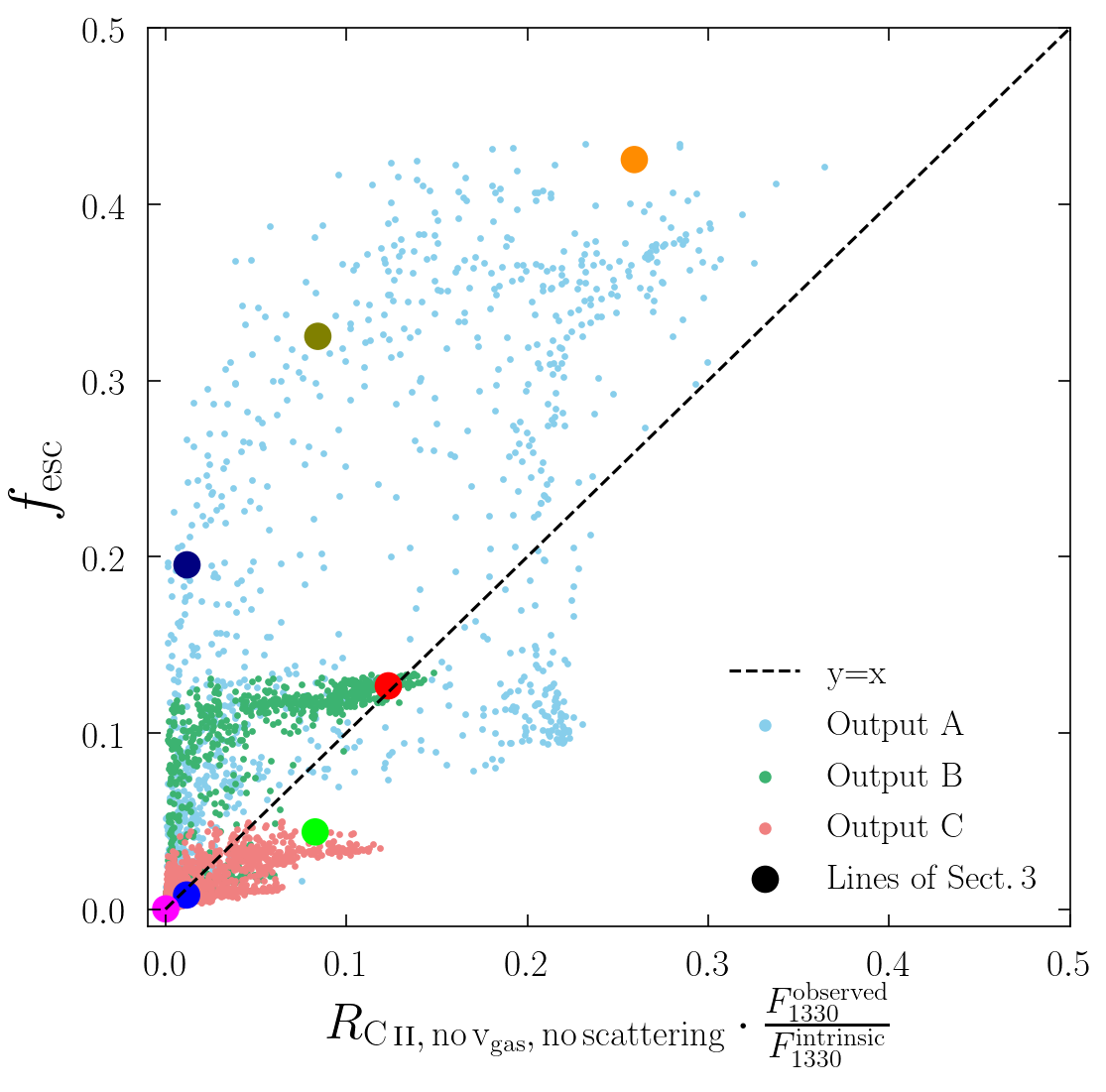}}
  \caption{As in the left panel of $\rm{Fig.} \, \ref{fig:scatter_all_dust}$, but omitting scattering and setting gas velocity to zero.}
  \label{fig:scatter_fesc_CII_nothing}
\end{figure}

Lastly, we highlight the extent to which processes during the radiative transfer of the line affect its properties; in particular the residual flux. We show in $\rm{Sect.} \,  \ref{sec:process}$ that the absorption lines are sensitive to the velocity distribution of the gas and sometimes to the effect of infilling, especially for directions with strong fluorescence. However, correcting for scattering and gas velocity dispersion does not improve the tightness of the relation between the residual flux of $\ciiline$ and the escape fraction of ionizing photons. $\rm{Figure} \, \ref{fig:scatter_fesc_CII_nothing}$ shows this relation with the scattering ignored and the gas velocity set to zero. We see that the relation has as much dispersion as in $\rm{Fig.} \, \ref{fig:scatter_all_dust}$. Moreover, the correction of scattering and gas velocity brings the points further away from the one-to-one relation, leading to $\ciiline$ being a lower limit of $\fesc$, similar to $\lyb$. In other words, $\ciiline$ globally saturates at column densities of gas that are too low for ionizing photons to be significantly absorbed, but the effect of infilling and the dispersion of gas velocity increase the residual flux of $\ciiline$ so that it gets closer to the value of $\fesc$.

% -------------------------------------------------------------------------------
\subsection{Another metallic line: \texorpdfstring{$\siiiline$}{Lg}} \label{sec:siii}

\begin{figure}
  \resizebox{\hsize}{!}{\includegraphics{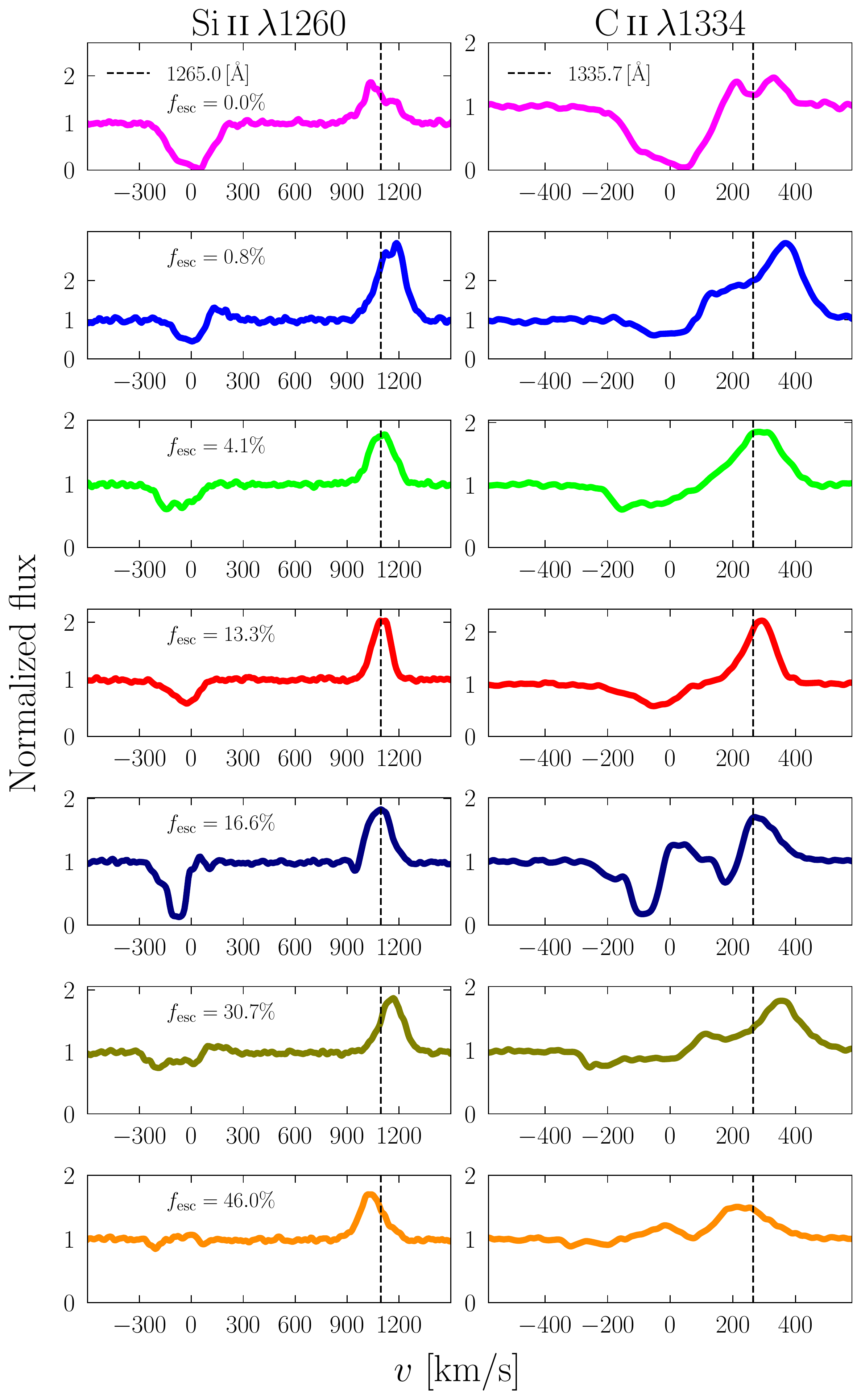}}
  \caption{Same as $\rm{Fig.} \, \ref{fig:spectra_fid}$ but with $\siiiline$ instead of $\lyb$. The vertical dashed lines highlight the wavelengths of the fluorescent channels.}
  \label{fig:spectra_fid_SiII}
\end{figure}

\begin{figure}
  \resizebox{\hsize}{!}{\includegraphics{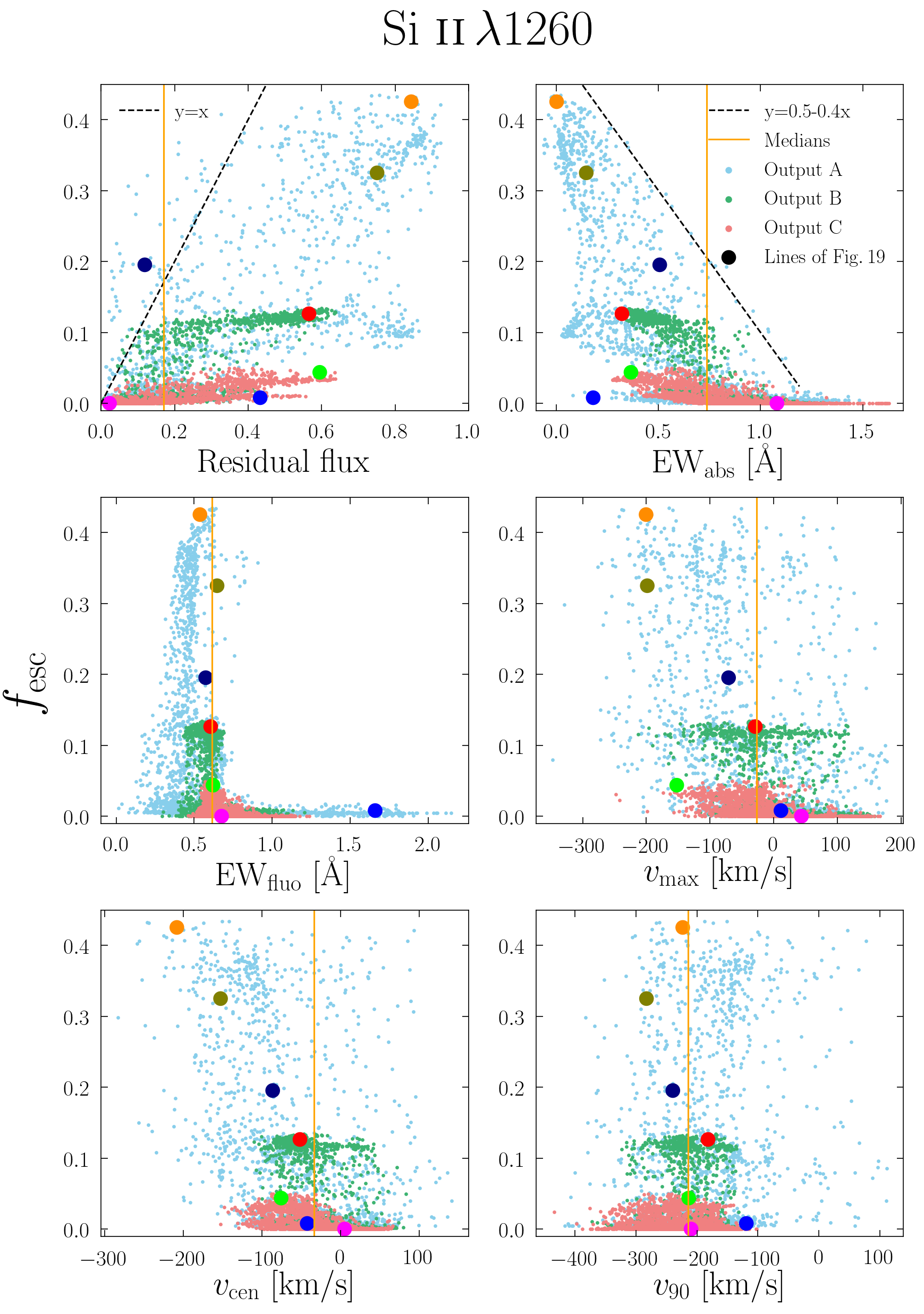}}
  \caption{Same as $\rm{Fig.} \, \ref{fig:scatter_fesc_all}$ but for $\siiiline$.}
  \label{fig:scatter_SiII}
\end{figure}

\begin{figure}
  \resizebox{\hsize}{!}{\includegraphics{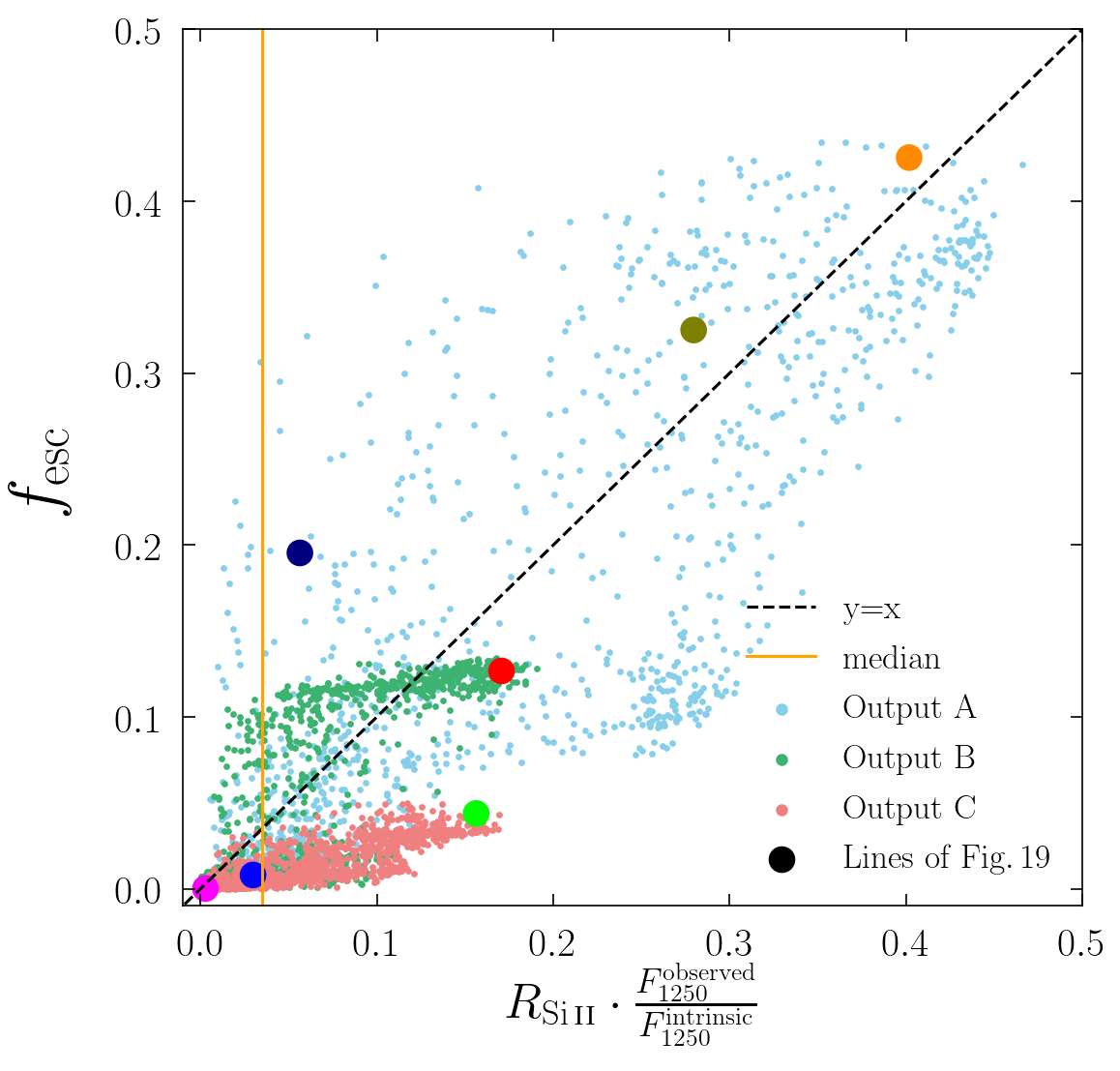}}
  \caption{Same as $\rm{Fig.} \, \ref{fig:scatter_all_dust}$ but for $\siiiline$.}
  \label{fig:scatter_SiII_dust}
\end{figure}

We find that $\ciiline$ traces the escape fractions of ionizing photons better than $\lyb$ does. We now show that all the results for $\ciiline$ apply to $\siiiline$, another LIS absorption line. The modeling of the distribution of $\rm{Si^+}$ is slightly less robust than that of $\rm{C}^+$, first because silicon depletes more onto dust, which we do not model here, and second because of charge transfer reactions ($\rm{Appendix} \, \ref{app:comp_fractions}$). However, the spectra are not very sensitive to a change in density, as shown in $\rm{Sect.} \, \ref{sec:comp_densities}$. First, $\rm{Fig.} \, \ref{fig:spectra_fid_SiII}$ shows the seven directions of observations that we selected in this paper, with the spectra of $\siiiline$ on the left and that of $\ciiline$ on the right. We see that they have similar properties: the residual flux and EWs are almost the same. This is confirmed in $\rm{Fig.} \, \ref{fig:scatter_SiII}$ where we plot the relation between the escape fractions and the different properties of the $\siiiline$ line. All the relations are similar to those seen in the $\cii$ plot in the left panel of $\rm{Fig.} \, \ref{fig:scatter_fesc_all}$. Finally, as $\rm{Fig.} \, \ref{fig:scatter_SiII_dust}$ shows, the relations after dust correction are also closer to the one-to-one relation.  $\siiiline$ is therefore as good a tracer of $\fesc$ as $\ciiline$, providing relatively good predictions when the (dust corrected) residual flux is large, but unfortunately predictions that often overestimate reality when the residual flux is smaller than $0.3$.

%%%%%%%%%%%%%%%%%%%%%%%%%%%%%%%%%%%%%%%%%%%%%%%%%%%%%%%%%%%%%%%%%%%%%%%%%%%%%%%%%
\section{Summary} \label{sec:summary}

 In this paper we present a method to post-process radiative hydrodynamics simulations in order to build mock observations of LIS UV absorption lines. We applied this method to a cosmological zoom-in simulation of a galaxy at $z=3$ with a stellar mass of $2.3\times 10^9 \, \rm{\Msun}$ in order to study the processes that affect the formation of absorption lines and to find correlations between the escape fraction of ionizing photons and the properties of the absorption lines. Our main results can be summarized as follows:

\begin{enumerate}
 \item Our procedure reproduces metal absorption lines with a diversity of shape and strength ($\rm{Fig.} \, \ref{fig:spectra_fid}$) that is similar to observations; such as for example those of \cite{Jaskot19}. The $\lyb$ absorption lines of our galaxy seem to differ from the ones in \cite{Steidel18} or \cite{Gazagnes20}, probably because our simulated galaxy is less massive and has a lower SFR. 
 
 \item Many properties of the observed radiation depend highly on the direction of observation. There are variations of up to more than one magnitude at $1500 \, \angstrom$ ($\rm{Table} \, \ref{table:outputs}$). The $\lyc$ escape fractions vary from $0\%$ to $47\%$ ($\rm{Fig.} \, \ref{fig:histo_fesc}$). The residual flux goes from $0\%$ to $95\%$ ($\rm{Fig.} \, \ref{fig:scatter_fesc_all}$).
 
\item The precise implementation of the radiation between 6 eV and 13.6 eV does not impact the mock absorption lines. There are dense regions where the carbon ionization fraction depends on this implementation, but they are very dusty and do not contribute to our spectra. However, using a UV background at all energies instead of the ionizing radiation from the simulation fails to reproduce the same observed spectra ($\rm{Fig.} \, \ref{fig:spectra_robust}$). This shows that radiative transfer of ionizing photons in the simulation is crucial for this kind of study.

\item The fluorescent lines $\ciistar$ and $\siiistar$ are clearly visible ($\rm{Fig.} \, \ref{fig:spectra_fid_SiII}$), with an EW of the same order as for the absorption. The shape and size of the fluorescent lines is sensitive to the modeling of the fine-level structure of the ground state of their respective ions ($\rm{Fig.} \, \ref{fig:spectra_CII_process_1}$). The presence of absorption by ions in the fine-level structure can lead to the formation of a P-Cygni profile in the fluorescence.

\item The infilling effect increases the residual fluxes by around $0.06$ on average. There are however directions where the effect is much stronger, with a change of up to more than $0.3$ ($\rm{Fig.} \, \ref{fig:resflux_processes}$). A fluorescent line with a large EW is an indicator that the infilling effect is stronger than on average ($\rm{Fig.} \, \ref{fig:noscatter_EW_fluo}$).  Infilling depends very little on the aperture size of the mock observation, as long as it encompasses more than $\approx 90 \%$ of the stellar continuum ($\rm{Figs.} \, \ref{fig:spectra_CII_process_1}$ and \ref{fig:spectra_lyb_process}).

\item The large-scale distribution of the gas velocity spreads the absorption over a larger wavelength range than if all the gas had the same velocity ($\rm{Figs.} \, \ref{fig:spectra_lyb_process}$ and \ref{fig:spectra_CII_process_2}). This spread leads to an increase in the residual flux of the lines ($\rm{Fig.} \, \ref{fig:resflux_processes}$), which is one factor that challenges the use of absorption lines as tracers of the escape fraction of ionizing photons. 
 
\item The properties of the absorption lines without dust correction correlate poorly with the $\lyc$ escape fraction ($\rm{Figs.} \, \ref{fig:scatter_fesc_all}$ and \ref{fig:scatter_SiII}). The residual fluxes of $\ciiline$ or $\siiiline$ cannot predict the escape fraction, but are good upper limits. A large EW of absorption, either of a metallic line or of $\lyb$, or a large EW of fluorescence, indicate an escape fraction smaller than $2\%$.
 
\item Multiplying the residual flux of $\ciiline$ or $\siiiline$ by the extinction of the continuum by dust makes it a better indicator of escape fraction, following globally the one-to-one relation, although with strong dispersion ($\rm{Figs.} \, \ref{fig:scatter_all_dust}$ and \ref{fig:scatter_SiII_dust}). This dust correction is similar to the ones made in the literature \citep[e.g.,][]{Steidel18,Gazagnes18,Chisholm18}. For directions with dust-corrected residual flux between $\sim 2 \%$ and $30 \%$, the escape fraction is often overestimated. Concerning $\lyb$, multiplying its residual flux by the extinction by dust makes it a good lower limit of the escape fraction, but not a direct proxy ($\rm{Fig.} \, \ref{fig:scatter_all_dust}$).

\item The covering fraction, defined as the fraction of photons that are facing $\Nhi\geq10^{17.2} \, \percs$, gives a reasonable estimate for the escape fraction of ionizing radiation for our simulated galaxies, although the scatter in the relation is not negligible ($\rm{Fig.} \, \ref{fig:fesc_vs_CF}$). 
 
\item Many geometrical and physical complexities make it difficult to use absorption lines as reliable tracers of the escape fraction. The different distributions of sources for ionizing photons and nonionizing UV continuum ($\rm{Fig.} \, \ref{fig:map_lum_all}$), the complex distribution of column densities in front of stars ($\rm{Fig.} \, \ref{fig:histo_CD}$), the different optical depth between the ionizing photons and the lines, the tangled distribution of $\rm{C}^+$ with respect to $\hzero$ ($\rm{Fig.} \, \ref{fig:ratio_CII_HI}$), and the effects of infilling and gas velocity distribution are not correctable and explain the dispersion in the scatter plots, and the fact that dust-corrected residual fluxes below ${\sim} 30 \%$ overestimate the escape fraction.

\end{enumerate}

In summary, metallic absorption lines can predict the escape fraction of ionizing photons accurately only in two extreme regimes:  when $\fesc = 0$, with saturated absorption lines, or when the escape fraction is very high, with a dust corrected residual flux larger than 30\%. As a result, even if it may be difficult to estimate the escape fraction of ionizing radiation from a galaxy, metallic absorption lines have the potential to rule out $\lyc$ emission from line-saturated galaxies, and to select promising candidates with high residual fluxes. 

We are happy to share the mock observations of our simulated galaxy and encourage those interested  to contact us.

%APPENDICES
%/////////////////////////////////////////////////////////////////////////////////////////////////////////////////////////////////////////////////////
\begin{appendix}

  %%%%%%%%%%%%%%%%%%%%%%%%%%%%%%%%%%%%%%%%%%%%%%%%%%%%%%%%%%%%%%%%%%%%%%%%%%%%%%%%%%%%
\section{Comparison of ionization fractions between \textsc{Krome} and Cloudy} \label{app:comp_fractions}
\begin{figure}
  \resizebox{\hsize}{!}{\includegraphics{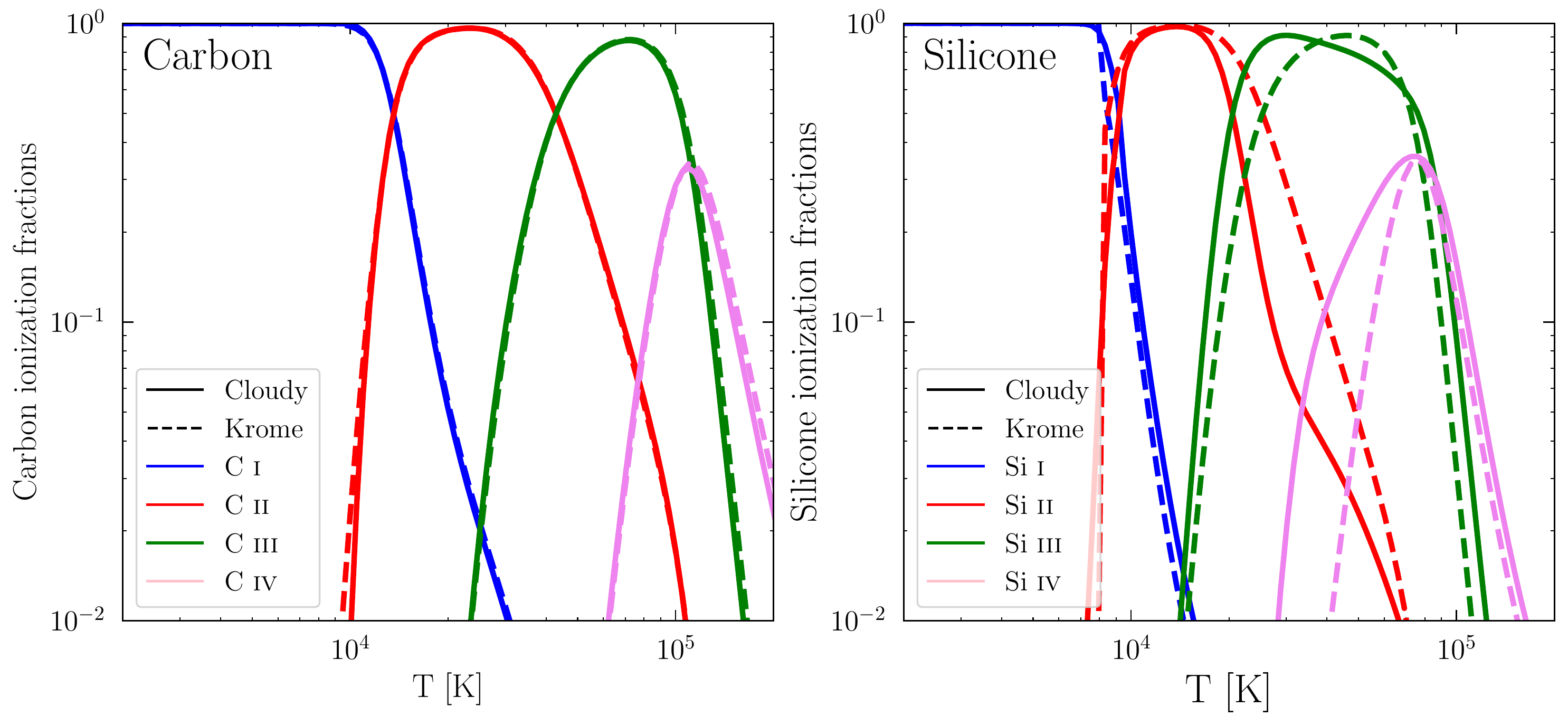}}
  \caption{Ionization fractions of carbon and silicon as a function of temperature. \textsc{Krome} reactions are recombinations from \cite{BadnellRR} and collisional ionization from \cite{Voronov97}.
  }
    \label{fig:all_ion}
\end{figure}

In $\rm{Fig.} \, \ref{fig:all_ion}$ we show a comparison of the ionization fractions of carbon and silicon as a function of temperature. We include here only recombinations and collisional ionizations, and no photo-reactions. The chemical network of \textsc{Krome} is chosen to match the results of Cloudy. We see that the results for carbon are coherent between the two codes. For silicon, the effects of charge transfer, which is included in Cloudy but not in \textsc{Krome}, are more important, and we do not recover the curves of Cloudy as well as with carbon. We still choose to omit those reactions in order to preserve the state of hydrogen and helium, as explained in $\rm{Sect.} \,  \ref{sec:comp_densities}$. 

  %%%%%%%%%%%%%%%%%%%%%%%%%%%%%%%%%%%%%%%%%%%%%%%%%%%%%%%%%%%%%%%%%%%%%%%%%%%%%%%%%%%%%
  \section{Computing photoionization rates} \label{app:photorate}
  
In \textsc{Krome}, the usual method to include photo-reactions is in three steps:
\begin{itemize}
\item Specify the photo-reaction in the chemical network, for example \ce{C+ + $\gamma$ -> C++ + e}.
\item Provide a file with a cross-section as a function of photon energy for this reaction.
\item Before calling \textsc{Krome} in a cell of a simulation (on-the-fly or in post-processing), specify a discretized spectral energy distribution.
\end{itemize}
\textsc{Krome} then computes the rate of the photo-reaction based on the cross-section and the radiation field, and uses it to compute the chemical evolution.
However, the photo-ionization rates of metallic ions can be easily computed from the \textsc{Ramses-RT} output (along with the modeling of the radiation below 13.6 eV, see $\rm{Sect.} \, \ref{sec:comp_densities}$) without using \textsc{Krome}, and so we bypass the method above, giving the more simple procedure:
\begin{itemize}
\item Specify the photo-reaction in the chemical network.
\item For every cell, compute the photoionization rate with the method below.
\item Give this photoionization rate to \textsc{Krome} via a routine that we added in the \textsc{Krome} code.
\end{itemize}

The photoionization rates are products of cross-sections and fluxes of photons integrated over frequency. The flux is given in every cell directly by the simulation. For cross-sections, we have to apply the same method as \textsc{Ramses-RT}, where the photoionization cross-sections are weighted by the SEDs of all stellar particles in the simulation. 

To begin with, we have a stellar SED library, BPASS in our case, with an SED for a grid of stellar ages and metallicities. For all SEDs, at each stellar age and metallicity in the grid, we compute the mean cross-section of photoionization for each bin of radiation of \textsc{Ramses-RT}, weighted by the number of photons, with this formula:
\be
\bar{\sigma}_{i, \rm{age},\rm{Z}}^X = \frac{\int_{\rm{bin} \: i} \, \frac{J_{\nu}^{\rm{age},\rm{Z}}}{h  \nu} \, \sigma_{\nu}^X \, \dd \nu}{\int_{\rm{bin} \: i} \, \frac{J_{\nu}^{\rm{age},\rm{Z}}}{h  \nu} \, \dd \nu},
\ee
where $\rm bin \: i$ is the $i^{\rm th}$ bin of radiation of the simulation, $X$ indicates an ion for which we compute the rate, $J_{\nu}^{\rm{age},\rm{Z}}$ is the luminosity as a function of frequency for a given stellar age and metallicity in BPASS, in units of $\rm erg \, s^{-1} \, Hz^{-1}$, $h\nu$ is the energy of a photon at frequency $\nu,$ and $\sigma_{\nu}^X$ is the photoionization cross-section as a function of frequency for species $X$, given by \cite{Verner96}. \\

From those mean cross-sections we compute the cross-sections of a given stellar particle, $\bar{\sigma}_{i,j_{\rm{star}}}^X$, by interpolating on age and metallicity. In a given domain of the simulation, where we want to compute the photoionization rates in every cell, we average those cross-sections on all stellar particles, weighing by their luminosities:
\be \label{eq:mean_sigma}
\bar{\sigma}_i^X = \frac{\sum_{j_{\rm{star}}=1}^{N_{\rm{stars}}} \, \bar{\sigma}_{i,j_{\rm{star}}}^X \cdot L_i^{j_{\rm{star}}}}{\sum_{j_{\rm{star}}=1}^{N_{\rm{stars}}} \, L_i^{j_{\rm{star}}}},
\ee
where $L_i^{j_{\rm{star}}}$ is the luminosity of the stellar particle $j_{\rm{star}}$ in the energy bin $i$. Finally, the photoionization rate in a given cell is:
\be
\Gamma_X^{\rm{cell}} [ \rm{s}^{-1} ] = \sum_{i=1}^{N_{\rm{bins}}} \bar{\sigma}_i^X \cdot F_i^{\rm{cell}},
\ee
where $F_i^{\rm cell}$ is the flux of photons in the cell, in units of $\rm{s^{-1} \, cm^{-2}}$, for the $i^{\rm th}$ bin of radiation.

\end{appendix}

\begin{acknowledgements}

The authors are grateful to the anonymous referee for her/his very careful and constructive report. We also thank J. Kerutt for her detailed revision of the manuscript. VM, AV and TG are supported by the ERC Starting Grant 757258 "TRIPLE". TK was supported in part by the National Research Foundation of Korea (NRF-2019K2A9A1A0609137711 and NRF-2020R1C1C100707911) and in part by the Yonsei University Future-leading Research Initiative (RMS2-2019-22-0216). This work was supported by the Programme National Cosmology et Galaxies (PNCG) of CNRS/INSU with INP and IN2P3, co-funded by CEA and CNES. The simulation could be run thanks to our access to computing resources at the Common Computing Facility (CCF) of the LABEX Lyon Institute of Origins (ANR-10-LABX-0066).

\end{acknowledgements}

\bibliography{references}

\begin{thebibliography}{115}
\expandafter\ifx\csname natexlab\endcsname\relax\def\natexlab#1{#1}\fi

\bibitem[{{Agertz} {et~al.}(2015){Agertz}, {Romeo}, \& {Grisdale}}]{Agertz15}
{Agertz}, O., {Romeo}, A.~B., \& {Grisdale}, K. 2015, \mnras, 449, 2156

\bibitem[{{Alexandroff} {et~al.}(2015){Alexandroff}, {Heckman}, {Borthakur},
  {Overzier}, \& {Leitherer}}]{Alexandroff15}
{Alexandroff}, R.~M., {Heckman}, T.~M., {Borthakur}, S., {Overzier}, R., \&
  {Leitherer}, C. 2015, \apj, 810, 104

\bibitem[{{Atek} {et~al.}(2015){Atek}, {Richard}, {Jauzac}, {Kneib},
  {Natarajan}, {Limousin}, {Schaerer}, {Jullo}, {Ebeling}, {Egami}, \&
  {Clement}}]{Atek15}
{Atek}, H., {Richard}, J., {Jauzac}, M., {et~al.} 2015, \apj, 814, 69

\bibitem[{{Ba{\~n}ados} {et~al.}(2018){Ba{\~n}ados}, {Venemans},
  {Mazzucchelli}, {Farina}, {Walter}, {Wang}, {Decarli}, {Stern}, {Fan},
  {Davies}, {Hennawi}, {Simcoe}, {Turner}, {Rix}, {Yang}, {Kelson}, {Rudie}, \&
  {Winters}}]{2018Natur.553..473B}
{Ba{\~n}ados}, E., {Venemans}, B.~P., {Mazzucchelli}, C., {et~al.} 2018, \nat,
  553, 473

\bibitem[{{Badnell}(2006)}]{BadnellRR}
{Badnell}, N.~R. 2006, \apjs, 167, 334

\bibitem[{{Barrow} {et~al.}(2018){Barrow}, {Wise}, {Aykutalp}, {O'Shea},
  {Norman}, \& {Xu}}]{Barrow18}
{Barrow}, K. S.~S., {Wise}, J.~H., {Aykutalp}, A., {et~al.} 2018, \mnras, 474,
  2617

\bibitem[{{Barrow} {et~al.}(2017){Barrow}, {Wise}, {Norman}, {O'Shea}, \&
  {Xu}}]{Barrow17}
{Barrow}, K. S.~S., {Wise}, J.~H., {Norman}, M.~L., {O'Shea}, B.~W., \& {Xu},
  H. 2017, \mnras, 469, 4863

\bibitem[{{Bassett} {et~al.}(2019){Bassett}, {Ryan-Weber}, {Cooke}, {Diaz},
  {Nanayakkara}, {Yuan}, {Spitler}, {Me{\v{s}}tri{\'c}}, {Garel}, {Sawicki},
  {Gwyn}, \& {Golob}}]{Bassett19}
{Bassett}, R., {Ryan-Weber}, E.~V., {Cooke}, J., {et~al.} 2019, \mnras, 483,
  5223

\bibitem[{{Bergvall} {et~al.}(2006){Bergvall}, {Zackrisson}, {Andersson},
  {Arnberg}, {Masegosa}, \& {{\"O}stlin}}]{Bergvall06}
{Bergvall}, N., {Zackrisson}, E., {Andersson}, B.~G., {et~al.} 2006, \aap, 448,
  513

\bibitem[{{Borthakur} {et~al.}(2014){Borthakur}, {Heckman}, {Leitherer}, \&
  {Overzier}}]{Borthakur14}
{Borthakur}, S., {Heckman}, T.~M., {Leitherer}, C., \& {Overzier}, R.~A. 2014,
  Science, 346, 216

\bibitem[{{Bosman} {et~al.}(2018){Bosman}, {Fan}, {Jiang}, {Reed}, {Matsuoka},
  {Becker}, \& {Haehnelt}}]{Bosman18}
{Bosman}, S. E.~I., {Fan}, X., {Jiang}, L., {et~al.} 2018, \mnras, 479, 1055

\bibitem[{{Carr} {et~al.}(2018){Carr}, {Scarlata}, {Panagia}, \&
  {Henry}}]{Carr18}
{Carr}, C., {Scarlata}, C., {Panagia}, N., \& {Henry}, A. 2018, \apj, 860, 143

\bibitem[{{Cen} \& {Kimm}(2015)}]{Cen15}
{Cen}, R. \& {Kimm}, T. 2015, \apjl, 801, L25

\bibitem[{{Chisholm} {et~al.}(2018){Chisholm}, {Gazagnes}, {Schaerer},
  {Verhamme}, {Rigby}, {Bayliss}, {Sharon}, {Gladders}, \&
  {Dahle}}]{Chisholm18}
{Chisholm}, J., {Gazagnes}, S., {Schaerer}, D., {et~al.} 2018, \aap, 616, A30

\bibitem[{{Chisholm} {et~al.}(2017){Chisholm}, {Orlitov{\'a}}, {Schaerer},
  {Verhamme}, {Worseck}, {Izotov}, {Thuan}, \& {Guseva}}]{Chisholm17}
{Chisholm}, J., {Orlitov{\'a}}, I., {Schaerer}, D., {et~al.} 2017, \aap, 605,
  A67

\bibitem[{{Corlies} {et~al.}(2020){Corlies}, {Peeples}, {Tumlinson}, {O'Shea},
  {Lehner}, {Howk}, {O'Meara}, \& {Smith}}]{Corlies20}
{Corlies}, L., {Peeples}, M.~S., {Tumlinson}, J., {et~al.} 2020, \apj, 896, 125

\bibitem[{{De Cia}(2018)}]{DeCia18}
{De Cia}, A. 2018, \aap, 613, L2

\bibitem[{{Dere}(2007)}]{Dere07}
{Dere}, K.~P. 2007, \aap, 466, 771

\bibitem[{{Dessauges-Zavadsky} {et~al.}(2010){Dessauges-Zavadsky}, {D'Odorico},
  {Schaerer}, {Modigliani}, {Tapken}, \& {Vernet}}]{Mirka10}
{Dessauges-Zavadsky}, M., {D'Odorico}, S., {Schaerer}, D., {et~al.} 2010, \aap,
  510, A26

\bibitem[{{Dijkstra}(2014)}]{Dijkstra14}
{Dijkstra}, M. 2014, \pasa, 31, e040

\bibitem[{{Dijkstra}(2017)}]{Dijkstra_Saas}
{Dijkstra}, M. 2017, arXiv e-prints, arXiv:1704.03416

\bibitem[{{Dijkstra} {et~al.}(2016){Dijkstra}, {Gronke}, \&
  {Venkatesan}}]{Dijkstra16}
{Dijkstra}, M., {Gronke}, M., \& {Venkatesan}, A. 2016, \apj, 828, 71

\bibitem[{{Eldridge} {et~al.}(2008){Eldridge}, {Izzard}, \& {Tout}}]{BPASS1}
{Eldridge}, J.~J., {Izzard}, R.~G., \& {Tout}, C.~A. 2008, \mnras, 384, 1109

\bibitem[{{Erb}(2015)}]{Erb15}
{Erb}, D.~K. 2015, \nat, 523, 169

\bibitem[{{Fan} {et~al.}(2006){Fan}, {Carilli}, \& {Keating}}]{Fan06}
{Fan}, X., {Carilli}, C.~L., \& {Keating}, B. 2006, \araa, 44, 415

\bibitem[{{Faucher-Gigu{\`e}re} {et~al.}(2009){Faucher-Gigu{\`e}re}, {Lidz},
  {Zaldarriaga}, \& {Hernquist}}]{UVB_simu}
{Faucher-Gigu{\`e}re}, C.-A., {Lidz}, A., {Zaldarriaga}, M., \& {Hernquist}, L.
  2009, \apj, 703, 1416

\bibitem[{{Feltre} {et~al.}(2018){Feltre}, {Bacon}, {Tresse}, {Finley},
  {Carton}, {Blaizot}, {Bouch{\'e}}, {Garel}, {Inami}, {Boogaard},
  {Brinchmann}, {Charlot}, {Chevallard}, {Contini}, {Michel-Dansac}, {Mahler},
  {Marino}, {Maseda}, {Richard}, {Schmidt}, \& {Verhamme}}]{Feltre18}
{Feltre}, A., {Bacon}, R., {Tresse}, L., {et~al.} 2018, \aap, 617, A62

\bibitem[{{Feltre} {et~al.}(2020){Feltre}, {Maseda}, {Bacon}, {Pradeep},
  {Leclercq}, {Kusakabe}, {Wisotzki}, {Hashimoto}, {Schmidt}, {Blaizot},
  {Brinchmann}, {Boogaard}, {Cantalupo}, {Carton}, {Inami}, {Kollatschny},
  {Marino}, {Matthee}, {Nanayakkara}, {Richard}, {Schaye}, {Tresse}, {Urrutia},
  {Verhamme}, \& {Weilbacher}}]{Feltre20}
{Feltre}, A., {Maseda}, M.~V., {Bacon}, R., {et~al.} 2020, \aap, 641, A118

\bibitem[{{Ferland} {et~al.}(2017){Ferland}, {Chatzikos}, {Guzm{\'a}n},
  {Lykins}, {van Hoof}, {Williams}, {Abel}, {Badnell}, {Keenan}, {Porter}, \&
  {Stancil}}]{Cloudy}
{Ferland}, G.~J., {Chatzikos}, M., {Guzm{\'a}n}, F., {et~al.} 2017, \rmxaa, 53,
  385

\bibitem[{{Finkelstein} {et~al.}(2019){Finkelstein}, {D'Aloisio},
  {Paardekooper}, {Ryan}, {Behroozi}, {Finlator}, {Livermore}, {Upton
  Sanderbeck}, {Dalla Vecchia}, \& {Khochfar}}]{Finkelstein19}
{Finkelstein}, S.~L., {D'Aloisio}, A., {Paardekooper}, J.-P., {et~al.} 2019,
  \apj, 879, 36

\bibitem[{{Finley} {et~al.}(2017){Finley}, {Bouch{\'e}}, {Contini}, {Paalvast},
  {Boogaard}, {Maseda}, {Bacon}, {Blaizot}, {Brinchmann}, {Epinat}, {Feltre},
  {Marino}, {Muzahid}, {Richard}, {Schaye}, {Verhamme}, {Weilbacher}, \&
  {Wisotzki}}]{Finley17}
{Finley}, H., {Bouch{\'e}}, N., {Contini}, T., {et~al.} 2017, \aap, 608, A7

\bibitem[{{Gazagnes} {et~al.}(2020){Gazagnes}, {Chisholm}, {Schaerer},
  {Verhamme}, \& {Izotov}}]{Gazagnes20}
{Gazagnes}, S., {Chisholm}, J., {Schaerer}, D., {Verhamme}, A., \& {Izotov}, Y.
  2020, \aap, 639, A85

\bibitem[{{Gazagnes} {et~al.}(2018){Gazagnes}, {Chisholm}, {Schaerer},
  {Verhamme}, {Rigby}, \& {Bayliss}}]{Gazagnes18}
{Gazagnes}, S., {Chisholm}, J., {Schaerer}, D., {et~al.} 2018, \aap, 616, A29

\bibitem[{{Giallongo} {et~al.}(2019){Giallongo}, {Grazian}, {Fiore}, {Kodra},
  {Urrutia}, {Castellano}, {Cristiani}, {Dickinson}, {Fontana}, {Menci},
  {Pentericci}, {Boutsia}, {Newman}, \& {Puccetti}}]{Giallongo19}
{Giallongo}, E., {Grazian}, A., {Fiore}, F., {et~al.} 2019, \apj, 884, 19

\bibitem[{{Gnedin}(2016)}]{Gnedin16}
{Gnedin}, N.~Y. 2016, \apjl, 825, L17

\bibitem[{{Gnedin} {et~al.}(2008){Gnedin}, {Kravtsov}, \& {Chen}}]{Gnedin08}
{Gnedin}, N.~Y., {Kravtsov}, A.~V., \& {Chen}, H.-W. 2008, \apj, 672, 765

\bibitem[{{Grassi} {et~al.}(2014){Grassi}, {Bovino}, {Schleicher}, {Prieto},
  {Seifried}, {Simoncini}, \& {Gianturco}}]{2014MNRAS.439.2386G}
{Grassi}, T., {Bovino}, S., {Schleicher}, D.~R.~G., {et~al.} 2014, \mnras, 439,
  2386

\bibitem[{{Grevesse} {et~al.}(2010){Grevesse}, {Asplund}, {Sauval}, \&
  {Scott}}]{Grevesse10}
{Grevesse}, N., {Asplund}, M., {Sauval}, A.~J., \& {Scott}, P. 2010, \apss,
  328, 179

\bibitem[{{Grimes} {et~al.}(2009){Grimes}, {Heckman}, {Aloisi}, {Calzetti},
  {Leitherer}, {Martin}, {Meurer}, {Sembach}, \& {Strickland}}]{Grimes09}
{Grimes}, J.~P., {Heckman}, T., {Aloisi}, A., {et~al.} 2009, \apjs, 181, 272

\bibitem[{{Grissom} {et~al.}(2014){Grissom}, {Ballantyne}, \&
  {Wise}}]{Grissom14}
{Grissom}, R.~L., {Ballantyne}, D.~R., \& {Wise}, J.~H. 2014, \aap, 561, A90

\bibitem[{{Gronke} {et~al.}(2020){Gronke}, {Ocvirk}, {Mason}, {Matthee},
  {Bosman}, {Sorce}, {Lewis}, {Ahn}, {Aubert}, {Dawoodbhoy}, {Iliev},
  {Shapiro}, \& {Yepes}}]{Gronke20}
{Gronke}, M., {Ocvirk}, P., {Mason}, C., {et~al.} 2020, arXiv e-prints,
  arXiv:2004.14496

\bibitem[{{Guillet} \& {Teyssier}(2011)}]{Guillet11}
{Guillet}, T. \& {Teyssier}, R. 2011, Journal of Computational Physics, 230,
  4756

\bibitem[{{Haardt} \& {Madau}(2012)}]{HM12}
{Haardt}, F. \& {Madau}, P. 2012, \apj, 746, 125

\bibitem[{{Habing}(1968)}]{Habing68}
{Habing}, H.~J. 1968, \bain, 19, 421

\bibitem[{{Hahn} \& {Abel}(2013)}]{MUSIC}
{Hahn}, O. \& {Abel}, T. 2013, {MUSIC: MUlti-Scale Initial Conditions}

\bibitem[{{Heckman} {et~al.}(2015){Heckman}, {Alexandroff}, {Borthakur},
  {Overzier}, \& {Leitherer}}]{Heckman15}
{Heckman}, T.~M., {Alexandroff}, R.~M., {Borthakur}, S., {Overzier}, R., \&
  {Leitherer}, C. 2015, \apj, 809, 147

\bibitem[{{Heckman} {et~al.}(2011){Heckman}, {Borthakur}, {Overzier},
  {Kauffmann}, {Basu-Zych}, {Leitherer}, {Sembach}, {Martin}, {Rich},
  {Schiminovich}, \& {Seibert}}]{Heckman11}
{Heckman}, T.~M., {Borthakur}, S., {Overzier}, R., {et~al.} 2011, \apj, 730, 5

\bibitem[{{Heckman} {et~al.}(2001){Heckman}, {Sembach}, {Meurer}, {Leitherer},
  {Calzetti}, \& {Martin}}]{Heckman01}
{Heckman}, T.~M., {Sembach}, K.~R., {Meurer}, G.~R., {et~al.} 2001, \apj, 558,
  56

\bibitem[{{Henry} {et~al.}(2018){Henry}, {Berg}, {Scarlata}, {Verhamme}, \&
  {Erb}}]{Henry18}
{Henry}, A., {Berg}, D.~A., {Scarlata}, C., {Verhamme}, A., \& {Erb}, D. 2018,
  \apj, 855, 96

\bibitem[{{Henyey} \& {Greenstein}(1941)}]{Henyey41}
{Henyey}, L.~G. \& {Greenstein}, J.~L. 1941, \apj, 93, 70

\bibitem[{{Hummels} {et~al.}(2017){Hummels}, {Smith}, \& {Silvia}}]{Hummels17}
{Hummels}, C.~B., {Smith}, B.~D., \& {Silvia}, D.~W. 2017, \apj, 847, 59

\bibitem[{{Iliev} {et~al.}(2014){Iliev}, {Mellema}, {Ahn}, {Shapiro}, {Mao}, \&
  {Pen}}]{Iliev14}
{Iliev}, I.~T., {Mellema}, G., {Ahn}, K., {et~al.} 2014, \mnras, 439, 725

\bibitem[{{Inoue} {et~al.}(2018){Inoue}, {Hasegawa}, {Ishiyama}, {Yajima},
  {Shimizu}, {Umemura}, {Konno}, {Harikane}, {Shibuya}, {Ouchi}, {Shimasaku},
  {Ono}, {Kusakabe}, {Higuchi}, \& {Lee}}]{2018PASJ...70...55I}
{Inoue}, A.~K., {Hasegawa}, K., {Ishiyama}, T., {et~al.} 2018, \pasj, 70, 55

\bibitem[{{Izotov} {et~al.}(2016{\natexlab{a}}){Izotov}, {Orlitov{\'a}},
  {Schaerer}, {Thuan}, {Verhamme}, {Guseva}, \& {Worseck}}]{Izotov16a}
{Izotov}, Y.~I., {Orlitov{\'a}}, I., {Schaerer}, D., {et~al.}
  2016{\natexlab{a}}, \nat, 529, 178

\bibitem[{{Izotov} {et~al.}(2016{\natexlab{b}}){Izotov}, {Schaerer}, {Thuan},
  {Worseck}, {Guseva}, {Orlitov{\'a}}, \& {Verhamme}}]{Izotov16b}
{Izotov}, Y.~I., {Schaerer}, D., {Thuan}, T.~X., {et~al.} 2016{\natexlab{b}},
  \mnras, 461, 3683

\bibitem[{{Izotov} {et~al.}(2018){Izotov}, {Worseck}, {Schaerer}, {Guseva},
  {Thuan}, {Fricke}, \& {Orlitov{\'a}}}]{Izotov18b}
{Izotov}, Y.~I., {Worseck}, G., {Schaerer}, D., {et~al.} 2018, \mnras, 478,
  4851

\bibitem[{{Jaskot} {et~al.}(2019){Jaskot}, {Dowd}, {Oey}, {Scarlata}, \&
  {McKinney}}]{Jaskot19}
{Jaskot}, A.~E., {Dowd}, T., {Oey}, M.~S., {Scarlata}, C., \& {McKinney}, J.
  2019, \apj, 885, 96

\bibitem[{{Jaskot} \& {Oey}(2013)}]{Jaskot13}
{Jaskot}, A.~E. \& {Oey}, M.~S. 2013, \apj, 766, 91

\bibitem[{{Jaskot} \& {Oey}(2014)}]{Jaskot14}
{Jaskot}, A.~E. \& {Oey}, M.~S. 2014, \apjl, 791, L19

\bibitem[{{Jenkins}(2009)}]{Jenkins09}
{Jenkins}, E.~B. 2009, \apj, 700, 1299

\bibitem[{{Jones} {et~al.}(2012){Jones}, {Stark}, \& {Ellis}}]{Jones12}
{Jones}, T., {Stark}, D.~P., \& {Ellis}, R.~S. 2012, \apj, 751, 51

\bibitem[{{Jones} {et~al.}(2013){Jones}, {Ellis}, {Schenker}, \&
  {Stark}}]{Jones13}
{Jones}, T.~A., {Ellis}, R.~S., {Schenker}, M.~A., \& {Stark}, D.~P. 2013,
  \apj, 779, 52

\bibitem[{{Katz} {et~al.}(2019){Katz}, {Galligan}, {Kimm}, {Rosdahl},
  {Haehnelt}, {Blaizot}, {Devriendt}, {Slyz}, {Laporte}, \& {Ellis}}]{Katz19}
{Katz}, H., {Galligan}, T.~P., {Kimm}, T., {et~al.} 2019, \mnras, 487, 5902

\bibitem[{{Katz} {et~al.}(2020){Katz}, {{\v{D}}urov{\v{c}}{\'\i}kov{\'a}},
  {Kimm}, {Rosdahl}, {Blaizot}, {Haehnelt}, {Devriendt}, {Slyz}, {Ellis}, \&
  {Laporte}}]{Katz20}
{Katz}, H., {{\v{D}}urov{\v{c}}{\'\i}kov{\'a}}, D., {Kimm}, T., {et~al.} 2020,
  \mnras, 498, 164

\bibitem[{{Kimm} {et~al.}(2019){Kimm}, {Blaizot}, {Garel}, {Michel-Dansac},
  {Katz}, {Rosdahl}, {Verhamme}, \& {Haehnelt}}]{Kimm19}
{Kimm}, T., {Blaizot}, J., {Garel}, T., {et~al.} 2019, \mnras, 486, 2215

\bibitem[{{Kimm} {et~al.}(2017){Kimm}, {Katz}, {Haehnelt}, {Rosdahl},
  {Devriendt}, \& {Slyz}}]{Kimm17}
{Kimm}, T., {Katz}, H., {Haehnelt}, M., {et~al.} 2017, \mnras, 466, 4826

\bibitem[{{Kimm} {et~al.}(2011){Kimm}, {Slyz}, {Devriendt}, \&
  {Pichon}}]{Kimm10}
{Kimm}, T., {Slyz}, A., {Devriendt}, J., \& {Pichon}, C. 2011, \mnras, 413, L51

\bibitem[{{Kretschmer} \& {Teyssier}(2020)}]{Kretschmer20}
{Kretschmer}, M. \& {Teyssier}, R. 2020, \mnras, 492, 1385

\bibitem[{{Kulkarni} {et~al.}(2019){Kulkarni}, {Keating}, {Haehnelt}, {Bosman},
  {Puchwein}, {Chardin}, \& {Aubert}}]{Kulkarni19}
{Kulkarni}, G., {Keating}, L.~C., {Haehnelt}, M.~G., {et~al.} 2019, \mnras,
  485, L24

\bibitem[{{Laursen} {et~al.}(2009){Laursen}, {Sommer-Larsen}, \&
  {Andersen}}]{Laursen09b}
{Laursen}, P., {Sommer-Larsen}, J., \& {Andersen}, A.~C. 2009, \apj, 704, 1640

\bibitem[{{Leitet} {et~al.}(2013){Leitet}, {Bergvall}, {Hayes}, {Linn{\'e}}, \&
  {Zackrisson}}]{Leitet13}
{Leitet}, E., {Bergvall}, N., {Hayes}, M., {Linn{\'e}}, S., \& {Zackrisson}, E.
  2013, \aap, 553, A106

\bibitem[{{Leitherer} {et~al.}(2016){Leitherer}, {Hernandez}, {Lee}, \&
  {Oey}}]{Leitherer16}
{Leitherer}, C., {Hernandez}, S., {Lee}, J.~C., \& {Oey}, M.~S. 2016, \apj,
  823, 64

\bibitem[{{Levermore}(1984)}]{Levermore84}
{Levermore}, C.~D. 1984, \jqsrt, 31, 149

\bibitem[{{Li} \& {Draine}(2001)}]{Li01}
{Li}, A. \& {Draine}, B.~T. 2001, \apj, 554, 778

\bibitem[{{Livermore} {et~al.}(2017){Livermore}, {Finkelstein}, \&
  {Lotz}}]{Livermore17}
{Livermore}, R.~C., {Finkelstein}, S.~L., \& {Lotz}, J.~M. 2017, \apj, 835, 113

\bibitem[{{Luridiana} {et~al.}(2015){Luridiana}, {Morisset}, \& {Shaw}}]{Pyneb}
{Luridiana}, V., {Morisset}, C., \& {Shaw}, R.~A. 2015, \aap, 573, A42

\bibitem[{{Ma} {et~al.}(2020){Ma}, {Quataert}, {Wetzel}, {Hopkins},
  {Faucher-Gigu{\`e}re}, \& {Kere{\v{s}}}}]{Ma20}
{Ma}, X., {Quataert}, E., {Wetzel}, A., {et~al.} 2020, \mnras, 498, 2001

\bibitem[{{Madau} \& {Haardt}(2015)}]{Madau15}
{Madau}, P. \& {Haardt}, F. 2015, \apjl, 813, L8

\bibitem[{{Mason} {et~al.}(2018){Mason}, {Treu}, {de Barros}, {Dijkstra},
  {Fontana}, {Mesinger}, {Pentericci}, {Trenti}, \& {Vanzella}}]{Mason18b}
{Mason}, C.~A., {Treu}, T., {de Barros}, S., {et~al.} 2018, \apjl, 857, L11

\bibitem[{{Michel-Dansac} {et~al.}(2020){Michel-Dansac}, {Blaizot}, {Garel},
  {Verhamme}, {Kimm}, \& {Trebitsch}}]{Rascas}
{Michel-Dansac}, L., {Blaizot}, J., {Garel}, T., {et~al.} 2020, \aap, 635, A154

\bibitem[{{Mitchell} {et~al.}(2020){Mitchell}, {Blaizot}, {Cadiou}, \&
  {Dubois}}]{Mitchell20}
{Mitchell}, P., {Blaizot}, J., {Cadiou}, C., \& {Dubois}, Y. 2020, arXiv
  e-prints, arXiv:2008.12790

\bibitem[{{Nakajima} \& {Ouchi}(2014)}]{Nakajima14}
{Nakajima}, K. \& {Ouchi}, M. 2014, \mnras, 442, 900

\bibitem[{{Ocvirk} {et~al.}(2016){Ocvirk}, {Gillet}, {Shapiro}, {Aubert},
  {Iliev}, {Teyssier}, {Yepes}, {Choi}, {Sullivan}, {Knebe}, {Gottl{\"o}ber},
  {D'Aloisio}, {Park}, {Hoffman}, \& {Stranex}}]{Ocvirk16}
{Ocvirk}, P., {Gillet}, N., {Shapiro}, P.~R., {et~al.} 2016, \mnras, 463, 1462

\bibitem[{{Ouchi} {et~al.}(2018){Ouchi}, {Harikane}, {Shibuya}, {Shimasaku},
  {Taniguchi}, {Konno}, {Kobayashi}, {Kajisawa}, {Nagao}, {Ono}, {Inoue},
  {Umemura}, {Mori}, {Hasegawa}, {Higuchi}, {Komiyama}, {Matsuda}, {Nakajima},
  {Saito}, \& {Wang}}]{2018PASJ...70S..13O}
{Ouchi}, M., {Harikane}, Y., {Shibuya}, T., {et~al.} 2018, \pasj, 70, S13

\bibitem[{{Pallottini} {et~al.}(2019){Pallottini}, {Ferrara}, {Decataldo},
  {Gallerani}, {Vallini}, {Carniani}, {Behrens}, {Kohandel}, \&
  {Salvadori}}]{Pallottini19}
{Pallottini}, A., {Ferrara}, A., {Decataldo}, D., {et~al.} 2019, \mnras, 487,
  1689

\bibitem[{{Parsa} {et~al.}(2018){Parsa}, {Dunlop}, \& {McLure}}]{Parsa18}
{Parsa}, S., {Dunlop}, J.~S., \& {McLure}, R.~J. 2018, \mnras, 474, 2904

\bibitem[{{Patr{\'\i}cio} {et~al.}(2016){Patr{\'\i}cio}, {Richard}, {Verhamme},
  {Wisotzki}, {Brinchmann}, {Turner}, {Christensen}, {Weilbacher}, {Blaizot},
  {Bacon}, {Contini}, {Lagattuta}, {Cantalupo}, {Cl{\'e}ment}, \&
  {Soucail}}]{Patricio16}
{Patr{\'\i}cio}, V., {Richard}, J., {Verhamme}, A., {et~al.} 2016, \mnras, 456,
  4191

\bibitem[{{Peeples} {et~al.}(2019){Peeples}, {Corlies}, {Tumlinson}, {O'Shea},
  {Lehner}, {O'Meara}, {Howk}, {Earl}, {Smith}, {Wise}, \&
  {Hummels}}]{Peeples19}
{Peeples}, M.~S., {Corlies}, L., {Tumlinson}, J., {et~al.} 2019, \apj, 873, 129

\bibitem[{{Prochaska} {et~al.}(2011){Prochaska}, {Kasen}, \&
  {Rubin}}]{Prochaska11}
{Prochaska}, J.~X., {Kasen}, D., \& {Rubin}, K. 2011, \apj, 734, 24

\bibitem[{{Puschnig} {et~al.}(2017){Puschnig}, {Hayes}, {{\"O}stlin},
  {Rivera-Thorsen}, {Melinder}, {Cannon}, {Menacho}, {Zackrisson}, {Bergvall},
  \& {Leitet}}]{Puschnig17}
{Puschnig}, J., {Hayes}, M., {{\"O}stlin}, G., {et~al.} 2017, \mnras, 469, 3252

\bibitem[{{Reddy} {et~al.}(2016){Reddy}, {Steidel}, {Pettini},
  {Bogosavljevi{\'c}}, \& {Shapley}}]{Reddy16}
{Reddy}, N.~A., {Steidel}, C.~C., {Pettini}, M., {Bogosavljevi{\'c}}, M., \&
  {Shapley}, A.~E. 2016, \apj, 828, 108

\bibitem[{{Rivera-Thorsen} {et~al.}(2015){Rivera-Thorsen}, {Hayes},
  {{\"O}stlin}, {Duval}, {Orlitov{\'a}}, {Verhamme}, {Mas-Hesse}, {Schaerer},
  {Cannon}, {Ot{\'\i}-Floranes}, {Sand berg}, {Guaita}, {Adamo}, {Atek},
  {Herenz}, {Kunth}, {Laursen}, \& {Melinder}}]{Rivera15}
{Rivera-Thorsen}, T.~E., {Hayes}, M., {{\"O}stlin}, G., {et~al.} 2015, \apj,
  805, 14

\bibitem[{{Robertson} {et~al.}(2015){Robertson}, {Ellis}, {Furlanetto}, \&
  {Dunlop}}]{Robertson15}
{Robertson}, B.~E., {Ellis}, R.~S., {Furlanetto}, S.~R., \& {Dunlop}, J.~S.
  2015, \apjl, 802, L19

\bibitem[{{Rosdahl} {et~al.}(2013){Rosdahl}, {Blaizot}, {Aubert}, {Stranex}, \&
  {Teyssier}}]{Joki2013}
{Rosdahl}, J., {Blaizot}, J., {Aubert}, D., {Stranex}, T., \& {Teyssier}, R.
  2013, \mnras, 436, 2188

\bibitem[{{Rosdahl} {et~al.}(2018){Rosdahl}, {Katz}, {Blaizot}, {Kimm},
  {Michel-Dansac}, {Garel}, {Haehnelt}, {Ocvirk}, \& {Teyssier}}]{Sphinx}
{Rosdahl}, J., {Katz}, H., {Blaizot}, J., {et~al.} 2018, \mnras, 479, 994

\bibitem[{{Rosdahl} \& {Teyssier}(2015)}]{RamsesRT}
{Rosdahl}, J. \& {Teyssier}, R. 2015, \mnras, 449, 4380

\bibitem[{{Scarlata} \& {Panagia}(2015)}]{Scarlata15}
{Scarlata}, C. \& {Panagia}, N. 2015, \apj, 801, 43

\bibitem[{{Schroeder} {et~al.}(2013){Schroeder}, {Mesinger}, \&
  {Haiman}}]{Schroeder13}
{Schroeder}, J., {Mesinger}, A., \& {Haiman}, Z. 2013, \mnras, 428, 3058

\bibitem[{{Shapley} {et~al.}(2003){Shapley}, {Steidel}, {Pettini}, \&
  {Adelberger}}]{Shapley03}
{Shapley}, A.~E., {Steidel}, C.~C., {Pettini}, M., \& {Adelberger}, K.~L. 2003,
  \apj, 588, 65

\bibitem[{{Smail} {et~al.}(2007){Smail}, {Swinbank}, {Richard}, {Ebeling},
  {Kneib}, {Edge}, {Stark}, {Ellis}, {Dye}, {Smith}, \& {Mullis}}]{Smail07}
{Smail}, I., {Swinbank}, A.~M., {Richard}, J., {et~al.} 2007, \apjl, 654, L33

\bibitem[{{Smith} {et~al.}(2015){Smith}, {Safranek-Shrader}, {Bromm}, \&
  {Milosavljevi{\'c}}}]{Smith_Colt}
{Smith}, A., {Safranek-Shrader}, C., {Bromm}, V., \& {Milosavljevi{\'c}}, M.
  2015, \mnras, 449, 4336

\bibitem[{{Sobacchi} \& {Mesinger}(2015)}]{2015MNRAS.453.1843S}
{Sobacchi}, E. \& {Mesinger}, A. 2015, \mnras, 453, 1843

\bibitem[{{Stanway} {et~al.}(2016){Stanway}, {Eldridge}, \& {Becker}}]{BPASS2}
{Stanway}, E.~R., {Eldridge}, J.~J., \& {Becker}, G.~D. 2016, \mnras, 456, 485

\bibitem[{{Steidel} {et~al.}(2018){Steidel}, {Bogosavljevi{\'c}}, {Shapley},
  {Reddy}, {Rudie}, {Pettini}, {Trainor}, \& {Strom}}]{Steidel18}
{Steidel}, C.~C., {Bogosavljevi{\'c}}, M., {Shapley}, A.~E., {et~al.} 2018,
  \apj, 869, 123

\bibitem[{{Steidel} {et~al.}(2010){Steidel}, {Erb}, {Shapley}, {Pettini},
  {Reddy}, {Bogosavljevi{\'c}}, {Rudie}, \& {Rakic}}]{Steidel10}
{Steidel}, C.~C., {Erb}, D.~K., {Shapley}, A.~E., {et~al.} 2010, \apj, 717, 289

\bibitem[{{Teyssier}(2002)}]{Ramses}
{Teyssier}, R. 2002, \aap, 385, 337

\bibitem[{{Toro} {et~al.}(1994){Toro}, {Spruce}, \& {Speares}}]{Toro94}
{Toro}, E.~F., {Spruce}, M., \& {Speares}, W. 1994, Shock Waves, 4, 25

\bibitem[{{Trebitsch} {et~al.}(2017){Trebitsch}, {Blaizot}, {Rosdahl},
  {Devriendt}, \& {Slyz}}]{Trebitsch17}
{Trebitsch}, M., {Blaizot}, J., {Rosdahl}, J., {Devriendt}, J., \& {Slyz}, A.
  2017, \mnras, 470, 224

\bibitem[{{Trebitsch} {et~al.}(2020){Trebitsch}, {Dubois}, {Volonteri},
  {Pfister}, {Cadiou}, {Katz}, {Rosdahl}, {Kimm}, {Pichon}, {Beckmann},
  {Devriendt}, \& {Slyz}}]{Obelisk}
{Trebitsch}, M., {Dubois}, Y., {Volonteri}, M., {et~al.} 2020, arXiv e-prints,
  arXiv:2002.04045

\bibitem[{{Vasei} {et~al.}(2016){Vasei}, {Siana}, {Shapley}, {Quider}, {Alavi},
  {Rafelski}, {Steidel}, {Pettini}, \& {Lewis}}]{Vasei16}
{Vasei}, K., {Siana}, B., {Shapley}, A.~E., {et~al.} 2016, \apj, 831, 38

\bibitem[{{Verhamme} {et~al.}(2015){Verhamme}, {Orlitov{\'a}}, {Schaerer}, \&
  {Hayes}}]{Verhamme15}
{Verhamme}, A., {Orlitov{\'a}}, I., {Schaerer}, D., \& {Hayes}, M. 2015, \aap,
  578, A7

\bibitem[{{Verhamme} {et~al.}(2017){Verhamme}, {Orlitov{\'a}}, {Schaerer},
  {Izotov}, {Worseck}, {Thuan}, \& {Guseva}}]{Verhamme17}
{Verhamme}, A., {Orlitov{\'a}}, I., {Schaerer}, D., {et~al.} 2017, \aap, 597,
  A13

\bibitem[{{Verner} {et~al.}(1996){Verner}, {Ferland}, {Korista}, \&
  {Yakovlev}}]{Verner96}
{Verner}, D.~A., {Ferland}, G.~J., {Korista}, K.~T., \& {Yakovlev}, D.~G. 1996,
  \apj, 465, 487

\bibitem[{{Voronov}(1997)}]{Voronov97}
{Voronov}, G.~S. 1997, Atomic Data and Nuclear Data Tables, 65, 1

\bibitem[{{Yoo} {et~al.}(2020){Yoo}, {Kimm}, \& {Rosdahl}}]{Yoo20}
{Yoo}, T., {Kimm}, T., \& {Rosdahl}, J. 2020, \mnras, 499, 5175

\end{thebibliography}
\bibliographystyle{aa} % style aa.bst

\end{document}